\documentclass[reprint,amsfonts,amsmath,amssymb,aps,pre,floatfix,10pt]{revtex4-1}

\usepackage{graphicx}
\usepackage{amsbsy}
\usepackage{latexsym}
\usepackage{color}
\usepackage{graphicx}
\usepackage{psfrag}
\usepackage[normalem]{ulem}
\usepackage{bm}
\usepackage{hyperref}
\usepackage{float}
\usepackage{overpic}

\usepackage[T1]{fontenc}

\newcommand\dd{\mathrm{d}}

\newcommand\x{\mathbf{x}}
\newcommand\n{\mathbf{n}}
\newcommand\Q{\mathbf{Q}}
\newcommand\Qvec{\mathbf{Q}}

\begin{document}
\title{Dispersion of Multiferroic Nanoparticles in a Bent-Core Nematic Liquid Crystal: Experimental and Theoretical Study}  
\author{Dhananjoy Mandal$^1$}
\author{Yiwei Wang$^2$}
\author{Supreet Kaur$^3$}
\author{Golam Mohiuddin$^4$}
\author{Apala Majumdar$^5$}
\author{Aloka Sinha$^1$}
\affiliation{$^1$Department of Physics, Indian Institute of Technology Delhi, New Delhi 110016 India}
\affiliation{$^2$Department of Mathematics, University of California, Riverside, Riverside, CA 92521, United States}
\affiliation{$^3$Department of Chemical Sciences, Indian Institute of Science Education and Research (IISER) Mohali, Sector-81, Knowledge City, Manauli 140306, India}
\affiliation{$^4$Department of Chemistry, University of Science $\And$ Technology Meghalaya, Ri$-$Bhoi, Meghalaya  793101 
India}
\affiliation{$^5$Department of Mathematics and Statistics, University of Strathclyde, Glasgow G11XQ, United Kingdom}

\begin{abstract}

A novel nanocomposite system has been prepared by dispersing multiferroic bismuth ferrite nanoparticles (BiFeO$_3$) in a bent-core nematic liquid crystal (8-F-OH) that exhibits cybotactic clusters. Transition temperature, optical textures, order parameter (\emph{S$_{m}$}), and dielectric spectroscopy experiments are performed in the doped system, and the results are compared with the pure one. The main experimental outcome is that the doped system has increased orientational order parameters, even though the cybotactic cluster size is reduced due to the incorporation of multiferroic BiFeO$_3$ nanoparticles. The transition temperature, as observed under a polarising optical microscope, clearly indicates a reduction of $1 - 2~ ^\circ{\rm C}$ in the doped system compared to the pure one, and we conjecture this is due to the disordering of the cybotactic cluster in the doped system. Based on the experimental findings, a Landau-de Gennes-type free energy model is developed. The model qualitatively explains the increased mean order parameter and the disordering of cybotactic clusters with increasing polarization value of nanoparticles. This is corroborated by experimental findings.  

\end{abstract}

\maketitle

\section{introduction}
Nematic liquid crystals (LCs) are anisotropic materials of nature, combining fluidity with the long-range orientational ordering of solid crystals. The nematic phase naturally has distinguished material directions, referred to as \emph{directors}, yielding direction-dependent optical and physical properties \cite{de1993physics_2}. Consequently, nematics have been hugely successful working materials for modern-day electro-optic and display devices \cite{collings2019introduction_1, de1993physics_2}. Bent-core liquid crystals are a new addition to the family of thermotropic liquid crystals; bent-core liquid crystals have less translational freedom owing to the bent shape of the molecules and the aromatic interactions, resulting in steric hindrance \cite{marino2012dielectric_3, ghosh2014ferroelectric_4}. Due to the bent shape of the molecules and transverse dipole moment, some of the molecules stack together in smectic-like ordered clusters, called cybotactic clusters, observed in the nematic and isotropic phases \cite{bailey2009rheological_5}. The locally polar cybotactic clusters are responsible for some interesting properties, such as biaxiality, macroscopic polarization, and ferroelectric-like fast switching \cite{takezoe2006bent_6, francescangeli2014cybotactic_7, keith2010nematic_8, kumar2018ferroelectric_9} of the bent-core nematic phase, usually not found in the calamitic nematic phase. There have been relatively few studies on the effects of nanodoping and external fields on the locally polar cybotactic clusters of bent-core nematic liquid crystals \cite{kumar2018quantum_10,khan2017elastic_11,kumar2018nanodoping_12}. Such studies can be promising for manipulating cybotactic clusters for novel technological applications.

Nano-doping is a well-known method for tailoring the physical properties of a liquid crystal matrix in a non-synthetic way. Ferroelectric nanoparticles exhibit spontaneous electric polarization (dipole moment per unit volume) and switching behaviour under an applied electric field. The dipole moment of the ferroelectric nanoparticle creates a local anisotropic electric field around the nanoparticle. The large local electric field produced by the ferroelectric nanoparticle polarises the nearby nematic molecules, inducing local dipoles, and this effective interaction is proportional to the square of the orientational order parameter \cite{li2006orientational_14}. There is a local order parameter associated with the molecules surrounding the ferroelectric nanoparticle, and this can enhance the overall order parameter and increase the isotropic to a nematic transition temperature \cite{lopatina2009theory_13}. Multiferroic nanoparticles have received recent attention due to their unique properties. Multiferroic materials, such as polycrystalline bismuth ferrite, exhibit ferroelectric and long-range antiferromagnetic ordering simultaneously \cite{kalinin2002potential_15,seidel2009conduction_16}. It has been reported that the incorporation of multiferroic BiFeO$_{3}$ nanoparticles in a liquid crystal can improve the electro-optic response of calamitic nematic and ferroelectric liquid crystals \cite{nayek2015superior_17, khan2020bismuth_18, ghosh2011effect_19}. In \cite{khan2017elastic_11}, a low concentration of BaTiO$_{3}$ nanoparticles are dispersed in a bent-core nematic liquid crystal and the dielectric and elastic properties of the pure and doped system are compared \cite{khan2017elastic_11}. The authors report that the doped system has positive elastic anisotropy whereas the pure system has negative elastic anisotropy \cite{khan2017elastic_11}. In \cite{derbali2020dielectric_33}, dispersion of ferroelectric LiNbO$_{3}$ nanoparticles in the cybotactic nematic phase suppresses the polarization of the nematic phase due to anti-parallel dipole configuration.

In this article, we synthesize bismuth ferrite (BiFeO$_3$) nanoparticles with size of 30 nm and estimate ferroelectric polarization of about 48 $\mu {\rm C}$/${\rm cm}^2$, with a very weak surface ferromagnetic moment \cite{castillo2013effect_20, Dhananjoy2023dielectric_21, selbach2007size_22, karpinsky2017thermodynamic_23}. Without an externally applied magnetic field, the inherent magnetic field generated by the magnetic dipoles of the nanoparticles is weak.  
Therefore, the magnetic moment of the multiferroic BiFeO$_3$ nanoparticle cannot affect the properties of the liquid crystal matrix notably \cite{reznikov2017ferromagnetic_24, emdadi2018behaviour_25, emdadi2018investigation_26}. We discard the magnetic moment of the nanoparticle for our work. It is assumed that within the nanoparticle, the electric polarization vector of multiferroic BiFeO$_3$ nanoparticles is fixed in one direction \cite{emdadi2018investigation_26}. The strong ferroelectric moment of the synthesized nanoparticle interacts with the dipole moment of the liquid crystal and has the potential to alter the order parameter of the host liquid crystal matrix through its local electric field.  It is reported that if the bent-core nematic liquid crystal molecule contains an electronegative group as a terminal unit, then the sample exhibits a large polarization value  \cite{mohiuddin2017observation_28,hiremath2016supramolecular_29}. The 8-F-OH compound also has electronegative fluorine ($-F$) as one terminal moiety \cite{zhang2022electric_27}. The studied bent-core compound (8-F-OH) exhibits a nematic phase with cybotactic clusters over a wide temperature range ($> 80 ^\circ {\rm C} $) and has an electric field-induced polar response with a macroscopic polarization $\sim$185 ${\rm nC/cm}^2$ \cite{zhang2022electric_27, kaur2023polar}. Such a large value of polarization is surely an indication of the strong local transverse dipole moment of cybotactic clusters \cite{ghosh2014ferroelectric_4, kumar2018nanodoping_12, vita2018polar_30}.

We prepare a suspension of 0.2 $ wt\%$ multiferroic bismuth ferrite (BiFeO$_3$) nanoparticle in a bent-core nematic liquid crystal (8-F-OH) sample. The experimental observations/measurements of transition temperatures, color textures, dielectric spectroscopy measurements, birefringence measurements, and order parameter calculations are performed and compared, for the pure and doped systems respectively, primarily to obtain a better understanding of the temperature-dependent effects of nano-doping on cybotactic clusters. It is observed that after nano-doping with multiferroic nanoparticles, the doped system has increased birefringence and order parameters compared to the pure system. Dielectric spectroscopy is a powerful and versatile technique to understand the growth of cybotactic clusters with temperature in the isotropic and the nematic phase of bent-core liquid crystal \cite{panarin2018formation_31}. Careful analysis of dielectric data in this article indicates that the number of molecules in the cybotactic cluster, or the cluster size, is significantly reduced in the doped system compared to the pure system. The dielectric strength and activation energy (\textit{E$_a$}) of the collective dielectric mode is also decreased in the doped system. These observations can be explained by possible interactions between locally polar cybotactic clusters and the multiferroic nanoparticles, specifically, the antiparallel dipole-dipole interaction between clusters and the nanoparticle domains. Collectively, the experiments suggest that the doped system has increased order parameters and reduced cluster size, compared to the pure sample.
 
We develop a mathematical model to complement the experiments. The nematic phase of bent-core liquid crystal is described by a two-state model, i.e., by two order parameters; the first order parameter, $S_g$, describes the state of ordering outside the cybotactic clusters, and the second order parameter, $S_c$ corresponds to the ordering within the cybotactic clusters with a coupling term between the ambient molecules and the clusters, captured by a phenomenological parameter $\gamma$ \cite{madhusudana2017two_36}. In Ref. \cite{patranabish2019one_37}, the authors generalize the model in \cite{madhusudana2017two_36} to account for the effects of spatial inhomogeneities and confinement on the order parameter profiles. In \cite{patranabish2021quantum_38}, the free energy in \cite{patranabish2019one_37} is embellished to incorporate the effects of dispersed quantum dots (QDs), and the model captures the experimentally observed trends of reduced order parameter and cluster size on QD-doping \cite{patranabish2021quantum_38}. In this study, we build on the work in \cite{madhusudana2017two_36} and \cite{patranabish2019one_37} to account for the effect of nano-doping on $S_g$ and $S_c$. Following previous work in the literature, we describe the effects of the dispersed nanoparticles in terms of a nanoparticle order parameter, $S_{NP}$, and there are additional interaction terms in the free energy between $S_{NP}, S_g$ and $S_c$. The mathematical model qualitatively captures the experimentally observed increment in the mean order parameter along with reduced cybotactic cluster order parameters ($S_c$) effectively, induced by large polarization values of the suspended nanoparticles. Our model is far from comprehensive but works at a grass-root level, and could be foundational for further rigorous work on the study of prototype order-disordered systems.

There have been very successful theories of dispersion of ferroelectric nanoparticles in calamitic nematic liquid crystal \cite{lopatina2009theory_13}. However, the existing literature on ferroelectric nano-doped bent-core systems has not given a clear foundation of what happens to the cybotactic cluster order and out-of-cluster order upon dispersion of ferroelectric nanoparticles in the cybotactic nematic phase. Here in our manuscript, we are giving a systematic experimental and theoretical foundation for some types of ferroelectric nano-doped bent core systems and the experimental and theoretical results are in tandem. The outcome of the present work (simultaneous ordering of out-of-cluster order and disordering of cybotactic cluster order upon ferroelectric nanoparticle dispersion in bent-core nematic liquid crystals) surely adds to the understanding of the ferroelectric nanoparticle dispersed two-state cybotactic nematic phase. Here, lies the novelty of the presented work.

\section{Experiments}
\subsection{Preparation of nanocomposites}
The nanocomposite is prepared with the bent-core nematic liquid crystal 8-F-OH and multiferroic bismuth ferrite (BiFeO$_3$) nanoparticle. The molecular structure and dipole moment of the 8-F-OH molecule is given in Ref. \cite{zhang2022electric_27}. Synthesis, phase confirmation, particle size, and polarization value of prepared BiFeO$_3$ nanoparticle are given in Ref. \cite{Dhananjoy2023dielectric_21}.  Initially, a suitable amount of LC compound is taken, and with this appropriate amount of BiFeO$_3$ NP is added such that the concentration of NP in the nanocomposites is 0.2 $wt\%$. Chloroform is used as a solvent. The mixture is ultrasonicated and shaken well for homogeneous dispersion of NPs. The prepared composite is kept at a higher temperature to evaporate the extra solvent and kept overnight at room temperature for complete evaporation of the solvent \cite{kumar2018nanodoping_12}. The nanocomposite does not contain any visible signature of aggregation of NP after the preparation procedures.

\subsection{Performed experiments}

The pure LC and the doped LC nanocomposites are filled in two different ITO-coated planar Instec cells (Instec Inc.) of thickness 5 $\mu m$ by capillary action in the isotropic phase. The heating-cooling cycle of the samples is performed using Instec HCS302 hot stage and MK1000 temperature controller (Instec). All measurements are performed when the samples are cooled from the isotropic to the nematic phase. The transition temperature and optical textures are recorded using a polarizing optical microscope (OLYMPUS BX-51P) with crossed polarizer analyzer configuration. The temperature controller (Instec. MK1000) has an accuracy of ±0.01 ºC. So, the experimental error in transition temperature is estimated to be ±0.01 ºC. The optical birefringence and order parameters are calculated using the optical transmission method. The transmission method is performed by keeping the sample between two crossed Glan-Thomson polarisers with the rubbing direction of liquid crystal molecules, making an angle $45^\circ$ with both polariser's pass axes for maximum output intensity \cite{chakraborty2015effect_39}. The sample is illuminated with a He-Ne laser having a wavelength of $\sim$ 633 $\mathrm{nm}$. The output power of the optical transmission experiment is measured using a Gentec PH100-Si-HAOD1 photodetector attached to a Gentec MAESTRO power meter. The sensitivity of the power meter as given in the user manual is 0.37958 A/W at wavelength 633 nm with a minimum range of 3.00 nW and a maximum range of 1 W. Finally, dielectric measurements are performed using an Agilent E4980A LCR meter in the frequency range of 20 Hz-2 MHz with 0.1 V applied RMS voltage.

\section{Results and Discussions}

\begin{figure*}[!t]
    \centering
        \includegraphics[width=1\linewidth]{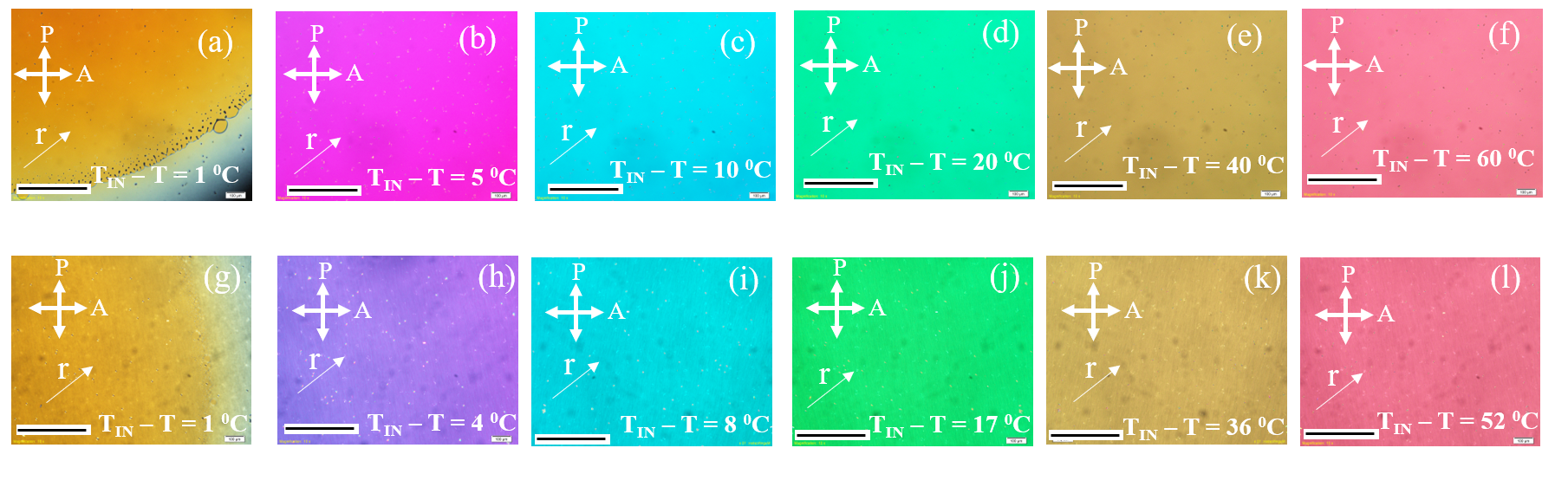}
    \caption{Birefringent color textures with temperature (reduced temperature T$_{IN}$ – T, T$_{IN}$ is isotropic to nematic transition temperature, T is the temperature at which texture is recorded) variation for pure bent-core 8-F-OH in (a)-(f) and 0.2 $wt\%$ BiFeO$_3$ dispersed 8-F-OH in (g)-(l), respectively, during the cooling process. Here, r represents the rubbing direction and the scale bar at the lower left side of each textures indicates a distance of 400 $\mu m$.}
    \label{1}
\end{figure*}

\subsection{Optical textures and transition temperatures}

The optical textures and the transition temperatures of liquid crystal phases and its composites are recorded while cooling the samples from the isotropic phase, using a polarizing optical microscope and temperature controller \cite{kumar2018quantum_10, khan2017elastic_11, kumar2018nanodoping_12}. The transition temperature and phase sequences of the pure 8-F-OH \cite{kaur2023polar} and doped 0.2 $wt\%$ BiFeO$_3$ nanoparticle dispersed 8-F-OH liquid crystal are shown below:
\begin{itemize}
\item Pure (8-F-OH): 

I $\xrightarrow{150.6 ^{\circ}{\rm C}}$  N$_{cyb}$ $\xrightarrow{72^{\circ}{\rm C}}$   Mx $\xrightarrow{55.6 ^{\circ}{\rm C}}$  Cr

\item Doped (0.2 $wt\%$ BiFeO$_3$+8-F-OH): 

I $\xrightarrow{149.3 ^{\circ}{\rm C}}$  N$_{cyb}$ $\xrightarrow{71.2 ^{\circ}{\rm C}}$  Mx $\xrightarrow{58.2 ^{\circ}{\rm C}}$ Cr
\end{itemize}
Here, "I" represents the isotropic phase, "N$_{cyb}$" refers to the nematic phase with cybotactic clusters, "Mx" denotes an unidentified liquid crystal phase comprising elongated cybotactic clusters, and "Cr" represents the crystalline phase.

The isotropic to nematic transition temperature in the doped system (T$_{\rm IN-Doped}$=149.3 $^{\circ}{\rm C}$) is decreased compared to the pure (T$_{\rm IN-Pure}$=150.6 $^{\circ}{\rm C}$) one. However, the total temperature range of the nematic phase of the doped system remains almost the same, but it is shifted to a lower temperature by $\sim$ 1$^{\circ}{\rm C}$ – 2$^{\circ}{\rm C}$. Even though there is an increment in the overall order parameter and an expected increment in isotropic to the nematic transition temperature \cite{lopatina2009theory_13}, the observed decrement in isotropic to nematic transition temperature in the doped system compared to the pure one can be explained by the fact that the phase stability of the nematic phase in the doped system is decreased due to decreased coupling between nematic director fluctuation and disordered smectic type cluster order \cite{mertelj2012critical}.

To understand this well, the local electric field (\textit{$\vec{E}$}) around the ferroelectric nanoparticle is \cite{nayek2015superior_17, basu2014soft_40}:
\begin{equation}
\label{local_field}
\vec{E}=\frac{PR^3}{3\epsilon_0 r^3}(2\cos \theta \hat{r} + \sin \theta \hat{\theta})
\end{equation}
Here \textit{P} is the polarization of the nanoparticle, which is the dipole moment per unit volume, \textit{R} is the radius of the nanoparticle, \textit{r} is the magnitude of position vector at which the electric field is calculated, and $\theta$ is the angle between the polarization vector (\textit{P}) and position vector (\textit{r}). The dipole moment of the nanoparticle is $p$ = $(4/3)\pi R^3P$. The electric field value at a few nano-meters from the surface of the nanoparticle is about $\sim 10^{10}$ $V/m$, which is almost four orders of magnitude larger than the average external electric field applied to a liquid crystal cell \cite{basu2014soft_40}. The nanoparticle creates a local electric field, and the molecular dipole moments align with the field, thus, creating an anisotropic nanoparticle domain \cite{kumar2018nanodoping_12, lopatina2009theory_13} as shown in Fig. 10 (b).

The nanoparticle domain will have different anisotropic properties compared to the bulk, due to the strong local electric field \cite{basu2014soft_40}. In our study, this effect will be pronounced because of the large polarization value of nanoparticles. We conjecture that there is anti-parallel dipole-dipole coupling between the nanoparticle domains and the cybotactic clusters, and this produces a certain disordering effect and for large values of the polarization of the nanoparticle or large nanoparticle-induced local anisotropy, this disordering effect dominates the ordering effects of the nanoparticle and the isotropic-nematic transition temperature decreases slightly. To support our conjecture, it is reported that the dispersion of strong ferroelectric LiNbO$_{3}$ nanoparticles with spontaneous polarization 71 $\mu C/cm^2$ decreases the transition temperature of the cybotactic nematic phase by $0.6 ^{\circ}{\rm C}$ to $2.7 ^{\circ}{\rm C}$ depending on the concentration of dispersed nanoparticles \cite{derbali2020dielectric_33}. To further support our claim, it is reported that the dispersion of 0.01 $wt$$\%$ of BaTiO$_{3}$ nanoparticle of size $\sim$ 22 $nm$ in a bent-core nematic liquid crystal decreases the isotropic to nematic transition temperature by nearly $\sim 5 ^{\circ}{\rm C}$ \cite{khan2017elastic_11}.

The polarizing optical microscope (POM) textures of the pure 8-F-OH and 0.2 $wt\%$ BiFeO$_3$ dispersed 8-F-OH nanocomposites are shown in Fig. 1. When a birefringent liquid crystal sample is placed between the crossed polarizer and the analyzer under a polarizing optical microscope (POM), the textural color of the sample indicates the birefringence value at that temperature. While recording the textures using POM, the rubbing direction of the liquid crystal cell is placed at $45^o$ with respect to both the polarizer and analyzer of the POM. The birefringence value ($\Delta n$) of a liquid crystal sample can be qualitatively estimated from the colors of the Michel-Levy chart \cite{hoffman2023michel_41}. The color of the POM textures changes in the nematic phase, with the variation of temperature indicating a temperature-dependent change in orientational order parameters since birefringence is proportional to the order parameter, $\Delta n \propto \emph{S$_{m}$}$ \cite{garbovskiy2017ferroelectric_42}. T$_{IN}$ denotes the isotropic to the nematic transition temperature, and T is the temperature at which the texture is taken. The temperature is scaled with (T$_{IN}$-T) to compare the textures of the pure and doped systems and their order parameters, since T$_{IN}$ is different for pure and doped systems.

Comparison of polarizing optical microscope texture pairs (d)-(j), (e)-(k), and (f)-(l) of pure and doped compounds respectively, indicates that a similar color sequence is obtained for the doped and pure system, but the color sequence is shifted to a larger temperature gap for the pure system, compared to its doped counterpart (e.g., T$_{IN}$-T = 17 $^{\circ}{\rm C}$ for the doped system and T$_{\rm IN}$-T = 20$^{\circ}{\rm C}$ for the pure system in (d)-(j) pair of texture).  Qualitatively, we deduce that for a given temperature gap (T$_{IN}$ - T), the (mean) order parameter is higher for a doped system compared to a pure system.

\subsection{Dielectric Studies}

One measures the complex dielectric permittivity of a liquid crystal sample as a function of a wide range of frequencies, using dielectric spectroscopy. If a sinusoidally varying electric field in the frequency range of Hz to MHz is applied to a dielectric material, several dielectric relaxation modes can appear depending on the different polarization processes present in the sample \cite{patranabish2021quantum_38}. To describe the amplitude and phase properties of electric polarization of a sample with an applied electric field of frequency \textit{f}, we use relative permittivity such that there is a relative permittivity ($\varepsilon^{\prime}(f)$) that describes the electric polarization variation which is in phase with the applied electric field and a relative permittivity ($\varepsilon^{\prime\prime}(f)$) that describe the electric polarization variation that is out of phase (shifted by $90^o$) with respect to the applied electric field. The complex dielectric permittivity is then represented as $\varepsilon^{\ast}(f)$ = $\varepsilon^{\prime}(f)$ - $i\varepsilon^{\prime\prime}(f)$, where $\varepsilon^{\prime}$ and $\varepsilon^{\prime\prime}$ are the real and imaginary parts of complex dielectric permittivity respectively \cite{haase2013relaxation_43}. The relaxation processes can be collective relaxation (collective response of some molecules) or molecular relaxation (an individual molecular response) observed in the spectrum of dielectric constant ($\varepsilon^\prime$ and $\varepsilon^{\prime\prime}$). At high frequency (MHz), there can be ITO (Indium Tin Oxide) electrode relaxation mode also \cite{patranabish2021quantum_38}.

The dielectric spectroscopy measurement is performed with the Agilent E4980A instrument with an applied r.m.s voltage of 0.1 volts with varying frequencies from 20 Hz to 2 MHz. Fig.2 and Fig. 3 represent the real ($\varepsilon^\prime$) and imaginary ($\varepsilon^{\prime\prime}$) part of dielectric permittivity, respectively, in both pure and doped systems. The maximum possible root mean square error in the dielectric measurements lies within ±1$\%$ of the experimental value.

It can be clearly seen from the dielectric absorption graph that the pure compound contains four distinct dielectric modes (M$_{1}$, M$_{2}$, M$_{3}$, and M$_{4}$). However, the doped system has only three dielectric modes (M$_{1}$, M$_{3}$, and M$_{4}$). The origin of these modes can be described as follows:
\begin{itemize}

\item M$_{1}$: Collective relaxation of cybotactic clusters. Since it is a collective process, the dielectric strength is much higher. The relaxation frequency varies from a few 10 Hz to a few 100 Hz. This mode is widely reported in the nematic phase of bent-core liquid crystal \cite{ghosh2014ferroelectric_4, shanker2014_44, shanker2012nematic_45, marino2012dielectric_46}.

\item M$_{2}$: Debye type molecular relaxation mode due to reorientation of short axis dipole moment. It is a relaxation process that is due to the rotation of transverse dipole moments about their long axis \cite{guo2010polar_47, merkel2006orientational_48, nagaraj2010electric, Dhananjoy2023dielectric_21}.

\item M$_{3}$: Molecular relaxation mode due to charge flip-flop around the short axis of the molecule or along the long axis of the molecule \cite{ghosh2014ferroelectric_4, patranabish2018cybotactic_49, tadapatri2010permittivity_50}.

\item M$_{4}$: High-frequency ITO relaxation mode.
\end{itemize}

The molecular modes M$_{2}$ and M$_{3}$ can be distinguished by looking at their relaxation frequencies/relaxation times. M$_{2}$ mode arises at around 10 kHz of frequencies, and M$_{3}$ mode arises in the MHz frequency range.

\begin{figure}[!htb]
  \centering
    \includegraphics[width=0.49\textwidth]{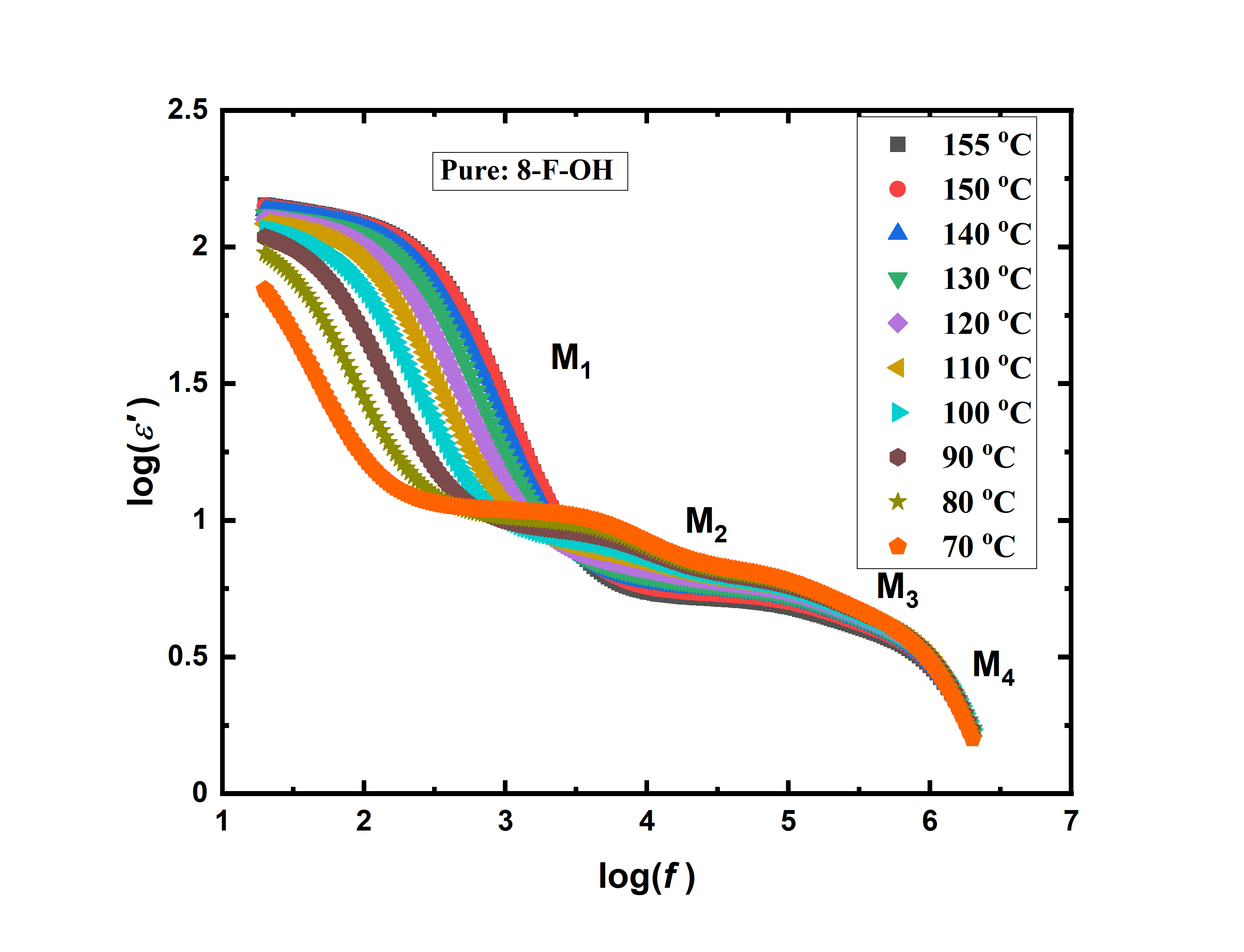}
    \includegraphics[width=0.49\textwidth]{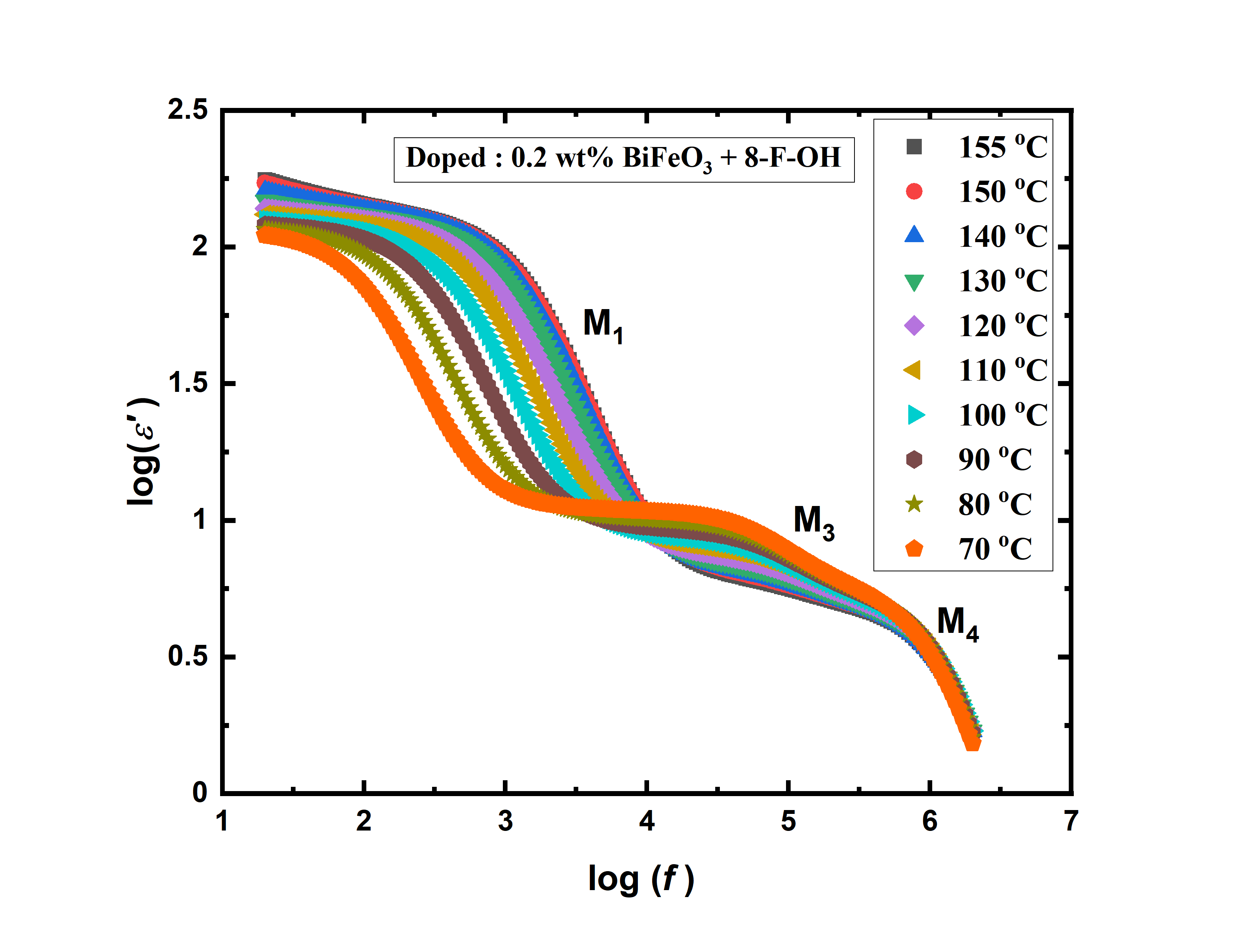} 
  \caption{Frequency-dependent real part of dielectric permittivity ($\varepsilon^{\prime}(f)$) of pure 8-F-OH (upper graph) and doped 8-F-OH + 0.2 $wt\%$ BiFeO$_3$ (lower graph) sample with variation of temperature}
\end{figure}

\begin{figure}[!htb]
  \centering
    \includegraphics[width=0.49\textwidth]{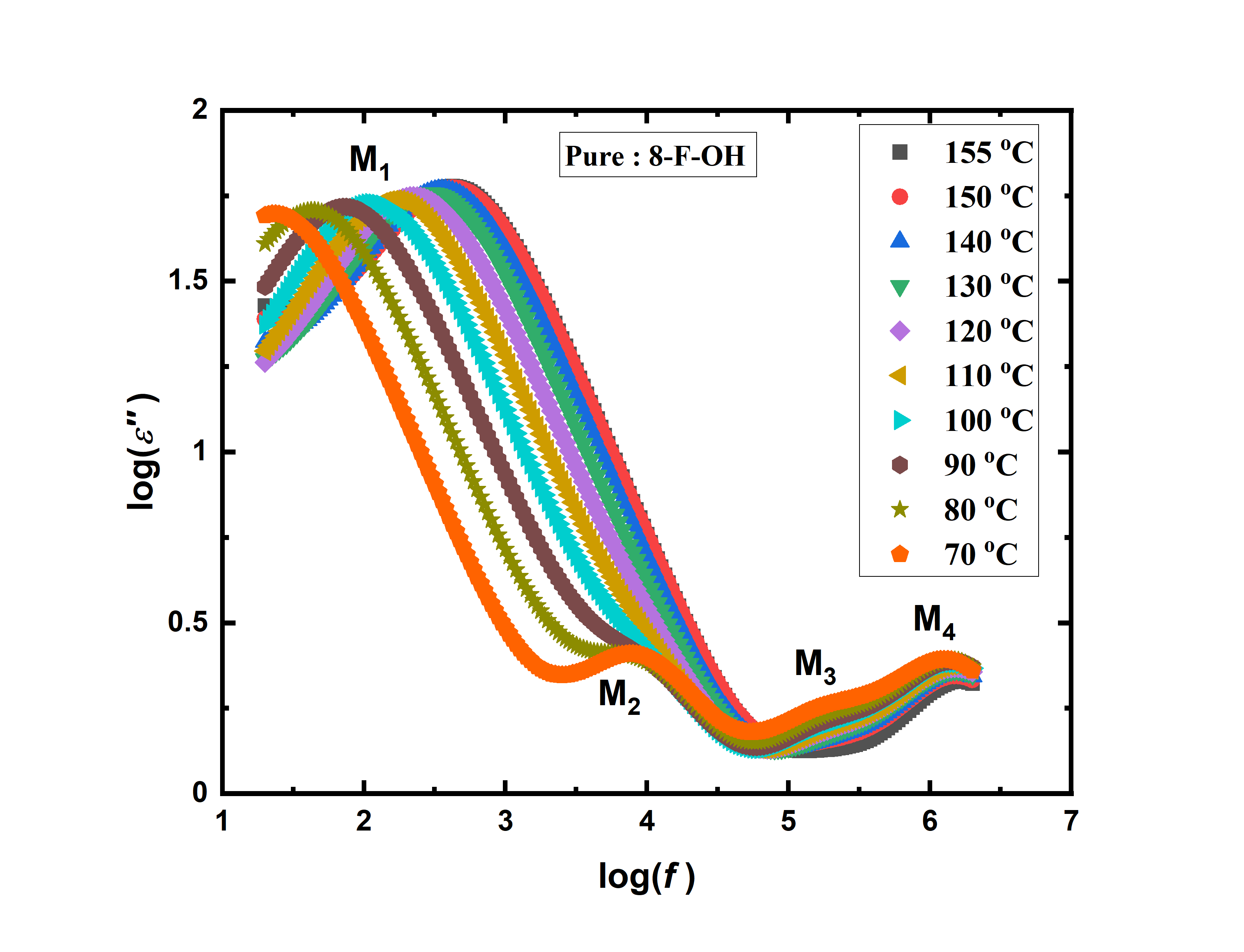}
    \includegraphics[width=0.49\textwidth]{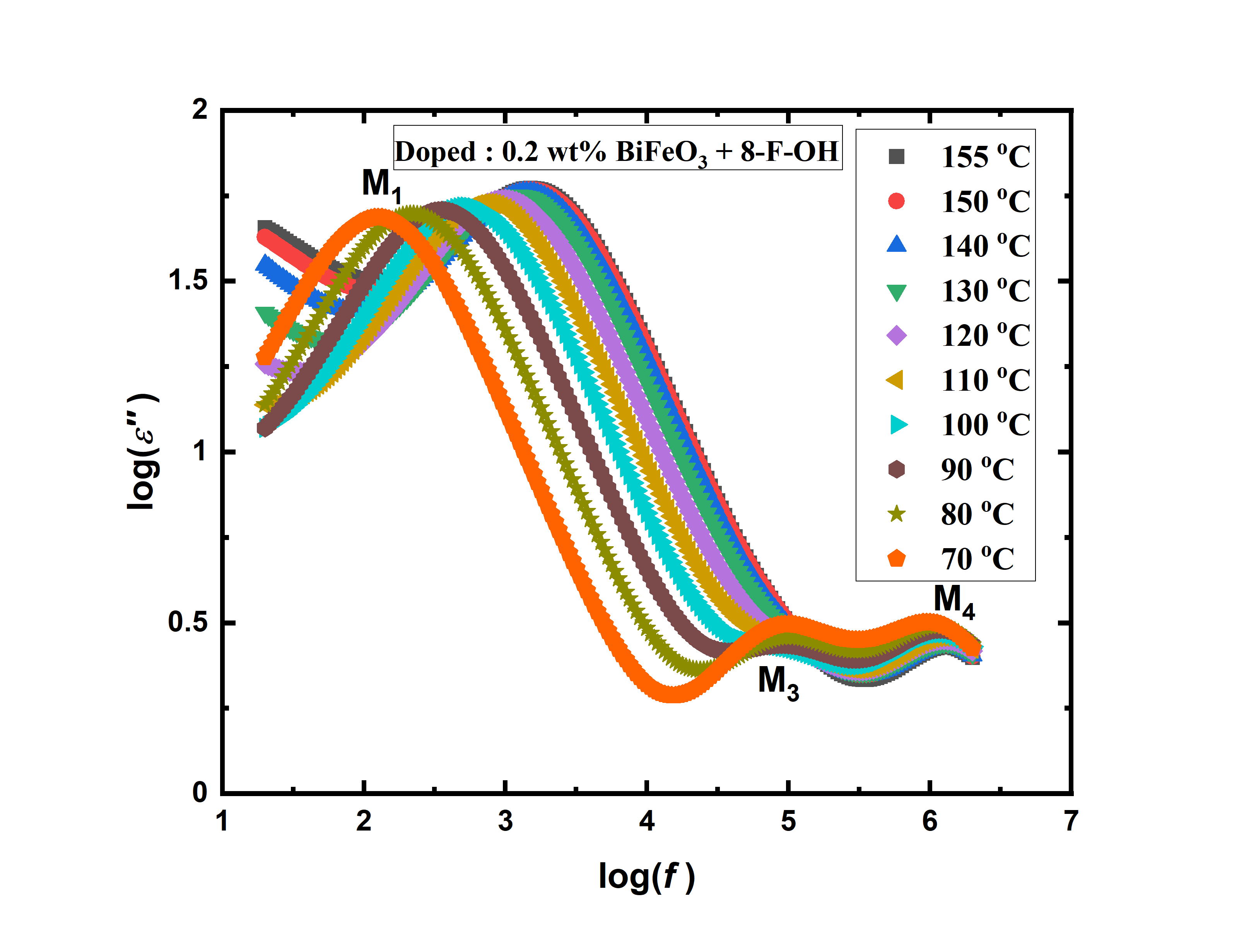} 
  \caption{Frequency-dependent imaginary part of dielectric permittivity ($\varepsilon^{\prime\prime}(f)$) of pure 8-F-OH (upper graph) and doped 8-F-OH + 0.2 $wt\%$ BiFeO$_3$ (lower graph) sample a with the variation of temperature}
\end{figure}

The M$_{2}$ molecular mode is absent in the doped system. We speculate that the absence is caused by the strong coupling of the liquid crystal molecules with the local electric field generated by the nanoparticles, which makes reorientation of short axis dipole moment difficult. To support our observation, it is reported that the relaxation mode observed at 8 kHz in the pure system is not present in the ferroelectric nanoparticle doped cybotactic nematic phase \cite{derbali2020dielectric_33}.

But, the M$_{3}$ molecular mode, which is due to charge fluctuation along the long molecular axis, is also strongly affected by the local electric field of the nanoparticle. The local electric field of the nanoparticle will create a dielectric hindrance to the M$_{3}$ molecular mode. Hence, in the doped system, the relaxation time for the M$_{3}$ molecular mode increases (relaxation frequency decreases). This kind of increment in the relaxation time of the molecular mode of ferroelectric nanoparticle dispersed liquid crystal is reported in Ref. \cite{ouskova2003dielectric_51}.

The fitting of dielectric data can give information on dielectric strength (dielectric increment $\delta\epsilon_{k}$ in $k^{th}$ relaxation process) and relaxation frequency for each collective and molecular dielectric mode. Proper analysis of these can help us to find out the cybotactic cluster size in both pure and doped systems. For this, the dielectric absorption data are fitted with the well-known Havrilliak-Negami equation \cite{havriliak1966complex_52, havriliak1967complex_53, chakraborty2019induced_54}, the form of which is given below:

{\small
\begin{equation}\label{HN_equation}
\varepsilon^{\prime\prime} = \frac{\sigma_0}{\epsilon_0 (2\pi f)^s} + \sum_{k=1}^{N} \frac{\delta\epsilon_{k}\sin(\beta_{k}\theta)}{[1+(2\pi f\tau_k)^{2\alpha_k}+2(2\pi f\tau_k)^{\alpha_{k}}\cos{(\alpha_{k}\pi/2)}]^{\beta_{k}/2}}
\end{equation}
}
where
\begin{equation}
\label{HN_equation_theta}
\theta = \tan^{-1}\Bigg[\frac{(2\pi f\tau_k)^{\alpha_{k}}\sin{(\alpha_{k}\pi/2)}}{1+(2\pi f\tau_k)^{\alpha_{k}}\cos{(\alpha_{k}\pi/2)}}\Bigg]
\end{equation}
Here $\sigma_0$ and $\epsilon_0$ are DC conductivity and free space permittivity, respectively, $\textit{k}$ is the number of relaxation modes present in the sample, $\tau_k$ is relaxation time for a particular k$^{th}$dielectric relaxation mode, $\alpha_k$ and $\beta_k$ are asymmetry and broadening parameters. The initial value of relaxation time is estimated by $\tau_k$=1/2$\pi$$f_k$, here, \textit{$f_k$} is the relaxation frequency at which dielectric mode has maxima in the absorption peak (Fig.4). $\alpha_k$ and $\beta_k$ are fitting parameters that decide the nature of the dielectric process \cite{haase2013relaxation_43}. For a perfectly Debye type dielectric process $\alpha_k$ = 1 and $\beta_k$ =1 \cite{chakraborty2019induced_54}.

\begin{figure}[!htb]
  \centering
    \includegraphics[width=0.49\textwidth]{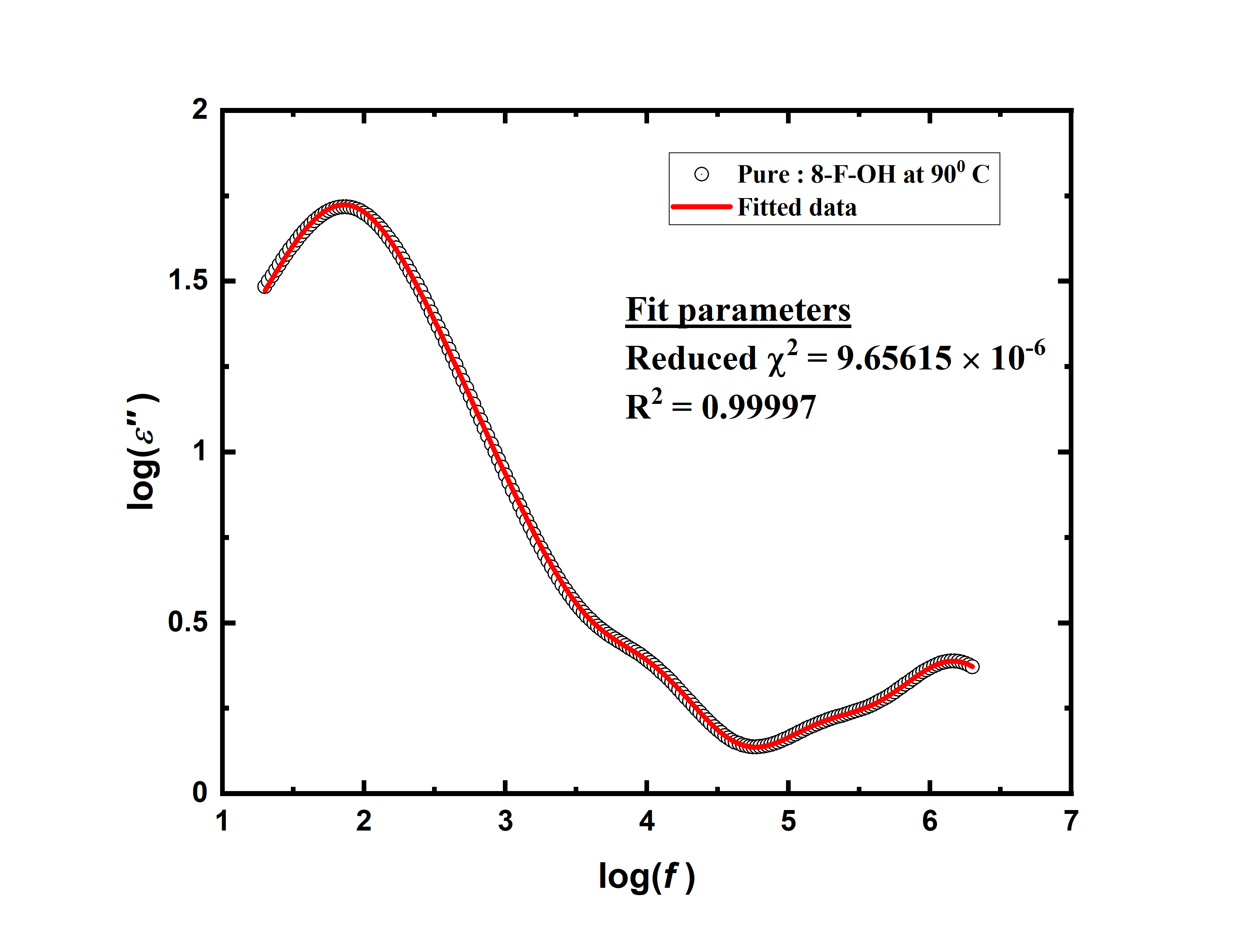}
    \includegraphics[width=0.49\textwidth]{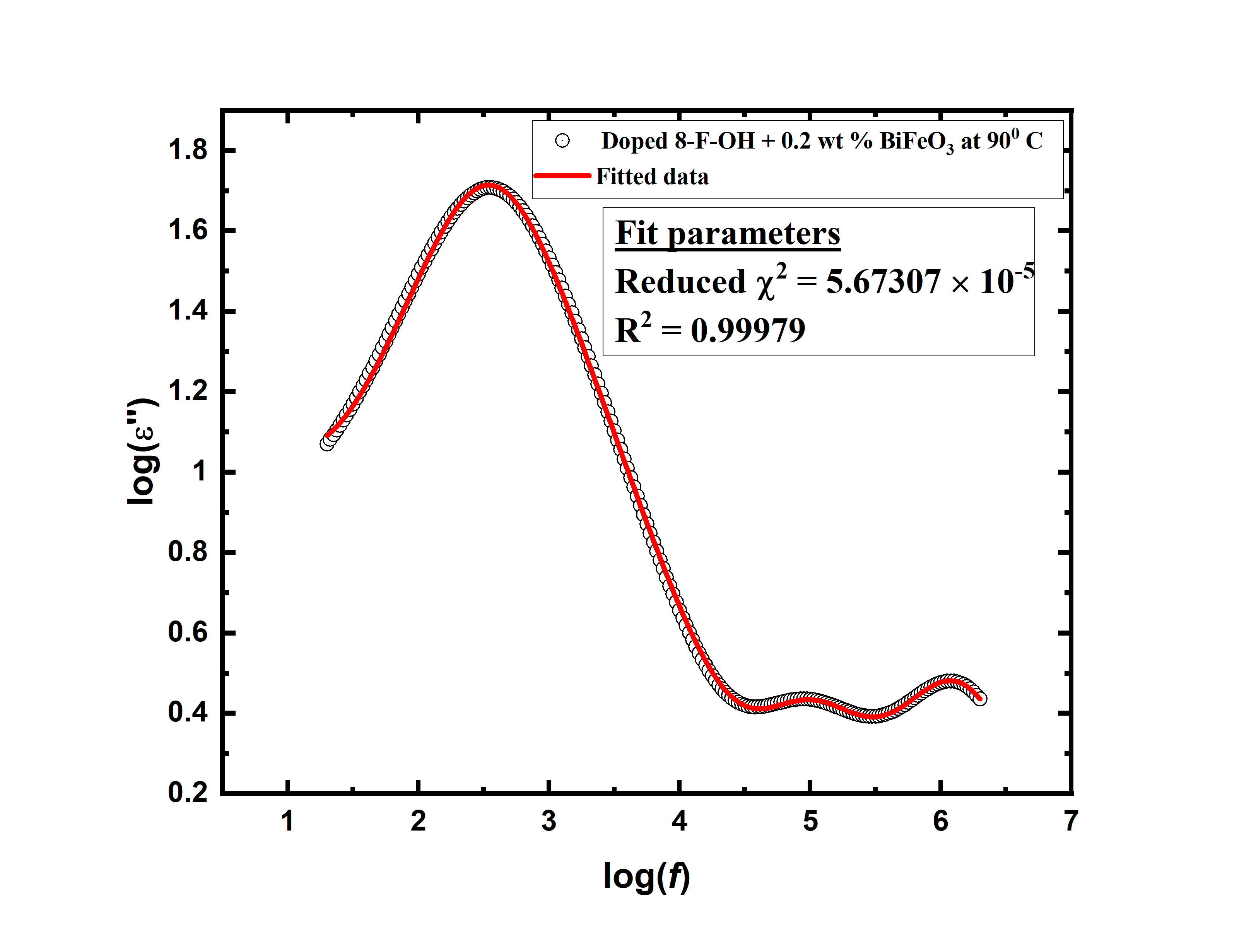} 
  \caption{Fitting of imaginary part of dielectric permittivity ($\varepsilon^{\prime\prime}(f)$) at $90^oC$ temperature for both pure 8-F-OH (upper graph) and doped 8-F-OH + 0.2 $wt\%$ BiFeO$_3$ (lower graph) sample by Havrilliak-Negami equation. The fitted value of reduced $\chi^2$ and $R^2$ is given so that reduced $\chi^2$ $\sim 0$ and $R^2$ $\sim 1$ represent a good fit.}
\end{figure}

The relative cybotactic cluster size can be calculated by the ratio of relaxation frequencies of collective cybotactic mode in the pure and doped systems respectively and taking the average with respect to temperature \cite{patranabish2021quantum_38}. The relative cluster size ($R_c$) is

\begin{equation}
\label{Relative cluster size}
R_c = \bigg \langle \frac{f_{M_{1}-Pure}}{f_{M_{1}-Doped}} \bigg \rangle_{T} = 0.23 = \frac{\text{cluster size of doped system}}{\text{cluster size of pure system}}
\end{equation}

In Ref. \cite{panarin2018formation_31}, it is reported that the absolute values of cybotactic cluster size, at a given temperature, in the pure bent-core nematic liquid crystal can be estimated as the ratio of relaxation frequency of the molecular mode (M$_{2}$) to the relaxation frequency of collective cybotactic cluster mode (M$_{1}$) at a fixed temperature.

\begin{equation}
\label{Pure cluster size}
N_{{\rm c-Pure}}(T) = \frac{f_{{\rm M_{2}-Pure}}(T)}{f_{{\rm M_{1}-Pure} }(T)}
\end{equation}

The M$_{2}$ molecular mode is not present throughout the nematic phase of the pure system. To estimate the value of relaxation frequency of M$_{2}$ mode for all relevant temperature ranges, we have extrapolated the Arrhenius plot for the M$_{2}$ mode to higher temperatures (since dielectric relaxation mode in liquid crystal follows Arrhenius law) where the mode is not present, and have obtained a value for the relaxation frequency of the M$_{2}$ mode. 

\begin{equation}
\label{Pure cluster size updated}
N_{{\rm c-Pure}}(T) = \frac{f_{{\rm M_{2}-Pure}}(T)_{{\rm extrapolated}}}{f_{{\rm M_{1}-Pure}}(T)}
\end{equation}

Then the cybotactic cluster size in the doped system, as a function of temperature, can be written as:

\begin{equation}
\label{Doped cluster size updated}
N_{{\rm c-Doped}}(T) = {R_{{\rm c}}(T)}\times N_{{\rm c-Pure}}(T)
\end{equation}

where $R_{c}(T)$ is the temperature-dependent relative cluster size \cite{patranabish2021quantum_38}. At each temperature, $R_{c}(T)$ is calculated by taking the ratio of relaxation frequencies of the M$_{1}$ mode, for the pure and doped sample respectively. By calculating these values, we get temperature-dependent $N_{c-Pure}$ and $N_{c-Doped}$ values, which define the cybotactic cluster sizes for the pure and doped systems respectively.

\begin{figure}[!htb]
    \centering
        \includegraphics[width=1\linewidth]{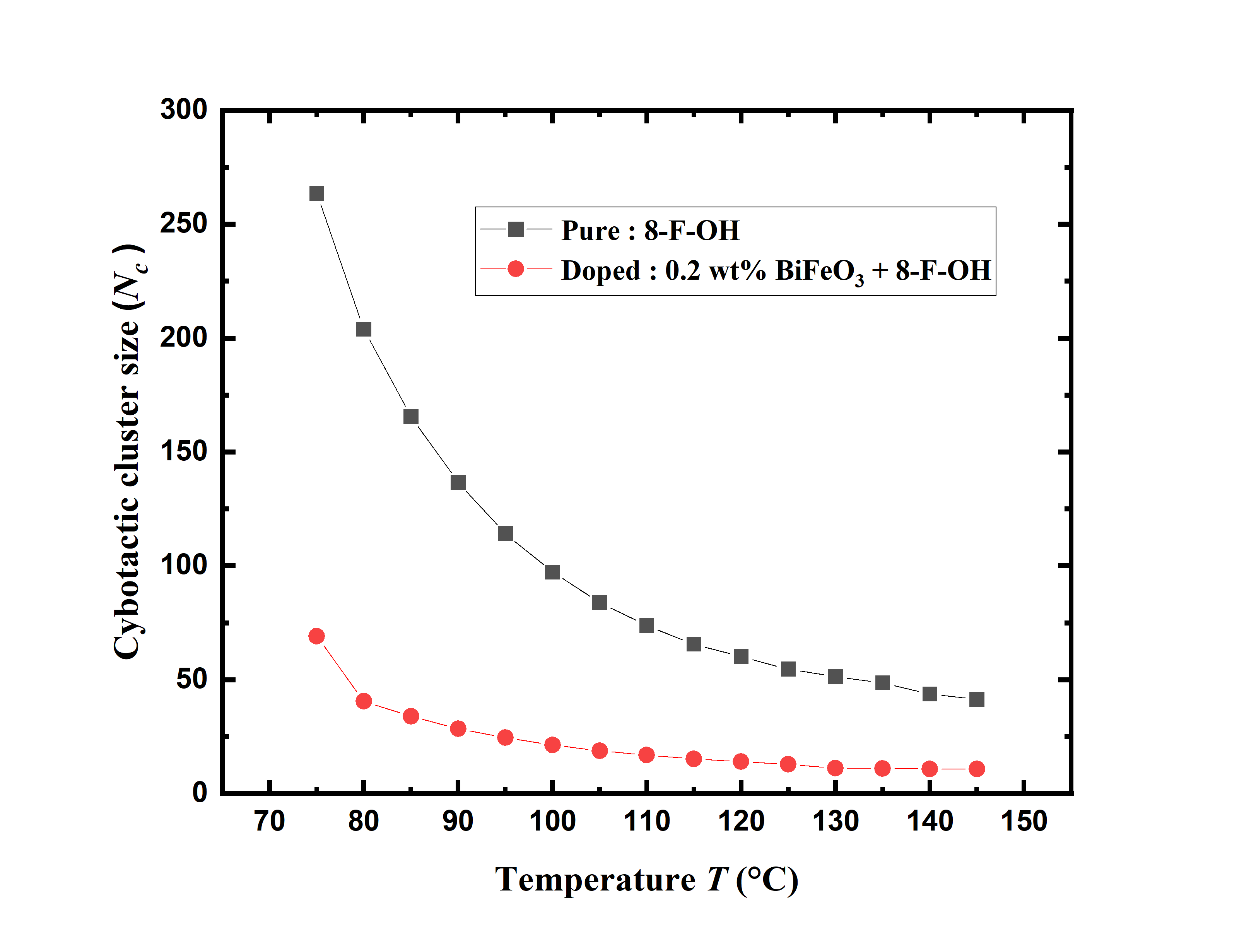}
    \caption{Temperature-dependent estimated values of the number of molecules in cybotactic clusters in pure 8-F-OH (black square) and doped 8-F-OH + 0.2 $wt\%$ BiFeO$_3$ (red circle).}
    \label{1}
\end{figure}

Fig. (5) indicates that the cluster size is significantly reduced in the doped system, and with decreasing temperature, the growth of the cybotactic cluster (no. of molecules present in a cluster) is higher in the pure sample compared to the doped sample. The reduction in the I-N transition temperature for the doped system, as compared to its pure counterpart, serves as additional evidence of the presence of disordered cybotactic clusters. Similar indications are gleaned from small-angle X-ray scattering (SAXS) measurements conducted on both the pure 8-F-OH and the doped 8-F-OH + 0.2 $wt\%$ BiFeO$_3$ system. A comprehensive analysis of the SAXS data is elaborated in the Supplementary data. The dielectric contribution ($\delta\epsilon_{1}$) of reduced cluster size in the doped system is less compared to the pure one, as indicated by the upper graph of Fig. 6.

\begin{figure}[!htb]
  \centering
    \includegraphics[width=0.49\textwidth]{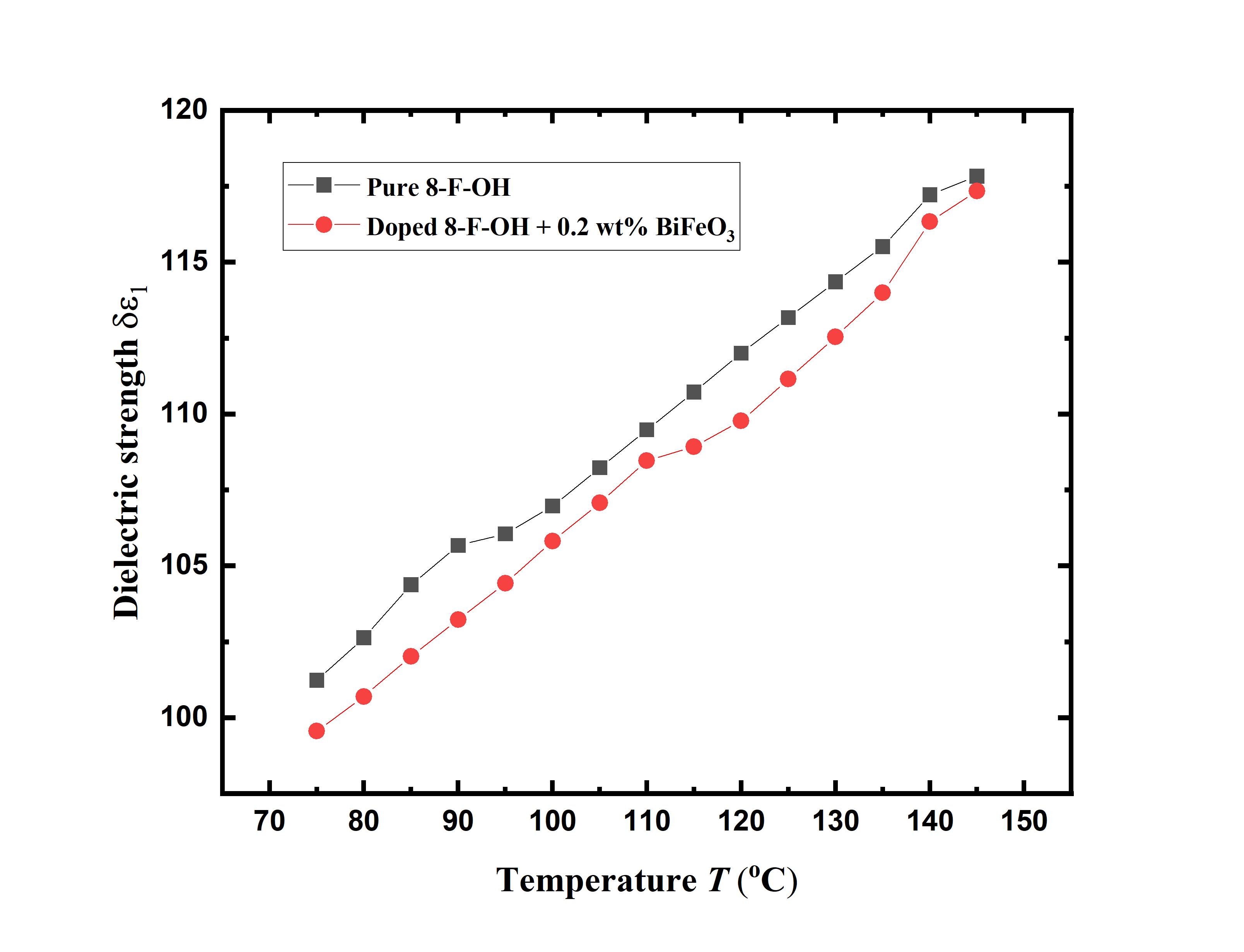}
    \includegraphics[width=0.49\textwidth]{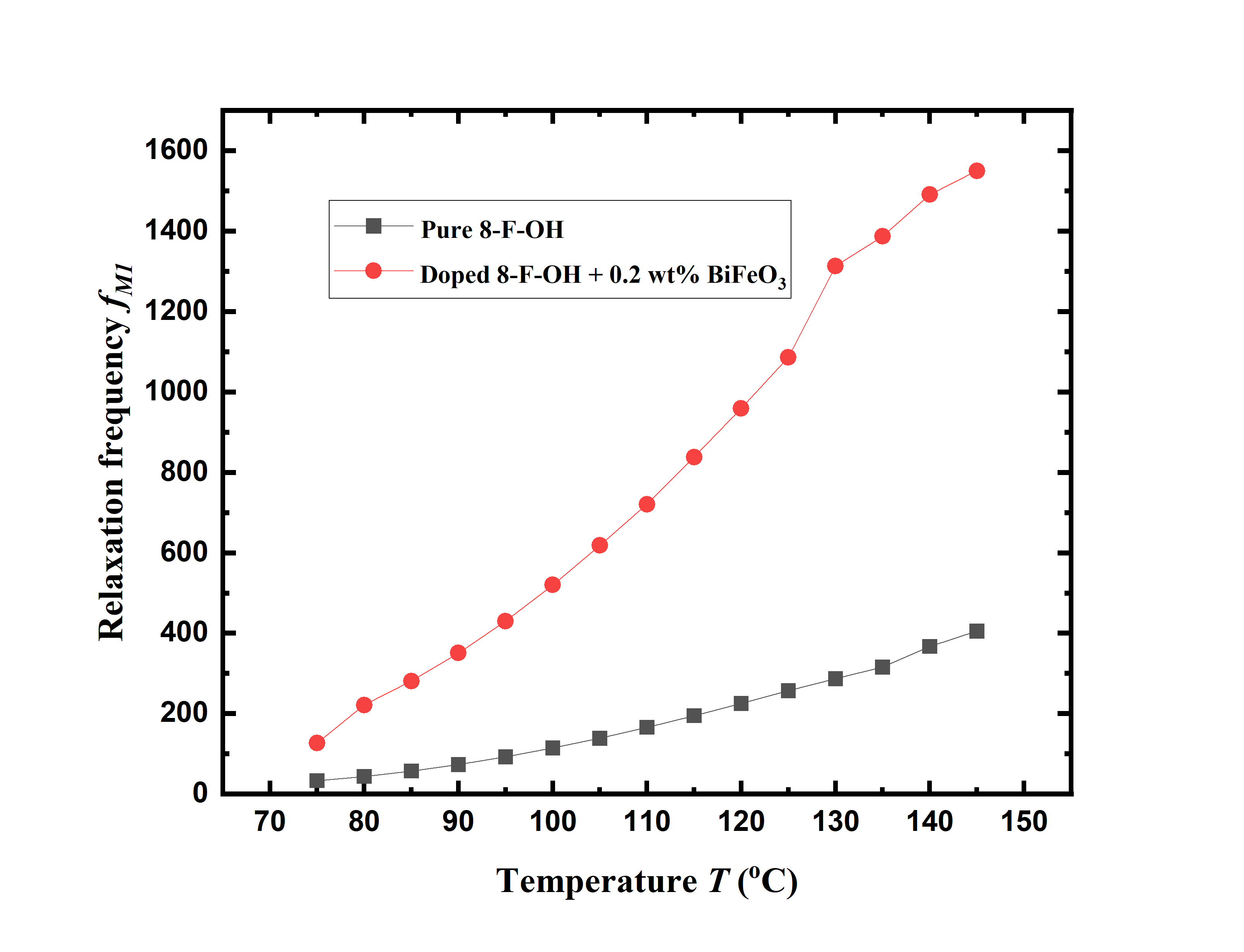} 
  \caption{Temperature variation of dielectric strength (upper graph) and relaxation frequency (lower graph) of collective cybotactic mode (M$_{1}$) in both pure and doped sample}
\end{figure}

Next, to find the activation energy of cybotactic clusters, we use an Arrhenius type equation, i.e., $f_{R}$ = $f_{0}\exp{\frac{-E_{a}}{k_{B}T}}$, where 
$f_{R}$ is the relaxation frequency of collective cybotactic mode, extracted from the Havrilliak-Negami fitting of dielectric absorption data, (\textit{E$_a$}) is the activation energy and $T$ is the absolute temperature and $f_{0}$ is a constant \cite{haase2013relaxation_43}. Activation energy is the amount of energy that is required to activate a dielectric relaxation process 
\cite{haase2013relaxation_43}. We can compute the activation energy of the cybotactic cluster mode, for the pure and doped samples, by plotting ln($f_{R}$) vs. 1/T plot and fitting it with the Arrhenius type plot. For a pure system, the activation energy (\textit{E$_a$}) = 45.91 ${\rm kJ/mol}$, and for the doped system, the activation energy becomes (\textit{E$_a$}) = 40.96 ${\rm kJ/mol}$ as obtained from Fig. 7. A reduced activation energy in the doped system compared to the pure implies that lower hindrance to the collective relaxation process. It is expected that a smaller cluster size would ordinarily increase the activation energy in the doped system \cite{patranabish2021quantum_38}, but the opposite effect is observed here, potentially due to the coupling between ferroelectric nanoparticles and cybotactic clusters. It's has been reported that dispersion of ferroelectric BaTiO$_{3}$ nanoparticle in ferroelectric liquid crystal decreases the activation energy of collective mode in the doped system compared to the pure one \cite{mikulko2009complementary_55}.

\begin{figure}[!htb]
  \centering
    \includegraphics[width=0.49\textwidth]{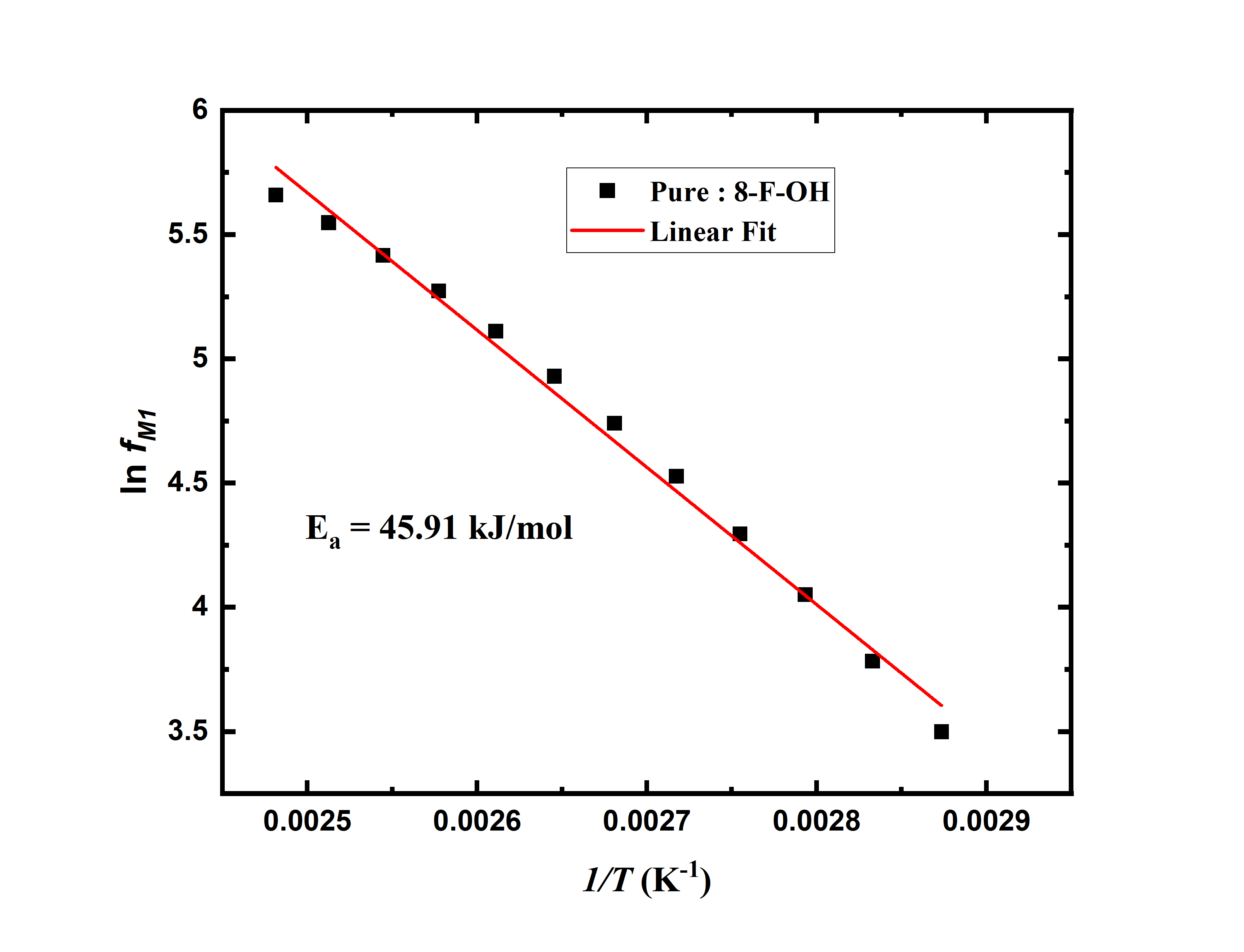}
    \includegraphics[width=0.49\textwidth]{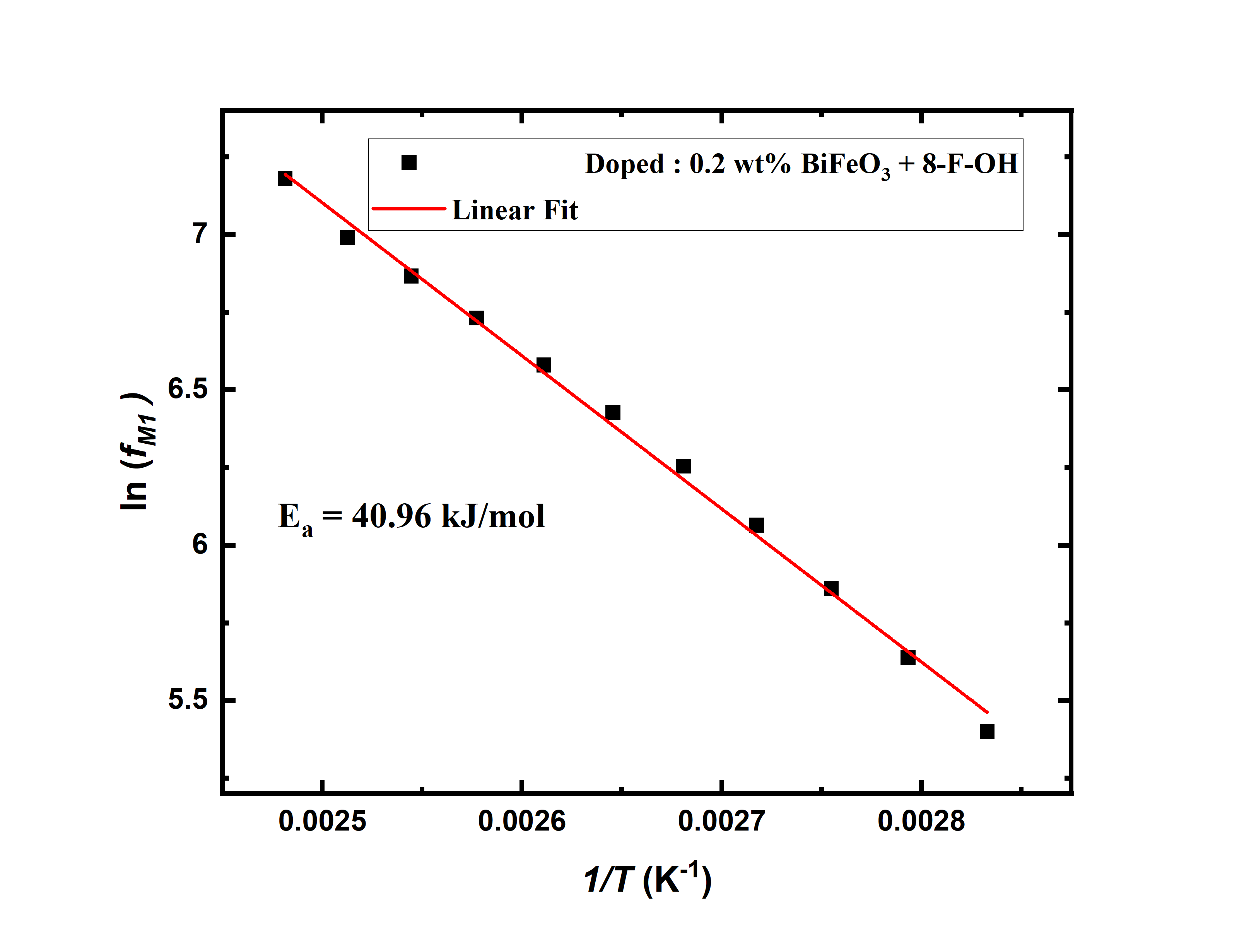} 
  \caption{Arrhenius fitting of collective relaxation frequency of M$_{1}$ mode in pure 8-F-OH (upper graph) and doped 8-F-OH + 0.2 $wt\%$ BiFeO$_3$ (lower graph). Activation energy is obtained from the slope of the linear fitting.}
\end{figure}

\subsection{Measurements of the Orientational Order Parameter}

The optical birefringence of a nematic liquid crystal sample can be written in terms of phase retardation ($\delta$) or output intensity when the laser light passes through a birefringent sample placed in between crossed polariser-analyser conditions. The optical birefringence ($\Delta n$) and order parameter (\emph{S$_{m}$}) of the pure and doped liquid crystal can be calculated by the well-known optical transmission method \cite{dierking2003textures_56}, given by
\begin{equation}
\label{Birefringence}
\Delta n = \frac{\lambda}{2\pi d}\delta 
= \frac{\lambda}{2\pi d}\cos ^{-1}\big\{1-2I_{t}\big\}
\end{equation}
and
\begin{equation}
\label{Phase retardation}
\delta = 2m\pi \pm \cos ^{-1}\big\{1-2I_{t}\big\},\;  m = 0, 1, 2... \ .
\end{equation}
Here $\lambda$ is the wavelength of incident laser light ($\sim$ 633  $nm$), $d$ is the thickness of the LC cell, $\delta$ is the phase retardation of the light on passing through the thickness of the LC sample, $I_{t} = I / I_0$ is the normalized intensity with $I$ being the intensity of the transmitted light and $I_{0}$ being the intensity of incident light, $\Delta n$ = $n_{e}$ – $n_{0}$ is the birefringence where $n_{e}$ is the extraordinary refractive index, and $n_{0}$ is the ordinary refractive index. The bent-shaped 8-F-OH molecules form a uniaxial nematic phase with the director along the average direction of long molecular axes. With this molecule, the refractive index along the long molecular axis is called extraordinary refractive index ($n_{e}$), and the refractive index along the short or transverse molecular axis is called ordinary refractive index ($n_{0}$) and the difference between the two ($n_{e}$ – $n_{0}$) is called birefringence ($\Delta n$) \cite{dierking2003textures_56}.

The phase retardation $\delta$ is calculated from the normalized output intensity by using equation (9). Now substituting the values of $\delta$ in equation (8), we calculate the birefringence value ($\Delta n$) of the sample for a certain thickness of sample and wavelength of laser light used. The temperature-dependent birefringence plot is obtained by calculating the birefringence value for different values of temperature. The error estimation of experimentally measured optical birefringence ($\Delta n$) is performed using root mean square error. It is found that the error in $\Delta n$ lies with ±2.1 $\%$ of experimentally calculated values.

The experimentally obtained variation of birefringence with temperature for a pure and doped sample is shown in Fig. 8. To estimate the orientational order parameters (\emph{S$_{m}$}), we fit the experimental birefringence graphs by Haller’s equation \cite{haller1975thermodynamic_57, prasad2010refractive_58, vita2016molecular_59},

\begin{equation}
\label{Haller equation}
\Delta n = \Delta n_{0}\Bigg(1-\frac{T}{T^{*}}\Bigg)^{\beta}
\end{equation}

Here $\Delta n_{0}$, $T^{*}$, and $\beta$ are adjustable fitting parameters. $\Delta n_{0}$ is the birefringence of the perfectly aligned sample. $\Delta n $ is the value of birefringence at the absolute temperature $\emph{T}$. $\beta $ is a critical exponent with a value of around 0.2 \cite{ramakrishna2010orientational_60}. $T^{*}$ is slightly higher than the isotropic to the nematic transition temperature \cite{ramakrishna2010orientational_60}.

\begin{figure}[!htb]
    \centering
        \includegraphics[width=1\linewidth]{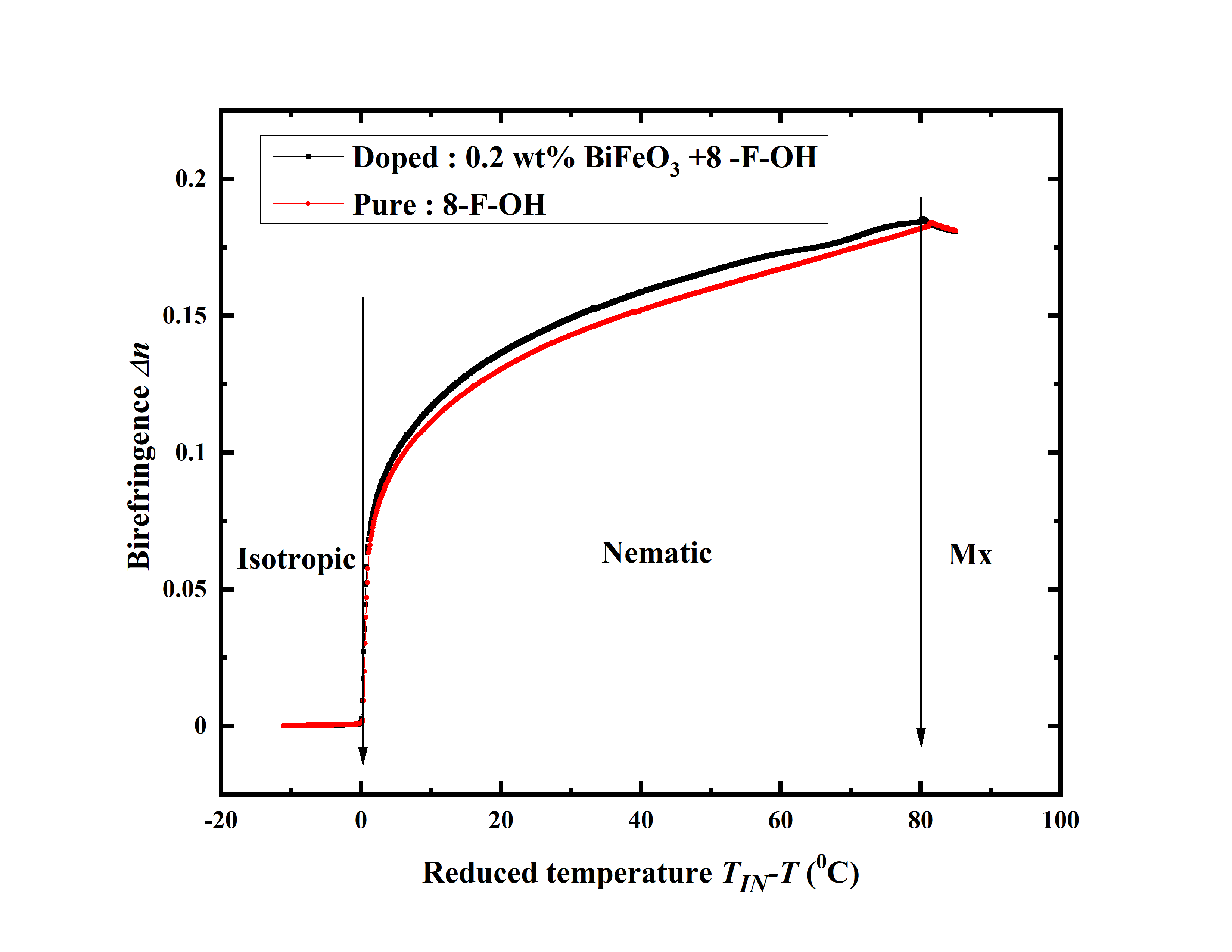}
    \caption{Comparison of birefringence ($\Delta n$) value of pure 8-F-OH and 0.2 $wt\%$ BiFeO$_3$ + 8-F-OH liquid crystal with temperature. In the isotropic region, the black curve is superimposed by the red curve.}
    \label{1}
\end{figure}

The orientational order parameter can be defined to be $\emph{S$_{m}$}$ = ${\Delta n}/{\Delta n_{0}}$ \cite{ramakrishna2010orientational_60}. The corresponding order parameter plot for both the pure and 0.2 $wt\%$ BiFeO$_3$ doped 8-F-OH sample is shown in Fig. 9. The values of parameters from the Hallers fitting are $\Delta n_{0}$ = 0.20552, $\beta$ = 0.23142   for the pure system, and $\Delta n_{0}$ = 0.20994, $\beta$ = 0.21514   for the doped system. The enhanced value of $\Delta n_{0}$ in the doped system signifies that the anisotropic liquid crystal molecules get polarised by the local electric field of ferroelectric nanoparticle, and this effectively enhances the birefringence value in the doped system. A comparative order parameter plot for the pure and doped compound suggests that the order parameter in the doped system is higher than its pure counterpart.

\begin{figure}[!htb]
    \centering
        \includegraphics[width=1\linewidth]{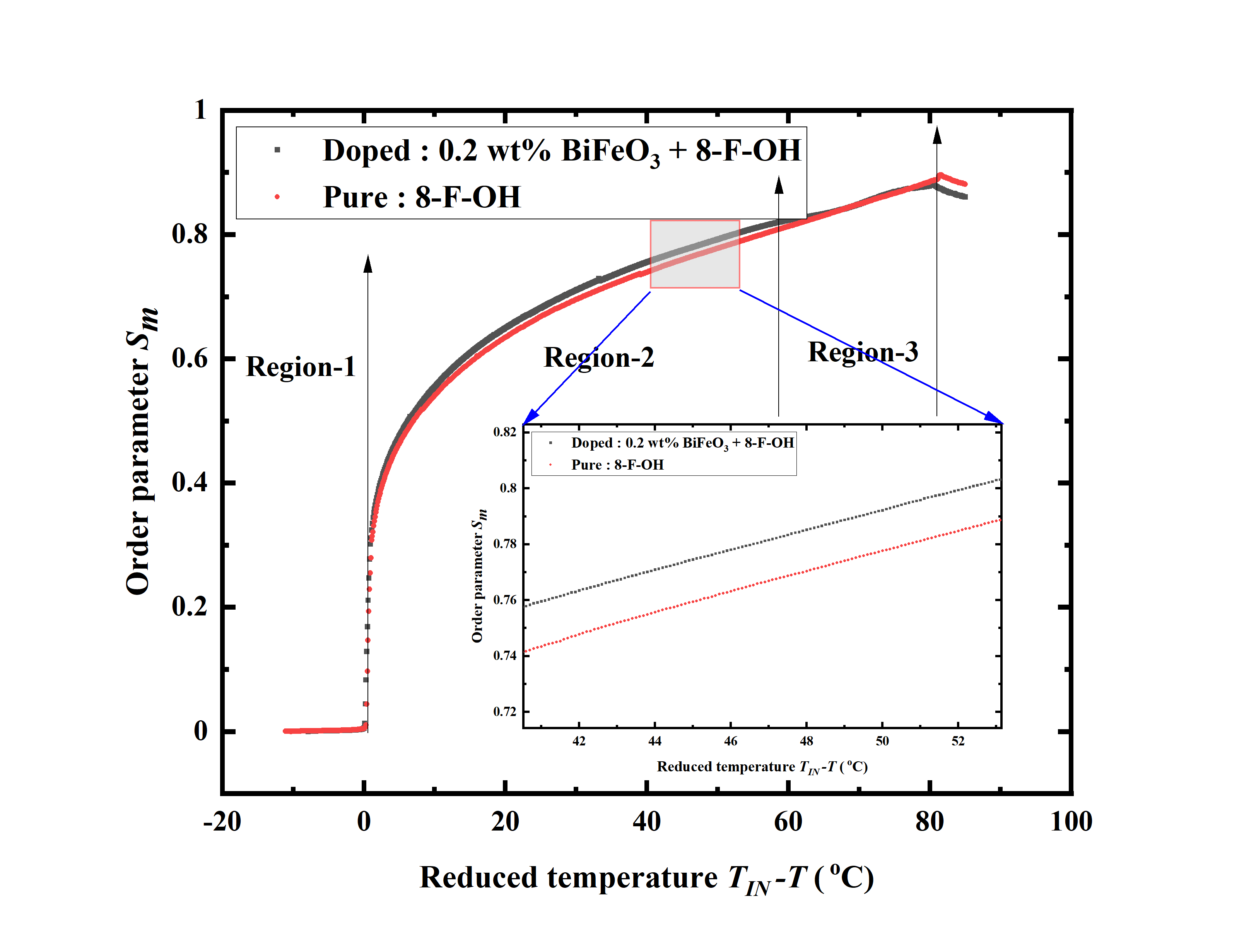}
    \caption{Temperature dependent orientational order parameter ($\emph{S$_{m}$}$) of the pure 8-F-OH and doped 8-F-OH + 0.2 $wt\%$ BiFeO$_3$ system. The zoomed region clearly shows that the doped system has enhanced order parameter compared to the pure one.}
    \label{fig:9}
\end{figure}

Consider three regions of Fig. 9 to understand the behavior of temperature-dependent orientational order parameters.

\underline{Region 1}: Randomly oriented ferroelectric nanoparticle domain in the isotropic and near I-N transition

In the isotropic phase, the order parameter is negligible or small in magnitude. Near the isotropic to nematic transition region, as the nematic domains start to grow, the order parameter increases rapidly. Near the phase transition region, the order parameter values are almost the same for pure and doped compounds. In this region, the random orientation of the molecules in the nanoparticle domains does not contribute to the effective order parameter due to the weak coupling between the nanoparticle and liquid crystal order. 

\underline{Region 2}: Strong Coupling between Ferroelectric Nanoparticle Domains and Bent-Core Nematic Medium 

As the temperature decreases, the order parameter in the nematic phase of the pure system increases gradually, and the molecules in the nanoparticle domain start to orient along the bulk nematic direction.  
We conjecture that the average orientation of the molecules in the nanoparticle domain is aligned with the bulk nematic director, denoted by $\mathbf{n}_g$, for energetic reasons, and this contributes to the enhanced mean order parameter in this region. In particular, the experimental results in Figure~\ref{fig:9} suggest a sharp increment in the mean order parameter of the doped system in this region, compared to the pure sample.

Cybotactic clusters are a nanometric size collection of molecules arranged in $Sm C$ type arrangement \cite{panarin2018formation_31, francescangeli2009ferroelectric_61}. The dipole moment (${\bm p}^{\prime}$) of the cybotactic cluster 
is perpendicular to the long molecular axes \cite{francescangeli2009ferroelectric_61}. In other words, if $\mathbf{n}_c$ denotes the cluster director, then the dipole moment of the cybotactic clusters, denoted by ${\bm p}^{\prime}$, will be orthogonal (or approximately orthogonal) to $\mathbf{n}_c$. We conjecture that in Region $2$ of Figure~\ref{fig:9}, for moderately low temperatures below the isotropic-nematic transition temperature,  we have anti-parallel dipole-dipole coupling between the dipole moment of ferroelectric nanoparticle (${\bm p}$), and dipole moment of cybotactic clusters (${\bm p}^{\prime}$) \cite{ghosh2011effect_19, derbali2020dielectric_33}. As a consequence, ${\bm p}^{\prime}$ is either parallel or anti-parallel $\bm p$, which in turn, is aligned with $\mathbf{n}_g$ (see the assumption in the preceding paragraph). Thus, $\mathbf{n}_g$ and $\mathbf{n}_c$ are approximately orthogonal to each other. A possible scenario is that the cluster molecules lie in the plane orthogonal to $\mathbf{n}_c$ i.e. as ${\bm p}^{\prime}$ reorients to align with ${\bm p}$, $\mathbf{n}_c$ aligns along the direction normal to the cluster and the molecules lie randomly in the plane of the cluster as shown in Fig. (10), creating a disordering effect within the cluster. This is somewhat captured by the mathematical model in the next section, and is in qualitative agreement with the dielectric spectroscopy measurements that suggest reduced cluster size for the doped system, compared to the pure system as represented by Fig. (5). This also qualitatively explains the observation that the rate of increase of the mean order parameter of the doped system decreases with decreasing temperature in Figure~\ref{fig:9}, and the gap between the (mean) order parameters of the doped and pure systems decreases with decreasing temperature i.e. as we move into the deep nematic phase.

\begin{figure*}[!t]
    \centering
        \includegraphics[width=1\linewidth]{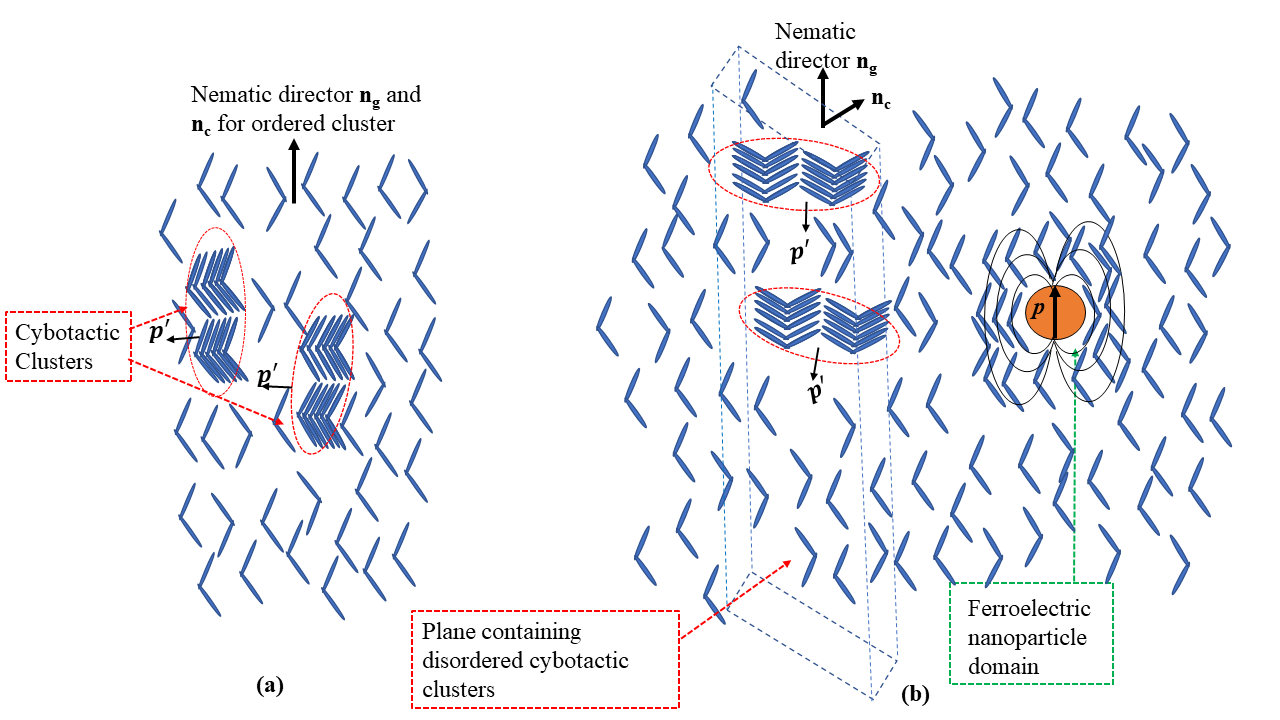}
    \caption{Schematic diagram of (a.) pure bent-core nematic liquid crystal having cybotactic clusters with dipole moment $p^{\prime}$, and (b.) ferroelectric nanoparticle doped bent-core nematic liquid crystal with anti-parallel dipole-dipole alignment between ferroelectric nanoparticle dipole $p$ and cybotactic cluster dipole $p^{\prime}$}
    \label{1}
\end{figure*}

\underline{Region 3}: The Low Temperature Regime

As the temperature is further decreased below the isotropic-nematic transition temperature, the rate of growth of the mean order parameter of the doped sample further decreases compared to the pure sample, and at some point, the curves of the doped and the pure sample overlap within the scope of our experimental measurements. We conjecture that as the temperature decreases, the number of cybotactic clusters increases as do the size of clusters and the cluster-cluster interactions dominate all other effects. 
Hence, the effects of the nano-doping are shielded by the cluster-cluster interactions or the interactions between the clusters and the bulk nematic, and there are no appreciable differences between the pure and doped samples deep in the nematic phase. 

For the Mx phase (enlarged or elongated cybotactic clusters), where there are elongated clusters, the order parameter value is decreased in the doped system compared to the pure one because, in a sea of cybotactic clusters, ferroelectric nanoparticles act as mere impurities.

\section{Theoretical Explanation}

In this section, we propose a simple phenomenological Landau-de Gennes (LdG) type model to study the dispersion of
multiferroic nanoparticles in a bent-core nematic (BC) system, building on the work in \cite{patranabish2019one_37}, which in turn is inspired by the model in \cite{madhusudana2017two_36}. The BC system has two types of molecules - the excited state (ES) molecules in cybotactic clusters and the surrounding ground-state (GS) molecules. In \cite{patranabish2019one_37}, the authors assume an LdG-type free energy density for the two types of molecules with relevant coupling terms, based on the assumption that the ordering state of the GS and ES molecules can be described by two different LdG order parameters, defined by $\Q_g$ and $\Q_c$ respectively. More precisely, $\Q_g$ and $\Q_c$ are symmetric, traceless $3\times 3$ matrices that contain information about the GS and ES ordering within their eigenvalues and eigenvectors, in the spirit of LdG order parameters \cite{patranabish2019one_37}. Both $\Q_g$ and $\Q_c$ are uniaxial by assumption, with two degenerate non-zero eigenvalues and one distinguished eigenvector corresponding to the non-degenerate eigenvalue. The distinguished eigenvectors are referred to as the GS and ES directors respectively. 
\begin{equation} \label{eq:Q}
\begin{aligned}
& \Q_g = \sqrt{\frac{3}{2} } S_g \left( \n_g \otimes \n_g - \frac{1}{3} \mathrm{I} \right), \\
& \Q_c =   \sqrt{\frac{3}{2} } S_c \left( \n_c \otimes \n_c - \frac{1}{3} \mathrm{I} \right). \\
\end{aligned}
\end{equation} Here, $\n_g$, $\n_c$ are unit-vectors and $\mathrm{I}$ is the $3\times 3$ identity matrix.
Here, $\n_g$ is the GS director and it models the preferred direction of averaged alignment of the GS molecules, and similar comments apply to $\n_c$ which is the cluster director. $S_g$ and $S_c$ are the corresponding order parameters associated with the GS and ES molecules respectively, which are scalar measures of the degree of ordering about $\n_g$ and $\n_c$ respectively. 
In \cite{patranabish2019one_37}, the authors assume that $\n_g$ and $\n_c$ are constant directors, and in the absence of nano-particles, they propose a phenomenological LdG-type energy for a confined BC sample as given below:
\begin{equation}\label{Energy1}
 \begin{aligned}
   \mathcal{F} & =  \int_{\Omega} (1 - a_x) \left( \frac{a_g}{2}(T - T^{*})S_g^2 - \frac{B_g}{3} S_g^3 + \frac{C_g}{4} S_g^4 - E_{el} S_g    \right) \\
   & + \frac{a_x}{N_c} \left( - (1 - a_x) \gamma S_g S_c + \frac{\alpha_c}{2} S_c^2 + \frac{\beta_c}{4} S_c^4  \right) \\
   & - a_xJ E_{el} S_c  + K_g |\nabla S_g|^2 + K_c |\nabla S_c|^2 \dd \x.  \\
   \end{aligned}
\end{equation}
The parameters $a_g$, $B_g$, $C_g$, $\alpha_c,$ and $\beta_c$ are the material-dependent parameters in the LdG free energy, and $T^*$ is the nematic supercooling temperature such that the isotropic phase of the GS phase is unstable for $T < T^*$. The parameter
$\gamma > 0$ is the coupling parameter between the GS molecules and
the clusters \cite{madhusudana2017two_36}.
$N_c$ is the number of
ES molecules in each cluster, and $a_x$ is the mole fraction of
the ES molecules. $J$ accounts for the shape anisotropy of ES
molecules. $E_{el} = \frac{1}{2} \epsilon_0 \Delta \epsilon E^2$ is the electric field energy where $\epsilon_0$
is the free-space permittivity,  $\Delta \epsilon$ is the dielectric anisotropy of the BC system, and $E$ is the applied electric field. We note that there are no cubic terms in $S_c$ above since there is no evidence of first-order isotropic-nematic phase transitions within cybotactic clusters to date. The energy \eqref{Energy1} can be improved, with more physical considerations, but works at a grass-root level for capturing the effects of nano-doping on BC systems.

To model the effects of the suspended ferroelectric nanoparticles, we add additional terms to the free energy in \eqref{Energy1}, keeping  $\n_g$ and $\n_c$ constant unit-vectors as in \cite{patranabish2019one_37}. 
In \cite{lopatina2009theory_13}, the authors show that at a low concentration $\rho_{NP}$ of nanoparticles, the interaction between the liquid crystal molecules and the nano-particles (or anisotropic nano-particle domains in our case) can be modeled by the following free energy density:
\begin{equation}
F_{int} =  - \frac{\Delta \epsilon \rho_{NP} p^2}{270 \pi \epsilon_0 \epsilon^2 R^3} \Q^{\rm LC} : \Q^{\rm NP},
\end{equation}
where 
\begin{equation}
  \Q^{\rm NP} = \frac{2}{3 p^2}  \langle {\bm p} \times {\bm p} \rangle - \frac{1}{2} {\bf I}.
\end{equation}
and ${\bf A} : {\bf B} = A_{ij} B_{ij}$. Here ${\bm p}$ is the electrostatic dipole moment of a spherical nanoparticle, $\langle \rangle$ denotes a spatial average over the distribution of nanoparticles in the nanoparticle domain, $p = {\frac{4}{3} \pi R^3} P$ and $P$ is the polarization. 
 We assume that $\langle {\bm p} \times {\bm p} \rangle$ is constant, i.e., $\langle {\bm p} \times {\bm p} \rangle = \n_{NP}\otimes \n_{NP}$ for some constant unit-vector $\n_{NP}$. Following \cite{lopatina2009theory_13}, we model the energetic contribution of the ferroelectric nanoparticles to the BC system by the interaction energy
\begin{equation}
  F_{int} =  \frac{a_x}{N_c} E_{l o c a l} S_c S_{N P}-\left(1-a_x\right) E_{l o c a l} S_g S_{N P},
\end{equation}
where 
\begin{equation}
  E_{local} =  \frac{\Delta \epsilon \rho_{NP} p^2}{180 \pi \epsilon_0 \epsilon^2 R^3},
\end{equation}
and $a_{\rm NP} = 5 k_B T \rho_{NP}$ \cite{lopatina2009theory_13}. 
Guided by the experimental results, we assume that $\n_{NP}$ and $\n_g$ prefer to be parallel to each other, and $\n_{NP}$ and $\n_c$ prefer to be perpendicular to each other.
The final form of free energy is then given by
\begin{equation}\label{FE1}
\begin{aligned} 
& \mathcal{F}[S_g, S_c, S_{NP}]= \int K_g\left|\nabla S_g\right|^2 +K_c\left|\nabla S_c\right|^2 \\
 & +\left(1-a_x\right)\Bigl\{\tfrac{a_g}{2}\left(T-T^*\right) S_g^2-\tfrac{B_g}{3} S_g^3+\tfrac{C_g}{4} S_g^4-E_{e l} S_g\Bigr\} \\ &+ \tfrac{a_x}{N_c}\left\{-\left(1-a_x\right) \gamma S_g S_c+\tfrac{\alpha_c}{2} S_c^2+\tfrac{\beta_c}{4} S_c^4\right\}-a_x J E_{e l} S_c \\
& +\tfrac{a_{N P}}{2} S_{N P}^2 + \tfrac{a_x}{N_c} E_{\rm  local} S_c S_{NP}-\left(1-a_x\right) E_{\rm local} S_g S_{NP} ~ \dd \x  \ , \end{aligned}
\end{equation}
where the term involving $S_{NP}^2$ is the internal ordering energy of the nanoparticle domain. In \eqref{FE1}, the coupling between $S_g$ and $S_{NP}$ has the opposite sign to the coupling between $S_c$ and $S_{NP}$ on the grounds of the assumed relative orientations of $\n_g$, $\n_{NP}$ and $\n_c$.

 As in \cite{lopatina2009theory_13}, we further assume that $S_{NP}$ can reach its equilibrium instantaneously, so the equilibrium value of $S_{NP}$ can be computed explicitly. This gives
\begin{equation}
\frac{\delta \mathcal{F}}{\delta \mathcal{S}_{NP}} = a_{NP} S_{NP} + \frac{a_x}{N_c} E_{\rm local} S_c - (1 - a_x) E_{\rm local} S_g = 0,
\end{equation}
or equivalently, the equilibrium value of $S_{NP}$ is given by
\begin{equation}\label{SNP}
S_{NP} =  \frac{E_{\rm local}}{a_{NP}} \left( (1 - a_x) S_g - \frac{a_x}{ N_c } S_c \right).
\end{equation}
Substituting (\ref{SNP}) into (\ref{FE1}), we obtain free energy solely in terms of $S_c$ and $S_g$, as given below:
\begin{equation}
  \begin{aligned}
  & \mathcal{F}[S_g, S_c] = \int K_g\left|\nabla S_g\right|^2 +K_c\left|\nabla S_c\right|^2 \\
  & +\left(1-a_x\right)\left\{ (\tfrac{a_g}{2}\left(T-T^*\right) + A') S_g^2-\tfrac{B_g}{3} S_g^3+\tfrac{C_g}{4} S_g^4-E_{e l} S_g\right\} \\ &+\tfrac{a_x}{N_c}\left\{-\left(1-a_x\right) (\gamma + C') S_g S_c + (\tfrac{\alpha_c}{2} + B') S_c^2+\tfrac{\beta_c}{4} S_c^4\right\} \\
  & -a_x J E_{e l} S_c ~ \dd \x,
  \end{aligned}
\end{equation}
where
\begin{equation}
  \begin{aligned}
    & A' = (1 - a_x) (- \frac{E_{\rm local}^2}{2 a_{NP}}), \quad B' =  \frac{a_x}{N_c} (- \frac{E_{\rm local}^2}{2 a_{NP}}), \\
    & C' = - \frac{E_{\rm local}^2}{a_{NP}}. \\
  \end{aligned}
\end{equation} 
Following \cite{patranabish2019one_37}, it is convenient to define
\begin{equation}
  \begin{aligned}
  & A = (1 -  a_x) ( a_g (T - T^*) + 2 A'),~~ B = (1 -  a_x) B_g,\\
  & C = (1 -  a_x) C_g,~~ D = a_x  (1 -  a_x) (\gamma + C') / N_c,  \\
  & E = (1 -  a_x) (E_{el}), ~~ M = (\alpha_c + 2 B') a_x / N_c, \\
  & N = \beta_c a_x / N_c, ~~ H =  a_x J E_{el};  \\
  \end{aligned}
\end{equation} 
and write the free energy as
\begin{equation}
  \begin{aligned}
    & \mathcal{F}= \int_{\Omega}\left(\frac{A}{2} S_g^2-\frac{B}{3} S_g^3+\frac{C}{4} S_g^4 -  E S_g\right) \\
    & + \left(-D S_g S_c+\frac{M}{2} S_c^2+\frac{N}{4} S_c^4 - H S_c \right) \\
    & + K_g\left|\nabla S_g\right|^2+K_c\left|\nabla S_c\right|^2 d \mathbf{x} \ .
    \end{aligned}
\end{equation}

As in \cite{patranabish2019one_37}, we take the domain $\Omega$ to be a one-dimensional domain of length $1$ micron; assuming invariance in two spatial directions so that order parameter profiles only depend on one spatial coordinate.
The free energy can be non-dimensionalized by letting
\begin{equation*}
\textstyle{  \overline{\mathbf{x}}=\mathbf{x} / x_s,~\bar{S}_g=\sqrt{\frac{27 C^2}{12 B^2}} S_g,~\overline{S_c}=\sqrt{\frac{27 C^2}{12 B^2}} S_c,~ \overline{\mathcal{F}}=\frac{27^2 C^3}{72 B^4 x_s^3} \mathcal{F}, }
\end{equation*}
where $x_s = 10^{-8} {\rm m}$ is a characteristic length scale and we could have chosen a different characteristic length scale. 
In the following, we drop all bars for convenience (so that $S_g$ and $S_c$ denote
the scaled order parameters), and the non-dimensionalized free energy is given by
\begin{equation} \label{eq:energy}
  \begin{aligned}
    & \mathcal{F}= \int_{\Omega_s}\left(\frac{t}{2} S_g^2-S_g^3+\frac{1}{2} S_g^4 - C_0 S_g \right) \\ 
    & +\left(-C_1 S_g S_c+C_2 S_c^2+C_3 S_c^4 - C_4 S_c \right) \\
    &+\kappa_g\left(\frac{d S_g}{d x}\right)^2+\kappa_c\left(\frac{d S_c}{d x}\right)^2 d \mathbf{x},
    \end{aligned}
\end{equation}
where
\begin{equation*}
  \begin{aligned}
    t &=\frac{27 A C}{6 B^2}, & C_0 & = \frac{27  E C^2}{4 B^3}   
    & C_1 &=\frac{27 C D}{6 B^2}, & C_2 & =\frac{27 C M}{12 B^2}, \\
    C_3 & =\frac{N}{2 C}, &
    C_4  & =   \frac{27  H C^2}{4 B^3 }  
    & \kappa_g &=\frac{27 C K_g}{6 B^2 x_s^2}, & \kappa_c &=\frac{27 C K_c}{6 B^2 x_s^2} \\
    \end{aligned}
\end{equation*} 
The parameter $t$ is a measure of the rescaled temperature and the polarization, and $C_1$ is a measure of the GS-cluster coupling.

\begin{figure*}[!t]
  \centering
  
  \begin{overpic}[width = 0.49\linewidth]{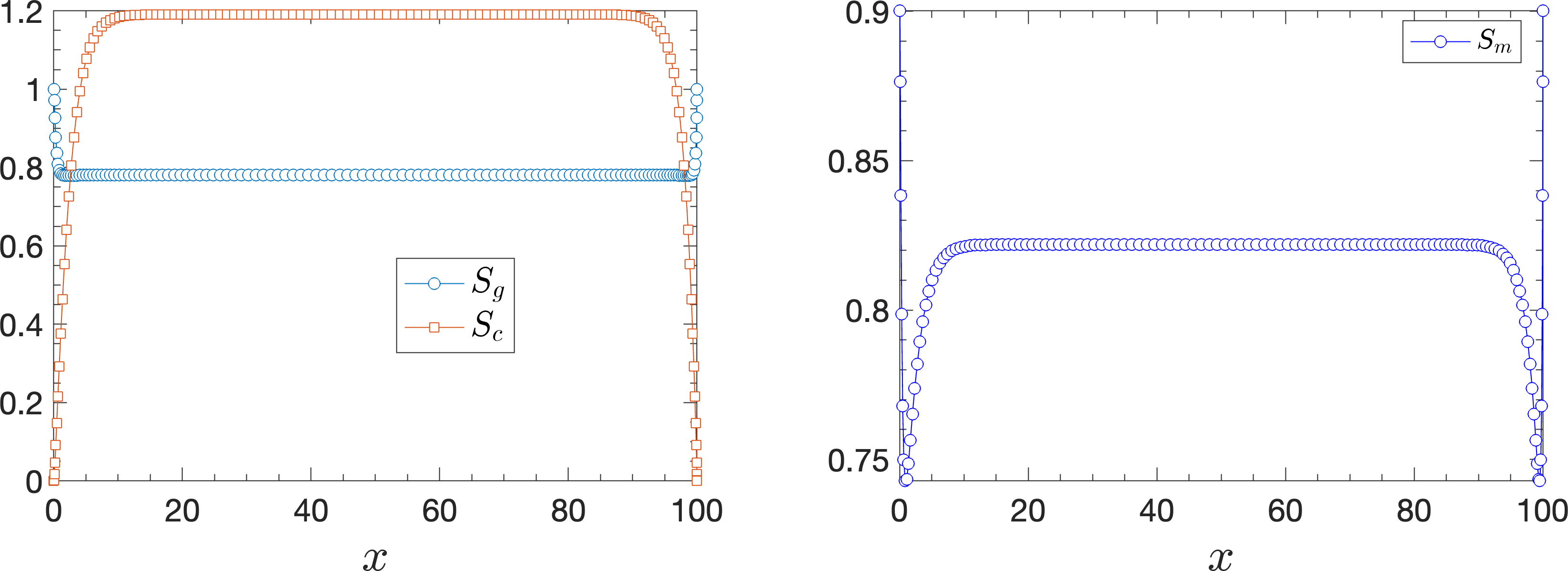}
    \put(40, -5){ \footnotesize (a) $P = 0 ~\rm{C}/\rm{m}^2$}
  \end{overpic}
  \hfill
  \begin{overpic}[width = 0.49\linewidth]{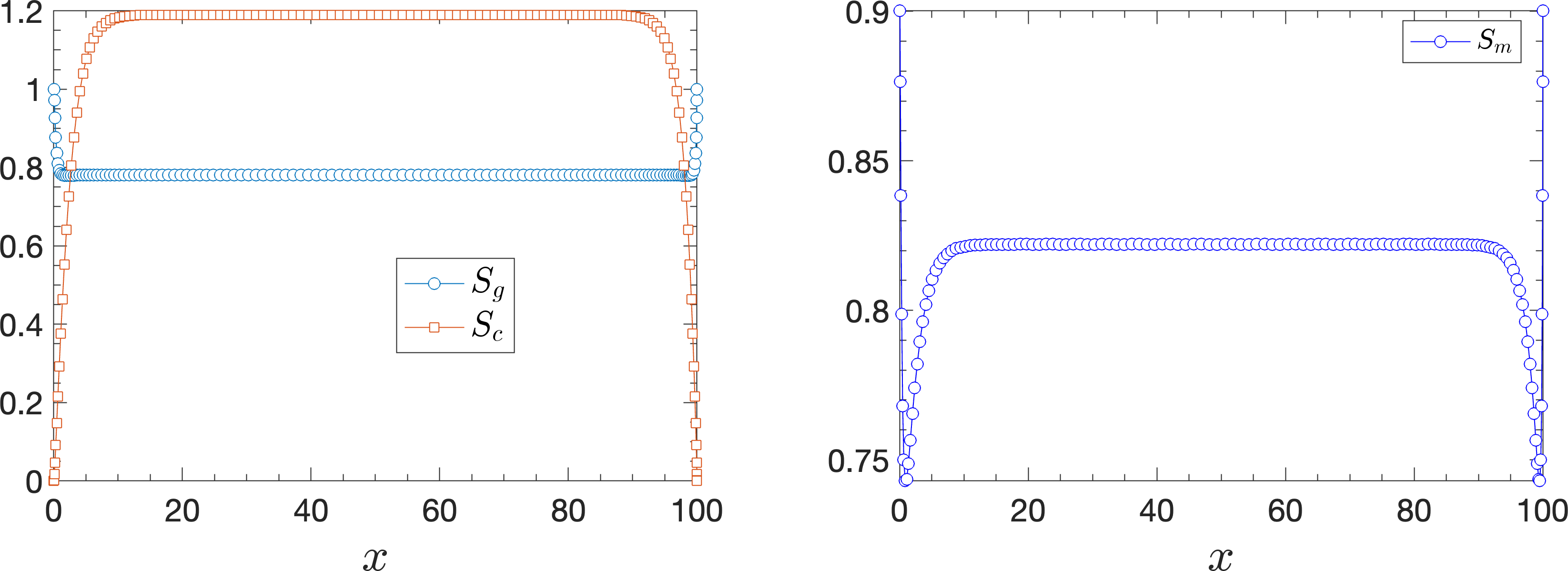}
    \put(40, -5){ \footnotesize (b) $P = 0.04 ~\rm{C}/\rm{m}^2$}
  \end{overpic}

  \vspace{2em}
  \begin{overpic}[width = 0.49\linewidth]{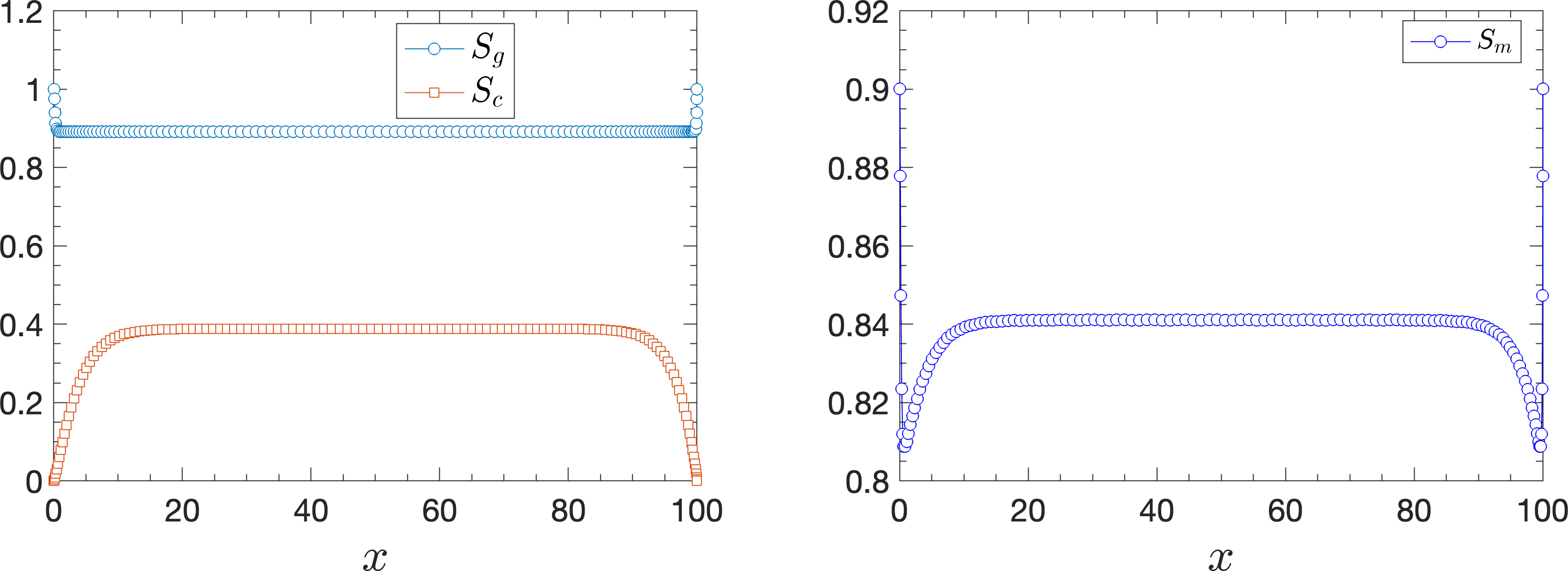}
    \put(40, -5){ \footnotesize (c) $P = 0.25 ~\rm{C}/\rm{m}^2$}
  \end{overpic}
   \hfill
   \begin{overpic}[width = 0.49\linewidth]{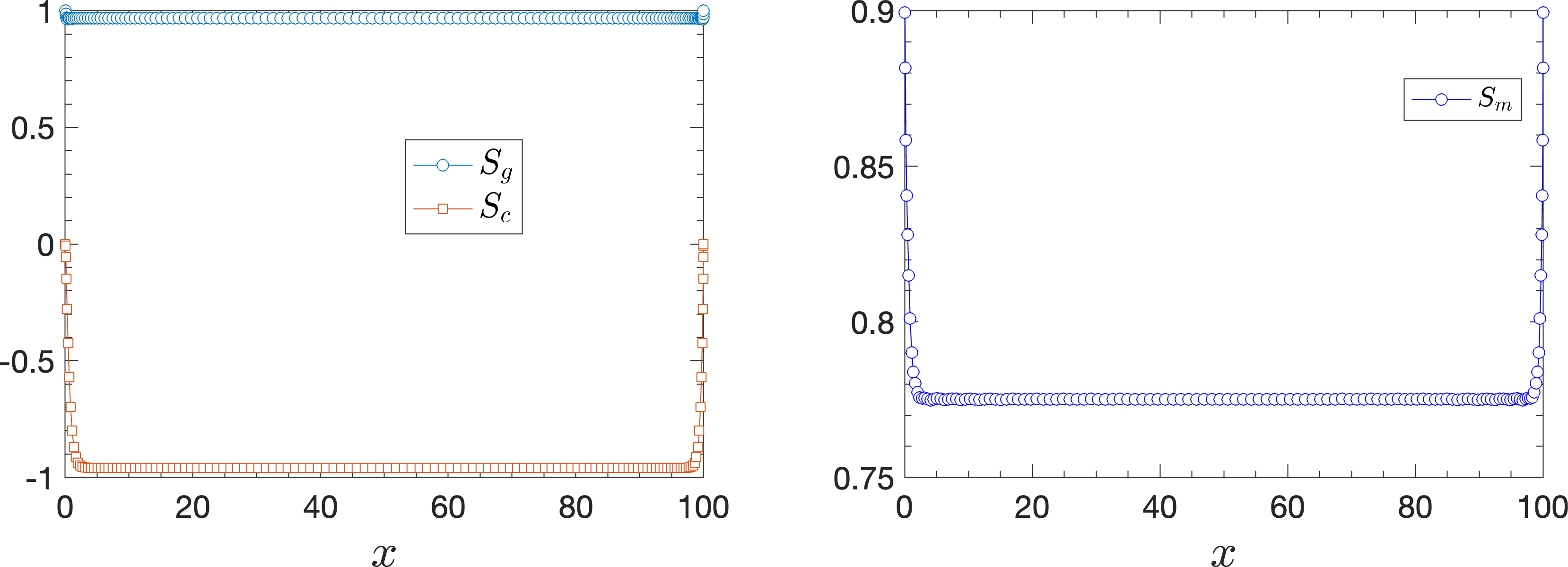}
    \put(40, -5){ \footnotesize (d) $P = 0.48 ~\rm{C}/\rm{m}^2$}
  \end{overpic}

  \vspace{1em}

  \caption{Numerical results for undoped and doped systems with $T = 393.15K$. }\label{undoped_1}
  \end{figure*}
  
The physically observable configurations or experimentally relevant configurations are parameterized by minimizing pairs $(S_g, S_c)$ of \eqref{eq:energy}, subject to the imposed boundary conditions for $S_g$ and $S_c$. The minimizing pair are solution of the corresponding Euler-Lagrange equations, given by:
\begin{equation}\label{EL_S}
\begin{aligned}
  \kappa_g \frac{d^2 S_g}{d x^2} &= 2 S_g^3 - 3  S_g^2 +  t  S_g - C_1 S_c - C_0  \\
  \kappa_c \frac{d^2 S_c}{d x^2} &=4 C_3 S_c^3+ 2 C_2 S_c - C_1 S_g - C_4.  \\
  \end{aligned}
\end{equation}
Since we work with constant $\n_g$, $\n_c$ (which is a strong modeling assumption), we only consider boundary conditions for $S_g$ and $S_c$ and by analogy with the work in \cite{patranabish2019one_37}, we impose 
\begin{equation}\label{BC_t}
\begin{aligned}
  & S_g = \begin{cases} 
    & (3+\sqrt{9-8 t}) / 4 \quad t \leq 1 \\ 
    & 0, \quad t > 1 \\
  \end{cases},  \\  
 &  S_c=0 \quad \text { on } x=0 \text { and } x=100. \\
  \end{aligned}
\end{equation}
This boundary condition is the minimizer of free energy without coupling ($C_1 = 0$) and polarization ($C_0 = C_4 = 0$).
Recall that $t$ depends on the polarization i.e. $t$ depends on $A' = (1 - a_x) (- E_{\rm local}^2)/ 2 a_{NP}$ and the polarization reduces the effective temperature $t$. This, in turn, implies that the boundary conditions for the doped systems with non-zero polarization are different to the boundary conditions for the undoped (pure) system. This is, of course, a modeling choice and one could fix the boundary conditions to be independent of the polarization induced by the nanoparticle domains, but we believe that our qualitative conclusions are independent of the fixed boundary conditions.

 \begin{figure*}[!t]
    \centering
    
    \begin{overpic}[width = 0.49\linewidth]{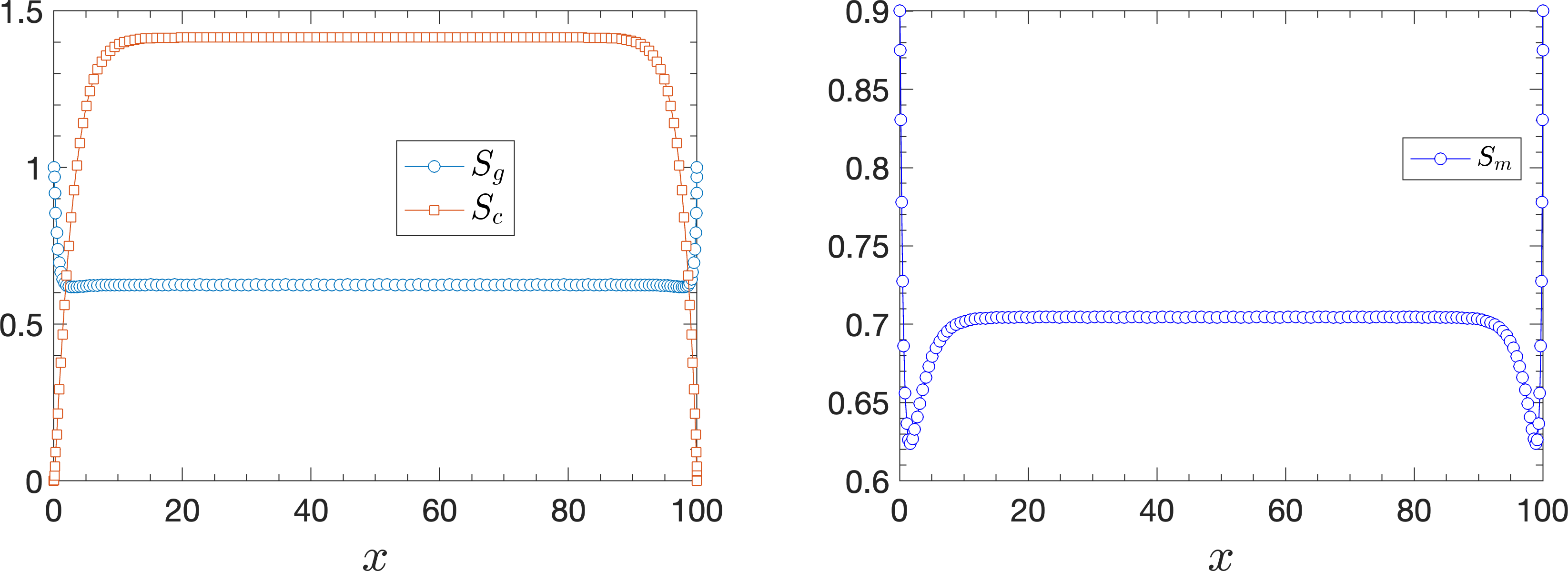}
      \put(40, -5){ \footnotesize (a) $P = 0 ~\rm{C}/\rm{m}^2$}
    \end{overpic}
    \hfill
    \begin{overpic}[width = 0.49\linewidth]{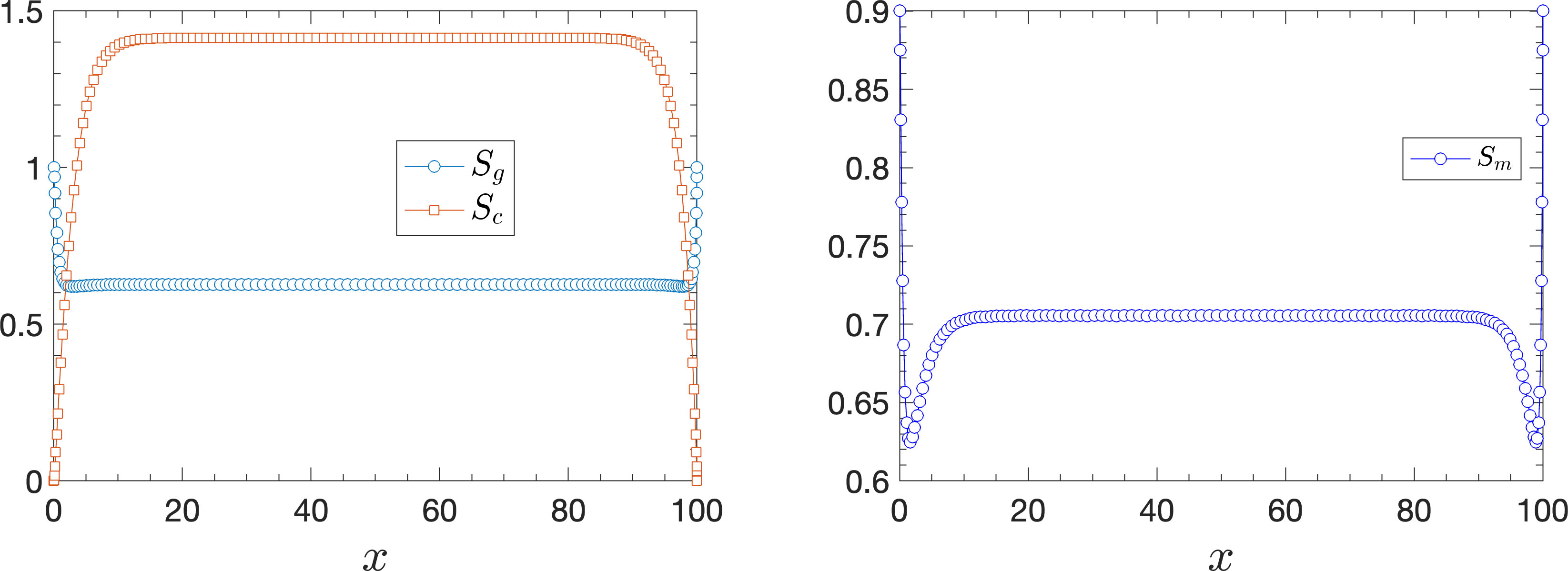}
      \put(40, -5){ \footnotesize (b) $P = 0.04 ~\rm{C}/\rm{m}^2$}
    \end{overpic}
  
    \vspace{2em}
    \begin{overpic}[width = 0.49\linewidth]{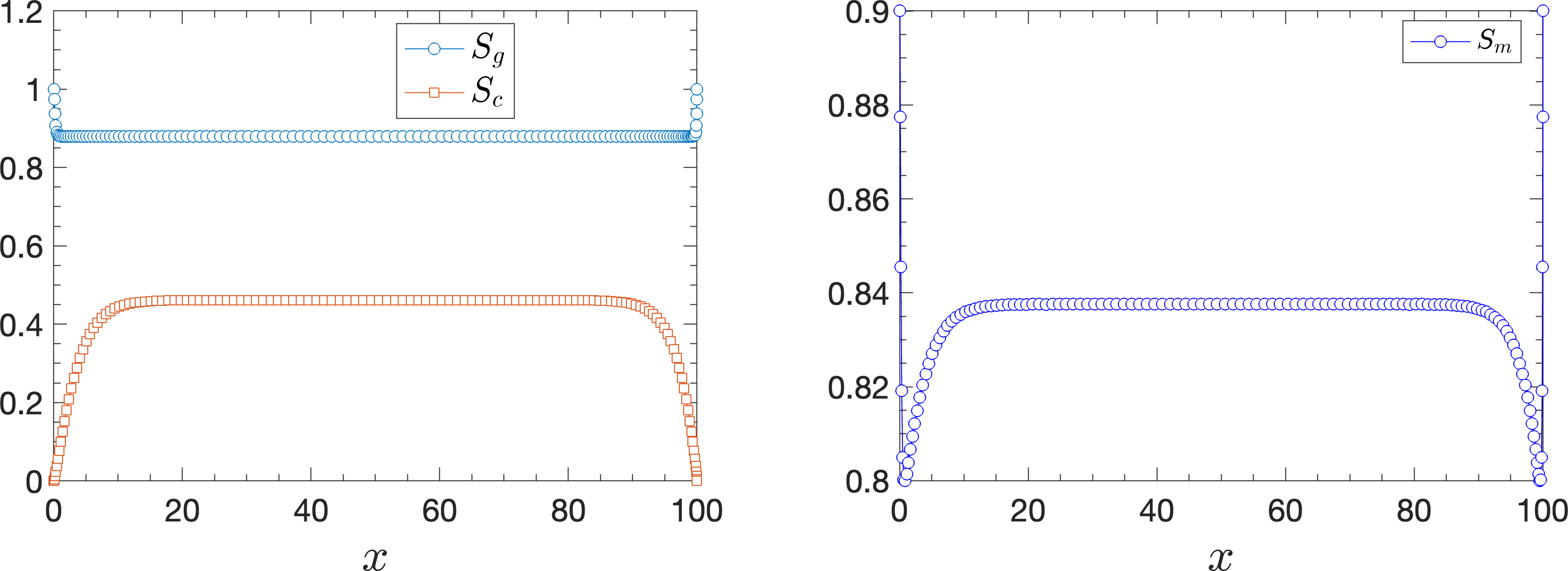}
      \put(40, -5){ \footnotesize (c) $P = 0.25 ~\rm{C}/\rm{m}^2$}
    \end{overpic}
     \hfill
     \begin{overpic}[width = 0.49\linewidth]{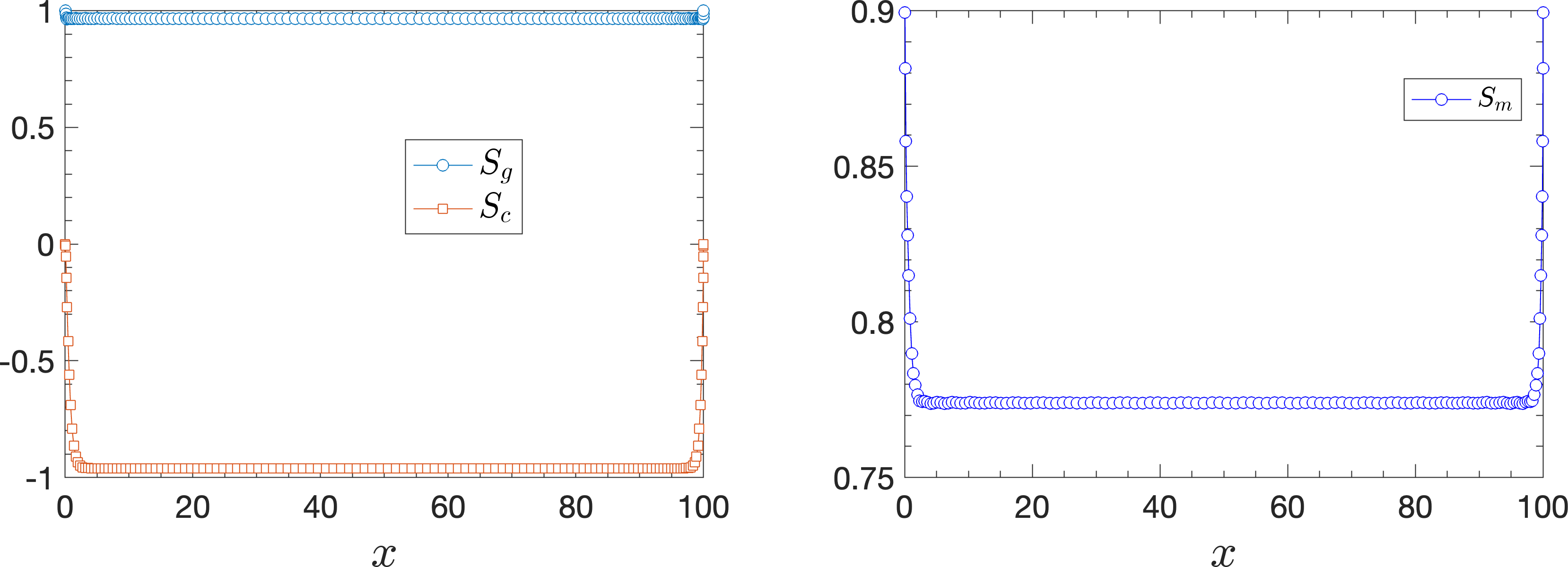}
      \put(40, -5){ \footnotesize (d) $P = 0.48 ~\rm{C}/\rm{m}^2$}
    \end{overpic}
  
    \vspace{1em}
  
    \caption{Numerical results for undoped and doped systems with $T = 413.15K$. }\label{undoped_T_413}
    \end{figure*}

Next, we numerically solve the Euler-Lagrange equations for $S_g$ and $S_c$, subject to the fixed boundary conditions, for different values of the absolute temperature and polarization. The Euler-Lagrange equation is solved by using a Chebyshev spectral method. The number of basis functions is set to 128. The Dirichlet boundary condition is enforced by an augmented Lagrangian method \cite{shen2011spectral_62}.
We do not comment on the physical validity of our parameter values in the numerical simulations, but these values do capture certain qualitative trends commensurate with the experimental observations, namely increased order parameters and reduced/disordered clusters with increasing polarization of the nanoparticle domains.
In the numerical calculations, we set $R = 15 \mathrm{nm}$,  $\rho_{NP} = \varphi_{NP} / (\tfrac{4}{3} \pi R^3)$, where $\varphi_{NP} = 0.00031135317$ is the volume fraction of nano-particles. The remaining parameter values are: \(K_{g}=\) \(K_{c}=K=15 {\rm p} N=15 \times 10^{-12} N \) (under one constant approximation); \(a_{g}=0.04, B_{g}=1.7, C_{g}=\) \(4.5, \alpha_{c}=0.22, \beta_{c}=4.0, \quad\) and \(\gamma=5.0 \quad\left(a_{g}, B_{g}, C_{g}, \gamma, \alpha_{c}\right.\), \(\beta_{c}\) in $10^6/4$ Syst{\`e}me
International units); \(T^{*}=418.15 \mathrm{~K}, N_{c}=50, J=\) \(1.2, E_{\mathrm{el}}=0 ~ \mathrm{J} / \mathrm{m}^{3}\) (so that $C_0 = C_4 = 0$).

We focus on the effects of $P$, which manifests in $t$, $C_1$ and $C_2$ in Eq. \ref{EL_S}, for three different temperatures, \(T=\) 423.15 K (for the \(T > T^*\) case), \(413.15 \mathrm{~K}\) (for the \(T < T^*\) case),  and \(393.15 \mathrm{~K}\) (for the \(T \ll T^*\) case). We conjecture that the model cannot capture Region $3$ of the experimental orientational order parameter plot in Figure~\ref{fig:9}, since there is no term that accounts for the temperature-dependence of the cybotactic clusters in \eqref{eq:energy}. To compare with the experimental results, we define the bulk mean order parameter $S_m$ to be
\begin{equation}
  S_m = (1 - a_x) S_g + a_x S_c,
\end{equation}
which is a weighted scalar order parameter. If $S_g \neq 0$ on the boundary, we scale the order parameters ($S_g$, $S_c$ and $S_m$) by the fixed boundary value of $S_g$ in \eqref{BC_t}. This is to ensure that the order parameter values remain in some physically sensible range, compared to the fixed boundary conditions. Recall from \cite{majumdar2010equilibrium_63} that LdG order parameters can be greater than $+1$, or less than $-1$, since these are phenomenological theories that are independent of the definition of the LdG $\Qvec$-tensor as the second moment of a probability distribution function for the molecular orientations. In fact, it is possible that these theories cease to be valid for low temperatures or large values of polarization, and hence, the reader is requested to focus on the trends in the numerical calculations as opposed to the precise numerical values of $S_g, S_c$ and $S_m$.

\subsection{$T = 393.15 K$}
This is a relatively low temperature, for which the isotropic phase with $S_g = S_c=0$ is energetically expensive. The values of $t$, $C_1, C_2$ depend on $P$, and we consider four different values of $P$ inspired by the experimental results. The values of $t, C_1, C_2$ have to be calculated separately for each value of $P$.

For $P = 0 ~\rm{C} / \rm{m}^2$, we have $t =-7.00692, C_1 =0.0700692, C_2 =0.00342561, C_3 =0.00395062$.
For $P = 0.04 ~ \rm{C} / \rm{m}^2$ ($t =-7.02526, C_1 =0.0700284, C_2 =0.00342551, C_3 =0.00395062$), the $S_g$ and $S_c$ profiles only change slightly compared to the case with $P = 0$. Indeed, $S_m$ only increases by about $2 \time 10^{-3}$ in the bulk, compared to the $P=0$ case. For $P = 0.25 ~ \rm{C} / \rm{m}^2$ ($t =-34.9954, C_1 =0.0078725, C_2 =0.00328739, C_3 =0.00395062$), $S_g$ increases sharply compared to the undoped case but $S_c$ decreases, consistent with the disordering of clusters with increasing polarization. However, $S_c$ remains non-negative in this case.  Finally, for $P = 0.48 ~ \rm{C}/\rm{m}^2$ ($t =-387.358, C_1 =-0.775156, C_2 =0.00154733, C_3 =0.00395062$), the effective temperature is very low and although $S_g$ increases to its maximum value almost everywhere and $|S_c|$ is large, $S_c$ is negative in the bulk, corroborating the experimental claim of disordered cybotactic clusters for very high values of the polarization used in this article. A negative value of $S_c$ indicates that the cluster molecules lie in a plane orthogonal to $\n_c$, with no preferred direction in the orthogonal plane containing the cluster. We speculate that disordered clusters correlate to smaller or reduced cluster sizes, as suggested by the dielectric spectroscopy results. Comparing the plots of $S_m$, the bulk values of $S_m$  (away from $x=0$ and $x=100$) do not change appreciably with $P$ and in fact, there is a slight decrease in the bulk value of $S_m$ for $P= 0.48 ~ \rm{C}/\rm{m}^2$, compared to the undoped system. Although not in perfect agreement with Region $2$ and $3$ of Figure~\ref{fig:9}, this indicates that the mean order parameters of the doped and undoped systems do not differ appreciably for very low temperatures, as suggested by Figure~\ref{fig:9}.

\begin{figure*}[!tbh]
  \centering
  
  \begin{overpic}[width = 0.49\linewidth]{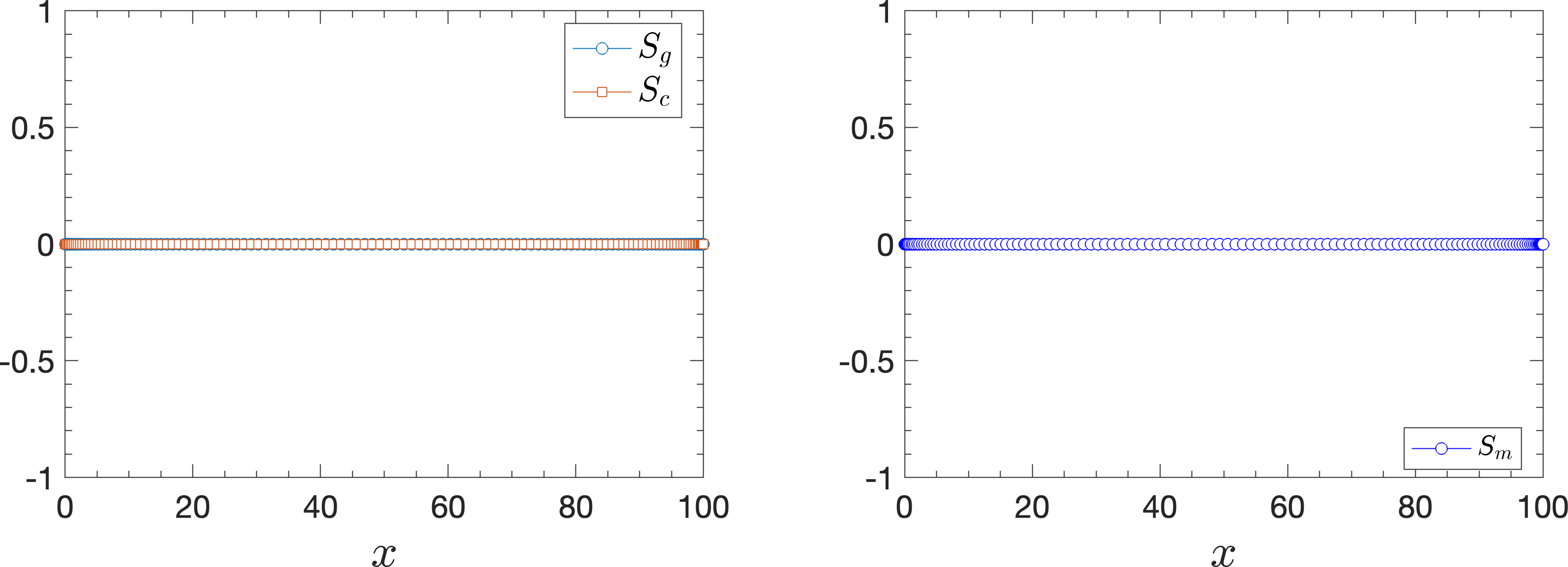}
    \put(40, -5){ \footnotesize (a) $P = 0 ~\rm{C}/\rm{m}^2$}
  \end{overpic}
  \hfill
  \begin{overpic}[width = 0.49\linewidth]{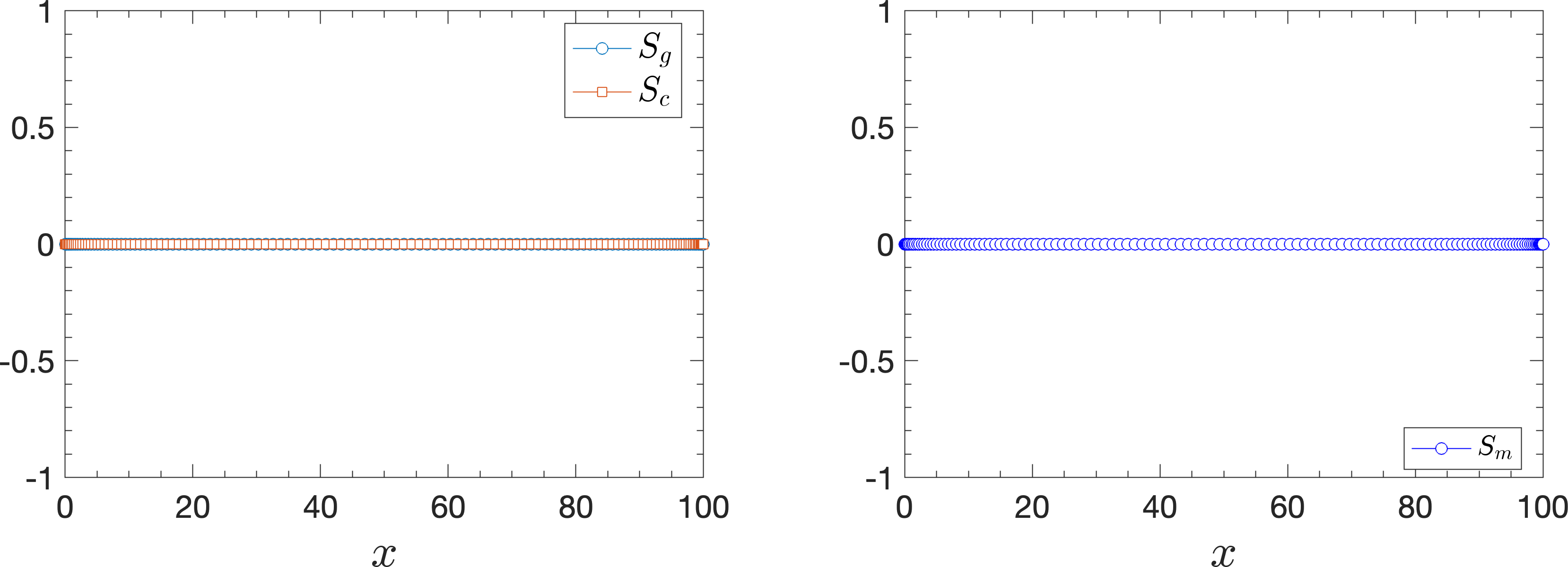}
    \put(40, -5){ \footnotesize (b) $P = 0.04 ~\rm{C}/\rm{m}^2$}
  \end{overpic}

  \vspace{2em}
  \begin{overpic}[width = 0.49\linewidth]{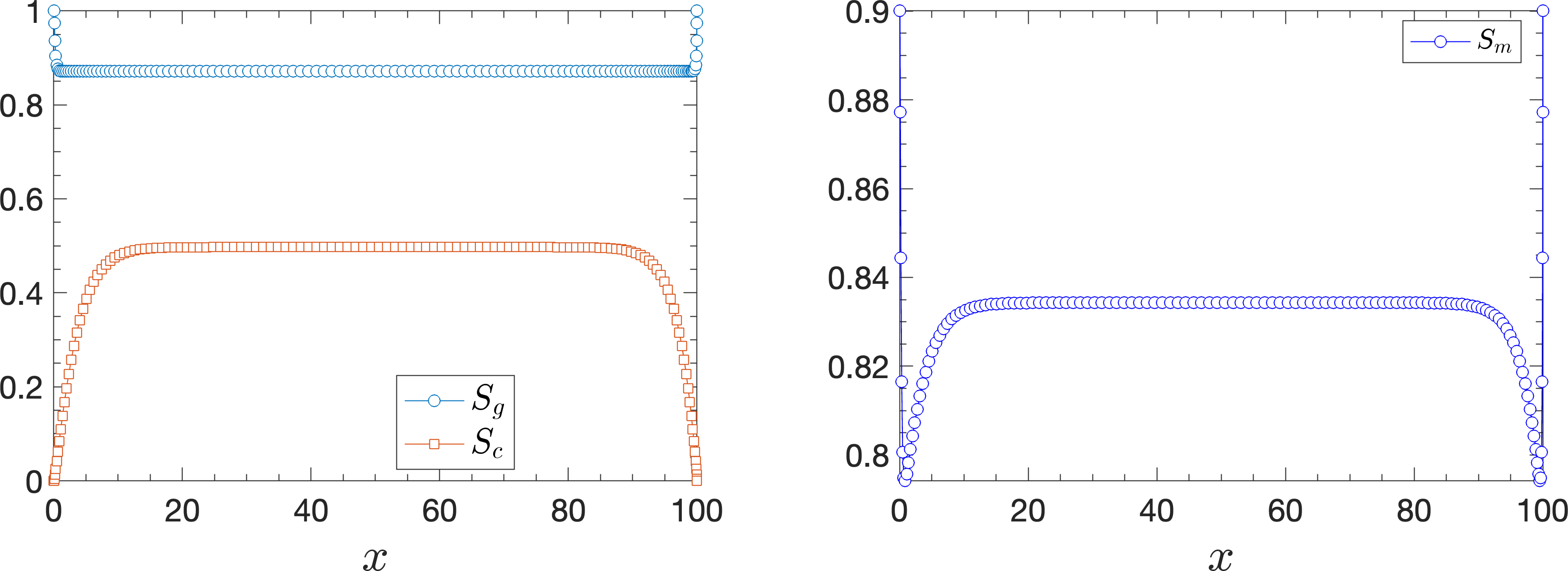}
    \put(40, -5){ \footnotesize (c) $P = 0.25 ~\rm{C}/\rm{m}^2$}
  \end{overpic}
   \hfill
   \begin{overpic}[width = 0.49\linewidth]{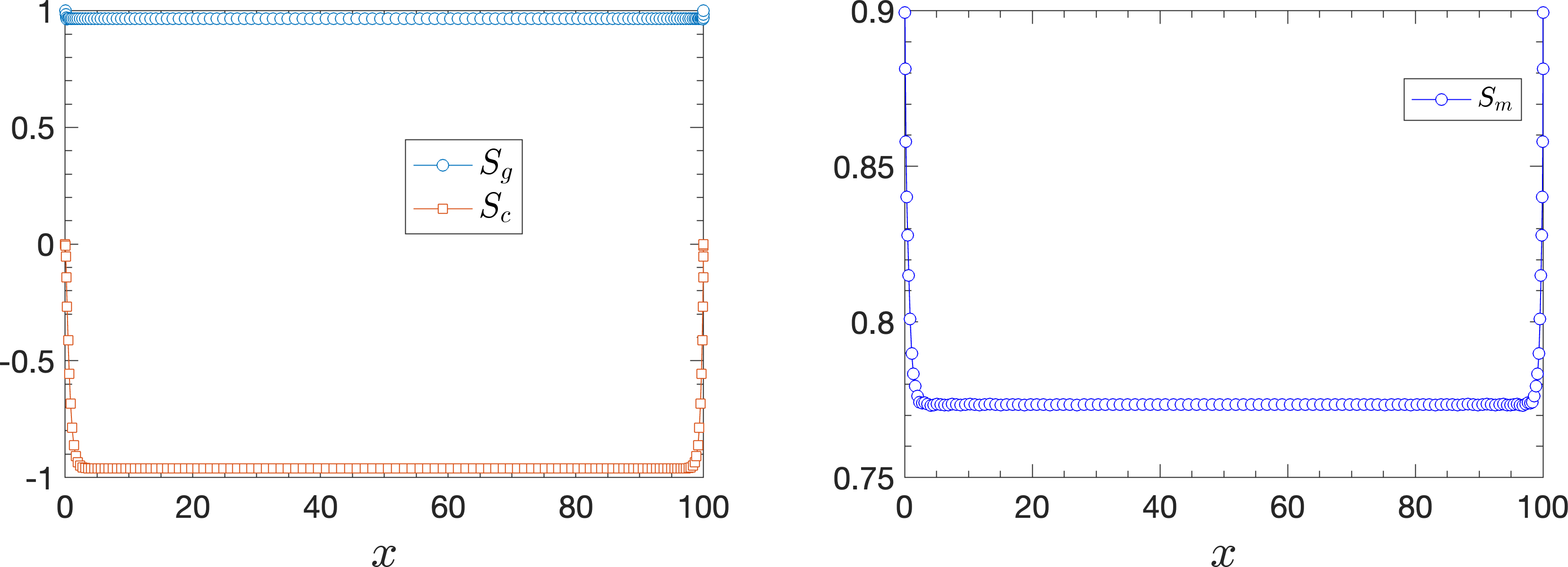}
    \put(40, -5){ \footnotesize (d) $P = 0.48 ~\rm{C}/\rm{m}^2$}
  \end{overpic}

  \vspace{1em}

  \caption{Numerical results for undoped system and doped systems with $T = 423.15K$. }\label{undoped_T_423}
  \end{figure*}


  \subsection{$T = 413.15 K$}

  The results are similar to the case with $T = 393.15 K$. 
  For $P =0 ~ \rm{C}/\rm{m}^2$, we have $t =-1.40138, C_1 =0.0700692, C_2 =0.00342561, C_3 =0.00395062$.
  For $P = 0.04 ~\rm{C}/\rm{m}^2$ ($t =-1.41884, C_1 =0.0700304, C_2 =0.00342552, C_3 =0.00395062$), the order parameter of $S_g$ and $S_c$ only change slightly compared to the undoped case. For $P = 0.25 ~\rm{C}/\rm{m}^2$ ($t =-28.035, C_1 =0.0108834, C_2 =0.00329408, C_3 =0.00395062$), we observe that $S_g$ and $S_m$ increase, $S_c$ decreases but remains non-negative.  For $P = 0.48 ~\rm{C}/\rm{m}^2$ ($t =-363.34, C_1 =-0.73424, C_2 =0.00163825, C_3 =0.00395062$), although $S_m$ increases and both $|S_g|$ and $|S_c|$ increase in magnitude, but $S_c$ becomes negative, for which we propose the same arguments as above. The qualitative trends remain the same - $S_g$ increases with $P$, $S_c$ decreases and becomes negative with increasing $P$ but for this value of $T$ (higher than in Figure~\ref{undoped_1}), $S_m$ also increases with $P$ and the bulk value of $S_m$ for $P = 0.48$ is approximately $0.78$, compared to approximately $S_m \approx 0.7$ for $P=0$. This is in qualitative agreement with the trends recorded in Region $2$ of Figure~\ref{fig:9}.

\subsection{$T = 423.15 K$}

Since $T =423.15 K > T^*$, we have $S_g = 0$ ($t =1.40138, C_1 =0.0700692, C_2 =0.00342561, C_3 =0.00395062$) for $P = 0 ~\rm{C}/\rm{m}^2$ and $P = 0.04 ~\rm{C}/\rm{m}^2$ ($t =1.38434, C_1 =0.0700313, C_2 =0.00342552, C_3 =0.00395062$).
 For $P = 0.25 ~ \rm{C}/\rm{m}^2$ ($t =-24.6028, C_1 =0.012282, C_2 =0.00329719, C_3 =0.00395062$), the effective temperature, $t$, decreases and consequently, $S_g > 0$, $S_c>0$ and $S_m$ increases compared to the smaller values of $P$.  For $P = 0.48 ~\rm{C}/\rm{m}^2$ ($t =-351.984, C_1 =-0.715232, C_2 =0.00168049, C_3 =0.00395062$), the effective temperature is low and we are essentially deep in the nematic phase. This manifests in elevated values of $S_m$, $|S_g|$ and $|S_c|$ and $S_c$ becomes negative in the interior. We also note that the boundary values of $S_g$ are non-zero for $P = 0.25 ~\rm{C}/\rm{m}^2$ and $P = 0.48 ~\rm{C}/\rm{m}^2$. These numerical results are not in agreement with Region $1$ of Figure~\ref{fig:9}, and this is again because there is no term in the free energy that captures the temperature dependence of the cybotactic clusters. We speculate that whilst the polarization does reduce the effective temperature and hence increase $S_g$, the coupling between the effective temperature and the polarization depends on the actual temperature i.e. for high temperatures, the polarization has a weaker effect on the effective temperature and hence, the order parameter plots of the doped and undoped systems do not differ much for relatively high temperatures or just below the $T_{IN}$ in Figure~\ref{fig:9}. This is not accounted for in the model at present.

 Intuitively, it is clear that as the effective temperature $t$ decreases and the BC system is pushed deep into the nematic phase, $S_g$ increases in the interior. It is perhaps less clear as to why $S_c$ becomes negative in this simplistic free energy model, with increasing $t$ and $P$. The answer lies in relative values of $C_1$ and $C_2$ in these simulations. Looking at \eqref{eq:energy}, $C_1$ is negative for non-zero $P$, so that the term $- C_1 S_g S_c$ favours negative $S_c$. The values of $C_2$ are at least two orders of magnitude smaller than those of $C_1$, and one can check that a negative value of $S_c$ (with large $|S_c|$) minimizes the polynomial, $- C_1 S_g S_c + C_2 S_c^2$ in \eqref{eq:energy}. This is far from proof but gives heuristic insights into the origin of the disordering mechanisms of cybotactic clusters or reduced cluster sizes induced by ferroelectric nanoparticle domains. The numerical results also highlight several limitations of the model, when compared to Figure~\ref{fig:9} e.g., we speculate that the model only captures some features of Region $2$ of Figure~\ref{fig:9} and we also note that the bulk value of $S_m$ is approximately independent of temperature for $P = 0.25 ~\rm{C}/\rm{m}^2$ and $P = 0.48 ~\rm{C}/\rm{m}^2$. From a modeling perspective, this is simply because larger values of $P$ dominate the definition of the effective temperature so that the actual value of $T$ becomes less significant for the value of $S_m$. This is however, inconsistent with Figure~\ref{fig:9} where there is a temperature-dependence of the order parameter plot, and again, this could imply that our model breaks down for large values of $P$, comparable to other phenomenological descriptions which can have limited regions of validity.

\section{Conclusion}
We study an experimental suspension of multiferroic nanoparticles in a bent-core nematic liquid crystal and study the temperature dependence of the optical textures, birefringence, and perform dielectric spectroscopy measurements. By extracting the order parameter plots from the birefringence data, and combining them with the results from the dielectric spectroscopy, we deduce that these nanoparticles enhance the overall ordering of the system but shrink the cybotactic clusters. The spectroscopy data suggests that the cybotactic clusters shrink or get smaller in the presence of the nanoparticles, compared to the pure system, somewhat corroborated by the order parameter plot in Figure~\ref{fig:9}. We develop a phenomenological LdG free energy to qualitatively explain the experimental data. The model has some merits -  it demonstrates how the polarization manifests in the effective temperature and suggests that nanoparticles with large values of $P$ effectively push the system to lower temperatures, offering a heuristic explanation of the enhanced order parameter of the doped system compared to the undoped system. The model also demonstrates that the cybotactic clusters get disordered with increasing polarization,  which is one possible explanation for the reduced cluster sizes for the doped system, compared to the undoped case. The model only captures the temperature-dependence of the mean order parameter roughly, but we do not have reliable parameter values for the phenomenological free energy and the free energy also neglects multiple physical considerations - spatially varying $\n_g$ and $\n_c$, the temperature-dependence of the polarization itself, more accurate modeling of the suspended nanoparticles and their interaction with the GS and cluster molecules, etc. Future work can include the development of more sophisticated mathematical models for such nano-doped systems that could guide new experiments and applications.

\begin{acknowledgements}
We acknowledge CRF and NRF, IIT Delhi for proving the characterization facility. We also acknowledge Defence Research and Development Organisation, India for funding support with a project grant DFTM/03/3203/P/01/JATC$-$P2QP$-$01.   D.M. acknowledges UGC for financial support under a full$-$time research fellowship. S.K. acknowledges CSIR (09/947(0254)/2020$-$EMR$-$I) for a Ph.D. fellowship. G.M. acknowledges SERB-SIRE grant SIR/2022/000175, University of Science $\And$ Technology Meghalaya (USTM), India, and the University of York, UK. D.M. and A.S. would like to thank Prof. S. K. Pal (IISER Mohali) for providing the liquid crystal sample. D.M. thanks Dr. Sourav Patranabish for the useful discussion during the preparation of the manuscript. Our work is inspired by \cite{patranabish2021quantum_38} and the authors acknowledge that.

S.K. and G.M. have designed, synthesized and characterized the 8-F-OH compound. D.M. and A.S. have performed the experiments and analyzed the results. Y.W. and A.M. have led the modeling section and numerical calculations, and have made significant contributions to the interpretation of the experimental results.

\end{acknowledgements}

\bibliographystyle{apsrev4-1}
\bibliography{biblio}

\begin{thebibliography}{63}%
\makeatletter
\providecommand \@ifxundefined [1]{%
 \@ifx{#1\undefined}
}%
\providecommand \@ifnum [1]{%
 \ifnum #1\expandafter \@firstoftwo
 \else \expandafter \@secondoftwo
 \fi
}%
\providecommand \@ifx [1]{%
 \ifx #1\expandafter \@firstoftwo
 \else \expandafter \@secondoftwo
 \fi
}%
\providecommand \natexlab [1]{#1}%
\providecommand \enquote  [1]{``#1''}%
\providecommand \bibnamefont  [1]{#1}%
\providecommand \bibfnamefont [1]{#1}%
\providecommand \citenamefont [1]{#1}%
\providecommand \href@noop [0]{\@secondoftwo}%
\providecommand \href [0]{\begingroup \@sanitize@url \@href}%
\providecommand \@href[1]{\@@startlink{#1}\@@href}%
\providecommand \@@href[1]{\endgroup#1\@@endlink}%
\providecommand \@sanitize@url [0]{\catcode `\\12\catcode `\$12\catcode `\&12\catcode `\#12\catcode `\^12\catcode `\_12\catcode `\%12\relax}%
\providecommand \@@startlink[1]{}%
\providecommand \@@endlink[0]{}%
\providecommand \url  [0]{\begingroup\@sanitize@url \@url }%
\providecommand \@url [1]{\endgroup\@href {#1}{\urlprefix }}%
\providecommand \urlprefix  [0]{URL }%
\providecommand \Eprint [0]{\href }%
\providecommand \doibase [0]{http://dx.doi.org/}%
\providecommand \selectlanguage [0]{\@gobble}%
\providecommand \bibinfo  [0]{\@secondoftwo}%
\providecommand \bibfield  [0]{\@secondoftwo}%
\providecommand \translation [1]{[#1]}%
\providecommand \BibitemOpen [0]{}%
\providecommand \bibitemStop [0]{}%
\providecommand \bibitemNoStop [0]{.\EOS\space}%
\providecommand \EOS [0]{\spacefactor3000\relax}%
\providecommand \BibitemShut  [1]{\csname bibitem#1\endcsname}%
\let\auto@bib@innerbib\@empty
\bibitem [{\citenamefont {De~Gennes}\ and\ \citenamefont {Prost}(1993)}]{de1993physics_2}%
  \BibitemOpen
  \bibfield  {author} {\bibinfo {author} {\bibfnamefont {P.-G.}\ \bibnamefont {De~Gennes}}\ and\ \bibinfo {author} {\bibfnamefont {J.}~\bibnamefont {Prost}},\ }\href@noop {} {\emph {\bibinfo {title} {The physics of liquid crystals}}},\ \bibinfo {number} {83}\ (\bibinfo  {publisher} {Oxford university press},\ \bibinfo {year} {1993})\BibitemShut {NoStop}%
\bibitem [{\citenamefont {Collings}\ and\ \citenamefont {Goodby}(2019)}]{collings2019introduction_1}%
  \BibitemOpen
  \bibfield  {author} {\bibinfo {author} {\bibfnamefont {P.~J.}\ \bibnamefont {Collings}}\ and\ \bibinfo {author} {\bibfnamefont {J.~W.}\ \bibnamefont {Goodby}},\ }\href@noop {} {\emph {\bibinfo {title} {Introduction to liquid crystals: chemistry and physics}}}\ (\bibinfo  {publisher} {Crc Press},\ \bibinfo {year} {2019})\BibitemShut {NoStop}%
\bibitem [{\citenamefont {Marino}\ \emph {et~al.}(2012{\natexlab{a}})\citenamefont {Marino}, \citenamefont {Ionescu}, \citenamefont {Marino},\ and\ \citenamefont {Scaramuzza}}]{marino2012dielectric_3}%
  \BibitemOpen
  \bibfield  {author} {\bibinfo {author} {\bibfnamefont {L.}~\bibnamefont {Marino}}, \bibinfo {author} {\bibfnamefont {A.~T.}\ \bibnamefont {Ionescu}}, \bibinfo {author} {\bibfnamefont {S.}~\bibnamefont {Marino}}, \ and\ \bibinfo {author} {\bibfnamefont {N.}~\bibnamefont {Scaramuzza}},\ }\href@noop {} {\bibfield  {journal} {\bibinfo  {journal} {Journal of Applied Physics}\ }\textbf {\bibinfo {volume} {112}},\ \bibinfo {pages} {114113} (\bibinfo {year} {2012}{\natexlab{a}})}\BibitemShut {NoStop}%
\bibitem [{\citenamefont {Ghosh}\ \emph {et~al.}(2014)\citenamefont {Ghosh}, \citenamefont {Begum}, \citenamefont {Turlapati}, \citenamefont {Roy}, \citenamefont {Das},\ and\ \citenamefont {Rao}}]{ghosh2014ferroelectric_4}%
  \BibitemOpen
  \bibfield  {author} {\bibinfo {author} {\bibfnamefont {S.}~\bibnamefont {Ghosh}}, \bibinfo {author} {\bibfnamefont {N.}~\bibnamefont {Begum}}, \bibinfo {author} {\bibfnamefont {S.}~\bibnamefont {Turlapati}}, \bibinfo {author} {\bibfnamefont {S.~K.}\ \bibnamefont {Roy}}, \bibinfo {author} {\bibfnamefont {A.~K.}\ \bibnamefont {Das}}, \ and\ \bibinfo {author} {\bibfnamefont {N.~V.}\ \bibnamefont {Rao}},\ }\href@noop {} {\bibfield  {journal} {\bibinfo  {journal} {Journal of Materials Chemistry C}\ }\textbf {\bibinfo {volume} {2}},\ \bibinfo {pages} {425} (\bibinfo {year} {2014})}\BibitemShut {NoStop}%
\bibitem [{\citenamefont {Bailey}\ \emph {et~al.}(2009)\citenamefont {Bailey}, \citenamefont {Fodor-Csorba}, \citenamefont {Gleeson}, \citenamefont {Sprunt},\ and\ \citenamefont {J{\'a}kli}}]{bailey2009rheological_5}%
  \BibitemOpen
  \bibfield  {author} {\bibinfo {author} {\bibfnamefont {C.}~\bibnamefont {Bailey}}, \bibinfo {author} {\bibfnamefont {K.}~\bibnamefont {Fodor-Csorba}}, \bibinfo {author} {\bibfnamefont {J.~T.}\ \bibnamefont {Gleeson}}, \bibinfo {author} {\bibfnamefont {S.~N.}\ \bibnamefont {Sprunt}}, \ and\ \bibinfo {author} {\bibfnamefont {A.}~\bibnamefont {J{\'a}kli}},\ }\href@noop {} {\bibfield  {journal} {\bibinfo  {journal} {Soft Matter}\ }\textbf {\bibinfo {volume} {5}},\ \bibinfo {pages} {3618} (\bibinfo {year} {2009})}\BibitemShut {NoStop}%
\bibitem [{\citenamefont {Takezoe}\ and\ \citenamefont {Takanishi}(2006)}]{takezoe2006bent_6}%
  \BibitemOpen
  \bibfield  {author} {\bibinfo {author} {\bibfnamefont {H.}~\bibnamefont {Takezoe}}\ and\ \bibinfo {author} {\bibfnamefont {Y.}~\bibnamefont {Takanishi}},\ }\href@noop {} {\bibfield  {journal} {\bibinfo  {journal} {Japanese journal of applied physics}\ }\textbf {\bibinfo {volume} {45}},\ \bibinfo {pages} {597} (\bibinfo {year} {2006})}\BibitemShut {NoStop}%
\bibitem [{\citenamefont {Francescangeli}\ \emph {et~al.}(2014)\citenamefont {Francescangeli}, \citenamefont {Vita},\ and\ \citenamefont {Samulski}}]{francescangeli2014cybotactic_7}%
  \BibitemOpen
  \bibfield  {author} {\bibinfo {author} {\bibfnamefont {O.}~\bibnamefont {Francescangeli}}, \bibinfo {author} {\bibfnamefont {F.}~\bibnamefont {Vita}}, \ and\ \bibinfo {author} {\bibfnamefont {E.~T.}\ \bibnamefont {Samulski}},\ }\href@noop {} {\bibfield  {journal} {\bibinfo  {journal} {Soft Matter}\ }\textbf {\bibinfo {volume} {10}},\ \bibinfo {pages} {7685} (\bibinfo {year} {2014})}\BibitemShut {NoStop}%
\bibitem [{\citenamefont {Keith}\ \emph {et~al.}(2010)\citenamefont {Keith}, \citenamefont {Lehmann}, \citenamefont {Baumeister}, \citenamefont {Prehm},\ and\ \citenamefont {Tschierske}}]{keith2010nematic_8}%
  \BibitemOpen
  \bibfield  {author} {\bibinfo {author} {\bibfnamefont {C.}~\bibnamefont {Keith}}, \bibinfo {author} {\bibfnamefont {A.}~\bibnamefont {Lehmann}}, \bibinfo {author} {\bibfnamefont {U.}~\bibnamefont {Baumeister}}, \bibinfo {author} {\bibfnamefont {M.}~\bibnamefont {Prehm}}, \ and\ \bibinfo {author} {\bibfnamefont {C.}~\bibnamefont {Tschierske}},\ }\href@noop {} {\bibfield  {journal} {\bibinfo  {journal} {Soft Matter}\ }\textbf {\bibinfo {volume} {6}},\ \bibinfo {pages} {1704} (\bibinfo {year} {2010})}\BibitemShut {NoStop}%
\bibitem [{\citenamefont {Kumar}\ and\ \citenamefont {Prasad}(2018)}]{kumar2018ferroelectric_9}%
  \BibitemOpen
  \bibfield  {author} {\bibinfo {author} {\bibfnamefont {J.}~\bibnamefont {Kumar}}\ and\ \bibinfo {author} {\bibfnamefont {V.}~\bibnamefont {Prasad}},\ }\href@noop {} {\bibfield  {journal} {\bibinfo  {journal} {The Journal of Physical Chemistry B}\ }\textbf {\bibinfo {volume} {122}},\ \bibinfo {pages} {2998} (\bibinfo {year} {2018})}\BibitemShut {NoStop}%
\bibitem [{\citenamefont {Kumar}\ \emph {et~al.}(2018{\natexlab{a}})\citenamefont {Kumar}, \citenamefont {Prasad},\ and\ \citenamefont {Manjunath}}]{kumar2018quantum_10}%
  \BibitemOpen
  \bibfield  {author} {\bibinfo {author} {\bibfnamefont {J.}~\bibnamefont {Kumar}}, \bibinfo {author} {\bibfnamefont {V.}~\bibnamefont {Prasad}}, \ and\ \bibinfo {author} {\bibfnamefont {M.}~\bibnamefont {Manjunath}},\ }\href@noop {} {\bibfield  {journal} {\bibinfo  {journal} {Journal of Molecular Liquids}\ }\textbf {\bibinfo {volume} {266}},\ \bibinfo {pages} {10} (\bibinfo {year} {2018}{\natexlab{a}})}\BibitemShut {NoStop}%
\bibitem [{\citenamefont {Khan}\ \emph {et~al.}(2017)\citenamefont {Khan}, \citenamefont {Turlapati}, \citenamefont {Rao},\ and\ \citenamefont {Ghosh}}]{khan2017elastic_11}%
  \BibitemOpen
  \bibfield  {author} {\bibinfo {author} {\bibfnamefont {R.~K.}\ \bibnamefont {Khan}}, \bibinfo {author} {\bibfnamefont {S.}~\bibnamefont {Turlapati}}, \bibinfo {author} {\bibfnamefont {N.~V.}\ \bibnamefont {Rao}}, \ and\ \bibinfo {author} {\bibfnamefont {S.}~\bibnamefont {Ghosh}},\ }\href@noop {} {\bibfield  {journal} {\bibinfo  {journal} {The European Physical Journal E}\ }\textbf {\bibinfo {volume} {40}},\ \bibinfo {pages} {1} (\bibinfo {year} {2017})}\BibitemShut {NoStop}%
\bibitem [{\citenamefont {Kumar}\ \emph {et~al.}(2018{\natexlab{b}})\citenamefont {Kumar}, \citenamefont {Debnath}, \citenamefont {Rao},\ and\ \citenamefont {Sinha}}]{kumar2018nanodoping_12}%
  \BibitemOpen
  \bibfield  {author} {\bibinfo {author} {\bibfnamefont {P.}~\bibnamefont {Kumar}}, \bibinfo {author} {\bibfnamefont {S.}~\bibnamefont {Debnath}}, \bibinfo {author} {\bibfnamefont {N.~V.}\ \bibnamefont {Rao}}, \ and\ \bibinfo {author} {\bibfnamefont {A.}~\bibnamefont {Sinha}},\ }\href@noop {} {\bibfield  {journal} {\bibinfo  {journal} {Journal of Physics: Condensed Matter}\ }\textbf {\bibinfo {volume} {30}},\ \bibinfo {pages} {095101} (\bibinfo {year} {2018}{\natexlab{b}})}\BibitemShut {NoStop}%
\bibitem [{\citenamefont {Li}\ \emph {et~al.}(2006)\citenamefont {Li}, \citenamefont {Buchnev}, \citenamefont {Cheon}, \citenamefont {Glushchenko}, \citenamefont {Reshetnyak}, \citenamefont {Reznikov}, \citenamefont {Sluckin},\ and\ \citenamefont {West}}]{li2006orientational_14}%
  \BibitemOpen
  \bibfield  {author} {\bibinfo {author} {\bibfnamefont {F.}~\bibnamefont {Li}}, \bibinfo {author} {\bibfnamefont {O.}~\bibnamefont {Buchnev}}, \bibinfo {author} {\bibfnamefont {C.~I.}\ \bibnamefont {Cheon}}, \bibinfo {author} {\bibfnamefont {A.}~\bibnamefont {Glushchenko}}, \bibinfo {author} {\bibfnamefont {V.}~\bibnamefont {Reshetnyak}}, \bibinfo {author} {\bibfnamefont {Y.}~\bibnamefont {Reznikov}}, \bibinfo {author} {\bibfnamefont {T.~J.}\ \bibnamefont {Sluckin}}, \ and\ \bibinfo {author} {\bibfnamefont {J.~L.}\ \bibnamefont {West}},\ }\href@noop {} {\bibfield  {journal} {\bibinfo  {journal} {Physical Review Letters}\ }\textbf {\bibinfo {volume} {97}},\ \bibinfo {pages} {147801} (\bibinfo {year} {2006})}\BibitemShut {NoStop}%
\bibitem [{\citenamefont {Lopatina}\ and\ \citenamefont {Selinger}(2009)}]{lopatina2009theory_13}%
  \BibitemOpen
  \bibfield  {author} {\bibinfo {author} {\bibfnamefont {L.~M.}\ \bibnamefont {Lopatina}}\ and\ \bibinfo {author} {\bibfnamefont {J.~V.}\ \bibnamefont {Selinger}},\ }\href@noop {} {\bibfield  {journal} {\bibinfo  {journal} {Physical review letters}\ }\textbf {\bibinfo {volume} {102}},\ \bibinfo {pages} {197802} (\bibinfo {year} {2009})}\BibitemShut {NoStop}%
\bibitem [{\citenamefont {Kalinin}\ \emph {et~al.}(2002)\citenamefont {Kalinin}, \citenamefont {Suchomel}, \citenamefont {Davies},\ and\ \citenamefont {Bonnell}}]{kalinin2002potential_15}%
  \BibitemOpen
  \bibfield  {author} {\bibinfo {author} {\bibfnamefont {S.~V.}\ \bibnamefont {Kalinin}}, \bibinfo {author} {\bibfnamefont {M.~R.}\ \bibnamefont {Suchomel}}, \bibinfo {author} {\bibfnamefont {P.~K.}\ \bibnamefont {Davies}}, \ and\ \bibinfo {author} {\bibfnamefont {D.~A.}\ \bibnamefont {Bonnell}},\ }\href@noop {} {\bibfield  {journal} {\bibinfo  {journal} {Journal of the American Ceramic Society}\ }\textbf {\bibinfo {volume} {85}},\ \bibinfo {pages} {3011} (\bibinfo {year} {2002})}\BibitemShut {NoStop}%
\bibitem [{\citenamefont {Seidel}\ \emph {et~al.}(2009)\citenamefont {Seidel}, \citenamefont {Martin}, \citenamefont {He}, \citenamefont {Zhan}, \citenamefont {Chu}, \citenamefont {Rother}, \citenamefont {Hawkridge}, \citenamefont {Maksymovych}, \citenamefont {Yu}, \citenamefont {Gajek} \emph {et~al.}}]{seidel2009conduction_16}%
  \BibitemOpen
  \bibfield  {author} {\bibinfo {author} {\bibfnamefont {J.}~\bibnamefont {Seidel}}, \bibinfo {author} {\bibfnamefont {L.~W.}\ \bibnamefont {Martin}}, \bibinfo {author} {\bibfnamefont {Q.}~\bibnamefont {He}}, \bibinfo {author} {\bibfnamefont {Q.}~\bibnamefont {Zhan}}, \bibinfo {author} {\bibfnamefont {Y.-H.}\ \bibnamefont {Chu}}, \bibinfo {author} {\bibfnamefont {A.}~\bibnamefont {Rother}}, \bibinfo {author} {\bibfnamefont {M.}~\bibnamefont {Hawkridge}}, \bibinfo {author} {\bibfnamefont {P.}~\bibnamefont {Maksymovych}}, \bibinfo {author} {\bibfnamefont {P.}~\bibnamefont {Yu}}, \bibinfo {author} {\bibfnamefont {M.}~\bibnamefont {Gajek}},  \emph {et~al.},\ }\href@noop {} {\bibfield  {journal} {\bibinfo  {journal} {Nature materials}\ }\textbf {\bibinfo {volume} {8}},\ \bibinfo {pages} {229} (\bibinfo {year} {2009})}\BibitemShut {NoStop}%
\bibitem [{\citenamefont {Nayek}\ and\ \citenamefont {Li}(2015)}]{nayek2015superior_17}%
  \BibitemOpen
  \bibfield  {author} {\bibinfo {author} {\bibfnamefont {P.}~\bibnamefont {Nayek}}\ and\ \bibinfo {author} {\bibfnamefont {G.}~\bibnamefont {Li}},\ }\href@noop {} {\bibfield  {journal} {\bibinfo  {journal} {Scientific reports}\ }\textbf {\bibinfo {volume} {5}},\ \bibinfo {pages} {10845} (\bibinfo {year} {2015})}\BibitemShut {NoStop}%
\bibitem [{\citenamefont {Khan}\ \emph {et~al.}(2020)\citenamefont {Khan}, \citenamefont {Prakash}, \citenamefont {Chauhan}, \citenamefont {Choudhary},\ and\ \citenamefont {Biradar}}]{khan2020bismuth_18}%
  \BibitemOpen
  \bibfield  {author} {\bibinfo {author} {\bibfnamefont {S.}~\bibnamefont {Khan}}, \bibinfo {author} {\bibfnamefont {J.}~\bibnamefont {Prakash}}, \bibinfo {author} {\bibfnamefont {S.}~\bibnamefont {Chauhan}}, \bibinfo {author} {\bibfnamefont {A.}~\bibnamefont {Choudhary}}, \ and\ \bibinfo {author} {\bibfnamefont {A.~M.}\ \bibnamefont {Biradar}},\ }\href@noop {} {\bibfield  {journal} {\bibinfo  {journal} {Journal of Applied Physics}\ }\textbf {\bibinfo {volume} {127}},\ \bibinfo {pages} {074102} (\bibinfo {year} {2020})}\BibitemShut {NoStop}%
\bibitem [{\citenamefont {Ghosh}\ \emph {et~al.}(2011)\citenamefont {Ghosh}, \citenamefont {Roy}, \citenamefont {Acharya}, \citenamefont {Chakrabarti}, \citenamefont {Zurowska},\ and\ \citenamefont {Dabrowski}}]{ghosh2011effect_19}%
  \BibitemOpen
  \bibfield  {author} {\bibinfo {author} {\bibfnamefont {S.}~\bibnamefont {Ghosh}}, \bibinfo {author} {\bibfnamefont {S.}~\bibnamefont {Roy}}, \bibinfo {author} {\bibfnamefont {S.}~\bibnamefont {Acharya}}, \bibinfo {author} {\bibfnamefont {P.}~\bibnamefont {Chakrabarti}}, \bibinfo {author} {\bibfnamefont {M.}~\bibnamefont {Zurowska}}, \ and\ \bibinfo {author} {\bibfnamefont {R.}~\bibnamefont {Dabrowski}},\ }\href@noop {} {\bibfield  {journal} {\bibinfo  {journal} {Europhysics Letters}\ }\textbf {\bibinfo {volume} {96}},\ \bibinfo {pages} {47003} (\bibinfo {year} {2011})}\BibitemShut {NoStop}%
\bibitem [{\citenamefont {Derbali}\ \emph {et~al.}(2020)\citenamefont {Derbali}, \citenamefont {Guesmi}, \citenamefont {Hamadi},\ and\ \citenamefont {Soltani}}]{derbali2020dielectric_33}%
  \BibitemOpen
  \bibfield  {author} {\bibinfo {author} {\bibfnamefont {M.}~\bibnamefont {Derbali}}, \bibinfo {author} {\bibfnamefont {A.}~\bibnamefont {Guesmi}}, \bibinfo {author} {\bibfnamefont {N.~B.}\ \bibnamefont {Hamadi}}, \ and\ \bibinfo {author} {\bibfnamefont {T.}~\bibnamefont {Soltani}},\ }\href@noop {} {\bibfield  {journal} {\bibinfo  {journal} {Journal of Molecular Liquids}\ }\textbf {\bibinfo {volume} {319}},\ \bibinfo {pages} {113768} (\bibinfo {year} {2020})}\BibitemShut {NoStop}%
\bibitem [{\citenamefont {Castillo}\ \emph {et~al.}(2013)\citenamefont {Castillo}, \citenamefont {Shvartsman}, \citenamefont {Gobeljic}, \citenamefont {Gao}, \citenamefont {Landers}, \citenamefont {Wende},\ and\ \citenamefont {Lupascu}}]{castillo2013effect_20}%
  \BibitemOpen
  \bibfield  {author} {\bibinfo {author} {\bibfnamefont {M.~E.}\ \bibnamefont {Castillo}}, \bibinfo {author} {\bibfnamefont {V.}~\bibnamefont {Shvartsman}}, \bibinfo {author} {\bibfnamefont {D.}~\bibnamefont {Gobeljic}}, \bibinfo {author} {\bibfnamefont {Y.}~\bibnamefont {Gao}}, \bibinfo {author} {\bibfnamefont {J.}~\bibnamefont {Landers}}, \bibinfo {author} {\bibfnamefont {H.}~\bibnamefont {Wende}}, \ and\ \bibinfo {author} {\bibfnamefont {D.}~\bibnamefont {Lupascu}},\ }\href@noop {} {\bibfield  {journal} {\bibinfo  {journal} {Nanotechnology}\ }\textbf {\bibinfo {volume} {24}},\ \bibinfo {pages} {355701} (\bibinfo {year} {2013})}\BibitemShut {NoStop}%
\bibitem [{\citenamefont {Mandal}\ \emph {et~al.}()\citenamefont {Mandal}, \citenamefont {Kaur}, \citenamefont {Mohiuddin}, \citenamefont {Pal},\ and\ \citenamefont {Sinha}}]{Dhananjoy2023dielectric_21}%
  \BibitemOpen
  \bibfield  {author} {\bibinfo {author} {\bibfnamefont {D.}~\bibnamefont {Mandal}}, \bibinfo {author} {\bibfnamefont {S.}~\bibnamefont {Kaur}}, \bibinfo {author} {\bibfnamefont {G.}~\bibnamefont {Mohiuddin}}, \bibinfo {author} {\bibfnamefont {S.~K.}\ \bibnamefont {Pal}}, \ and\ \bibinfo {author} {\bibfnamefont {A.}~\bibnamefont {Sinha}},\ }\href@noop {} {\bibinfo  {journal} {Unpublished work}\ }\BibitemShut {NoStop}%
\bibitem [{\citenamefont {Selbach}\ \emph {et~al.}(2007)\citenamefont {Selbach}, \citenamefont {Tybell}, \citenamefont {Einarsrud},\ and\ \citenamefont {Grande}}]{selbach2007size_22}%
  \BibitemOpen
\bibfield  {journal} {  }\bibfield  {author} {\bibinfo {author} {\bibfnamefont {S.~M.}\ \bibnamefont {Selbach}}, \bibinfo {author} {\bibfnamefont {T.}~\bibnamefont {Tybell}}, \bibinfo {author} {\bibfnamefont {M.-A.}\ \bibnamefont {Einarsrud}}, \ and\ \bibinfo {author} {\bibfnamefont {T.}~\bibnamefont {Grande}},\ }\href@noop {} {\bibfield  {journal} {\bibinfo  {journal} {Chemistry of materials}\ }\textbf {\bibinfo {volume} {19}},\ \bibinfo {pages} {6478} (\bibinfo {year} {2007})}\BibitemShut {NoStop}%
\bibitem [{\citenamefont {Karpinsky}\ \emph {et~al.}(2017)\citenamefont {Karpinsky}, \citenamefont {Eliseev}, \citenamefont {Xue}, \citenamefont {Silibin}, \citenamefont {Franz}, \citenamefont {Glinchuk}, \citenamefont {Troyanchuk}, \citenamefont {Gavrilov}, \citenamefont {Gopalan}, \citenamefont {Chen} \emph {et~al.}}]{karpinsky2017thermodynamic_23}%
  \BibitemOpen
  \bibfield  {author} {\bibinfo {author} {\bibfnamefont {D.~V.}\ \bibnamefont {Karpinsky}}, \bibinfo {author} {\bibfnamefont {E.~A.}\ \bibnamefont {Eliseev}}, \bibinfo {author} {\bibfnamefont {F.}~\bibnamefont {Xue}}, \bibinfo {author} {\bibfnamefont {M.~V.}\ \bibnamefont {Silibin}}, \bibinfo {author} {\bibfnamefont {A.}~\bibnamefont {Franz}}, \bibinfo {author} {\bibfnamefont {M.~D.}\ \bibnamefont {Glinchuk}}, \bibinfo {author} {\bibfnamefont {I.~O.}\ \bibnamefont {Troyanchuk}}, \bibinfo {author} {\bibfnamefont {S.~A.}\ \bibnamefont {Gavrilov}}, \bibinfo {author} {\bibfnamefont {V.}~\bibnamefont {Gopalan}}, \bibinfo {author} {\bibfnamefont {L.-Q.}\ \bibnamefont {Chen}},  \emph {et~al.},\ }\href@noop {} {\bibfield  {journal} {\bibinfo  {journal} {npj Computational Materials}\ }\textbf {\bibinfo {volume} {3}},\ \bibinfo {pages} {20} (\bibinfo {year} {2017})}\BibitemShut {NoStop}%
\bibitem [{\citenamefont {Reznikov}\ \emph {et~al.}(2017)\citenamefont {Reznikov}, \citenamefont {Glushchenko},\ and\ \citenamefont {Garbovskiy}}]{reznikov2017ferromagnetic_24}%
  \BibitemOpen
  \bibfield  {author} {\bibinfo {author} {\bibfnamefont {Y.}~\bibnamefont {Reznikov}}, \bibinfo {author} {\bibfnamefont {A.}~\bibnamefont {Glushchenko}}, \ and\ \bibinfo {author} {\bibfnamefont {Y.}~\bibnamefont {Garbovskiy}},\ }in\ \href@noop {} {\emph {\bibinfo {booktitle} {Liquid crystals with nano and microparticles}}}\ (\bibinfo  {publisher} {World Scientific},\ \bibinfo {year} {2017})\ pp.\ \bibinfo {pages} {657--693}\BibitemShut {NoStop}%
\bibitem [{\citenamefont {Emdadi}\ \emph {et~al.}(2018{\natexlab{a}})\citenamefont {Emdadi}, \citenamefont {Poursamad}, \citenamefont {Sahrai},\ and\ \citenamefont {Moghaddas}}]{emdadi2018behaviour_25}%
  \BibitemOpen
  \bibfield  {author} {\bibinfo {author} {\bibfnamefont {M.}~\bibnamefont {Emdadi}}, \bibinfo {author} {\bibfnamefont {J.}~\bibnamefont {Poursamad}}, \bibinfo {author} {\bibfnamefont {M.}~\bibnamefont {Sahrai}}, \ and\ \bibinfo {author} {\bibfnamefont {F.}~\bibnamefont {Moghaddas}},\ }\href@noop {} {\bibfield  {journal} {\bibinfo  {journal} {Molecular Physics}\ }\textbf {\bibinfo {volume} {116}},\ \bibinfo {pages} {1650} (\bibinfo {year} {2018}{\natexlab{a}})}\BibitemShut {NoStop}%
\bibitem [{\citenamefont {Emdadi}\ \emph {et~al.}(2018{\natexlab{b}})\citenamefont {Emdadi}, \citenamefont {Poursamad}, \citenamefont {Sahrai},\ and\ \citenamefont {Moghadas}}]{emdadi2018investigation_26}%
  \BibitemOpen
  \bibfield  {author} {\bibinfo {author} {\bibfnamefont {M.}~\bibnamefont {Emdadi}}, \bibinfo {author} {\bibfnamefont {J.}~\bibnamefont {Poursamad}}, \bibinfo {author} {\bibfnamefont {M.}~\bibnamefont {Sahrai}}, \ and\ \bibinfo {author} {\bibfnamefont {F.}~\bibnamefont {Moghadas}},\ }\href@noop {} {\bibfield  {journal} {\bibinfo  {journal} {Brazilian Journal of Physics}\ }\textbf {\bibinfo {volume} {48}},\ \bibinfo {pages} {433} (\bibinfo {year} {2018}{\natexlab{b}})}\BibitemShut {NoStop}%
\bibitem [{\citenamefont {Mohiuddin}\ \emph {et~al.}(2017)\citenamefont {Mohiuddin}, \citenamefont {Begum}, \citenamefont {Rao}, \citenamefont {Kaur}, \citenamefont {Punjani}, \citenamefont {Khan}, \citenamefont {Ghosh},\ and\ \citenamefont {Pal}}]{mohiuddin2017observation_28}%
  \BibitemOpen
  \bibfield  {author} {\bibinfo {author} {\bibfnamefont {G.}~\bibnamefont {Mohiuddin}}, \bibinfo {author} {\bibfnamefont {N.}~\bibnamefont {Begum}}, \bibinfo {author} {\bibfnamefont {N.~V.~S.}\ \bibnamefont {Rao}}, \bibinfo {author} {\bibfnamefont {S.}~\bibnamefont {Kaur}}, \bibinfo {author} {\bibfnamefont {V.}~\bibnamefont {Punjani}}, \bibinfo {author} {\bibfnamefont {R.~K.}\ \bibnamefont {Khan}}, \bibinfo {author} {\bibfnamefont {S.}~\bibnamefont {Ghosh}}, \ and\ \bibinfo {author} {\bibfnamefont {S.~K.}\ \bibnamefont {Pal}},\ }\href@noop {} {\bibfield  {journal} {\bibinfo  {journal} {Liquid Crystals}\ }\textbf {\bibinfo {volume} {44}},\ \bibinfo {pages} {2247} (\bibinfo {year} {2017})}\BibitemShut {NoStop}%
\bibitem [{\citenamefont {Hiremath}\ \emph {et~al.}(2016)\citenamefont {Hiremath}, \citenamefont {Nair},\ and\ \citenamefont {Rao}}]{hiremath2016supramolecular_29}%
  \BibitemOpen
  \bibfield  {author} {\bibinfo {author} {\bibfnamefont {U.~S.}\ \bibnamefont {Hiremath}}, \bibinfo {author} {\bibfnamefont {G.~G.}\ \bibnamefont {Nair}}, \ and\ \bibinfo {author} {\bibfnamefont {D.~S.}\ \bibnamefont {Rao}},\ }\href@noop {} {\bibfield  {journal} {\bibinfo  {journal} {Liquid Crystals}\ }\textbf {\bibinfo {volume} {43}},\ \bibinfo {pages} {711} (\bibinfo {year} {2016})}\BibitemShut {NoStop}%
\bibitem [{\citenamefont {Zhang}\ \emph {et~al.}(2022)\citenamefont {Zhang}, \citenamefont {Xiang}, \citenamefont {Ding}, \citenamefont {Hao}, \citenamefont {Kaur}, \citenamefont {Mohiuddin}, \citenamefont {Pal}, \citenamefont {Salamon}, \citenamefont {{\'E}ber},\ and\ \citenamefont {Buka}}]{zhang2022electric_27}%
  \BibitemOpen
  \bibfield  {author} {\bibinfo {author} {\bibfnamefont {J.}~\bibnamefont {Zhang}}, \bibinfo {author} {\bibfnamefont {Y.}~\bibnamefont {Xiang}}, \bibinfo {author} {\bibfnamefont {X.}~\bibnamefont {Ding}}, \bibinfo {author} {\bibfnamefont {L.}~\bibnamefont {Hao}}, \bibinfo {author} {\bibfnamefont {S.}~\bibnamefont {Kaur}}, \bibinfo {author} {\bibfnamefont {G.}~\bibnamefont {Mohiuddin}}, \bibinfo {author} {\bibfnamefont {S.~K.}\ \bibnamefont {Pal}}, \bibinfo {author} {\bibfnamefont {P.}~\bibnamefont {Salamon}}, \bibinfo {author} {\bibfnamefont {N.}~\bibnamefont {{\'E}ber}}, \ and\ \bibinfo {author} {\bibfnamefont {{\'A}.}~\bibnamefont {Buka}},\ }\href@noop {} {\bibfield  {journal} {\bibinfo  {journal} {Journal of Molecular Liquids}\ }\textbf {\bibinfo {volume} {366}},\ \bibinfo {pages} {120239} (\bibinfo {year} {2022})}\BibitemShut {NoStop}%
\bibitem [{\citenamefont {Kaur}\ \emph {et~al.}(2023)\citenamefont {Kaur}, \citenamefont {Mohiuddin}, \citenamefont {Zhang}, \citenamefont {Chakraborty}, \citenamefont {Ding}, \citenamefont {Verma}, \citenamefont {Sinha}, \citenamefont {Xiang},\ and\ \citenamefont {Pal}}]{kaur2023polar}%
  \BibitemOpen
  \bibfield  {author} {\bibinfo {author} {\bibfnamefont {S.}~\bibnamefont {Kaur}}, \bibinfo {author} {\bibfnamefont {G.}~\bibnamefont {Mohiuddin}}, \bibinfo {author} {\bibfnamefont {J.}~\bibnamefont {Zhang}}, \bibinfo {author} {\bibfnamefont {S.}~\bibnamefont {Chakraborty}}, \bibinfo {author} {\bibfnamefont {X.}~\bibnamefont {Ding}}, \bibinfo {author} {\bibfnamefont {D.}~\bibnamefont {Verma}}, \bibinfo {author} {\bibfnamefont {A.}~\bibnamefont {Sinha}}, \bibinfo {author} {\bibfnamefont {Y.}~\bibnamefont {Xiang}}, \ and\ \bibinfo {author} {\bibfnamefont {S.~K.}\ \bibnamefont {Pal}},\ }\href@noop {} {\bibfield  {journal} {\bibinfo  {journal} {Journal of Molecular Liquids}\ }\textbf {\bibinfo {volume} {387}},\ \bibinfo {pages} {122626} (\bibinfo {year} {2023})}\BibitemShut {NoStop}%
\bibitem [{\citenamefont {Vita}\ \emph {et~al.}(2018)\citenamefont {Vita}, \citenamefont {Adamo},\ and\ \citenamefont {Francescangeli}}]{vita2018polar_30}%
  \BibitemOpen
  \bibfield  {author} {\bibinfo {author} {\bibfnamefont {F.}~\bibnamefont {Vita}}, \bibinfo {author} {\bibfnamefont {F.~C.}\ \bibnamefont {Adamo}}, \ and\ \bibinfo {author} {\bibfnamefont {O.}~\bibnamefont {Francescangeli}},\ }\href@noop {} {\bibfield  {journal} {\bibinfo  {journal} {Journal of Molecular Liquids}\ }\textbf {\bibinfo {volume} {267}},\ \bibinfo {pages} {564} (\bibinfo {year} {2018})}\BibitemShut {NoStop}%
\bibitem [{\citenamefont {Panarin}\ \emph {et~al.}(2018)\citenamefont {Panarin}, \citenamefont {Sreenilayam}, \citenamefont {Vij}, \citenamefont {Lehmann},\ and\ \citenamefont {Tschierske}}]{panarin2018formation_31}%
  \BibitemOpen
  \bibfield  {author} {\bibinfo {author} {\bibfnamefont {Y.~P.}\ \bibnamefont {Panarin}}, \bibinfo {author} {\bibfnamefont {S.~P.}\ \bibnamefont {Sreenilayam}}, \bibinfo {author} {\bibfnamefont {J.~K.}\ \bibnamefont {Vij}}, \bibinfo {author} {\bibfnamefont {A.}~\bibnamefont {Lehmann}}, \ and\ \bibinfo {author} {\bibfnamefont {C.}~\bibnamefont {Tschierske}},\ }\href@noop {} {\bibfield  {journal} {\bibinfo  {journal} {Beilstein Journal of Nanotechnology}\ }\textbf {\bibinfo {volume} {9}},\ \bibinfo {pages} {1288} (\bibinfo {year} {2018})}\BibitemShut {NoStop}%
\bibitem [{\citenamefont {Madhusudana}(2017)}]{madhusudana2017two_36}%
  \BibitemOpen
  \bibfield  {author} {\bibinfo {author} {\bibfnamefont {N.}~\bibnamefont {Madhusudana}},\ }\href@noop {} {\bibfield  {journal} {\bibinfo  {journal} {Physical Review E}\ }\textbf {\bibinfo {volume} {96}},\ \bibinfo {pages} {022710} (\bibinfo {year} {2017})}\BibitemShut {NoStop}%
\bibitem [{\citenamefont {Patranabish}\ \emph {et~al.}(2019)\citenamefont {Patranabish}, \citenamefont {Wang}, \citenamefont {Sinha},\ and\ \citenamefont {Majumdar}}]{patranabish2019one_37}%
  \BibitemOpen
  \bibfield  {author} {\bibinfo {author} {\bibfnamefont {S.}~\bibnamefont {Patranabish}}, \bibinfo {author} {\bibfnamefont {Y.}~\bibnamefont {Wang}}, \bibinfo {author} {\bibfnamefont {A.}~\bibnamefont {Sinha}}, \ and\ \bibinfo {author} {\bibfnamefont {A.}~\bibnamefont {Majumdar}},\ }\href@noop {} {\bibfield  {journal} {\bibinfo  {journal} {Physical Review E}\ }\textbf {\bibinfo {volume} {99}},\ \bibinfo {pages} {012703} (\bibinfo {year} {2019})}\BibitemShut {NoStop}%
\bibitem [{\citenamefont {Patranabish}\ \emph {et~al.}(2021)\citenamefont {Patranabish}, \citenamefont {Wang}, \citenamefont {Sinha},\ and\ \citenamefont {Majumdar}}]{patranabish2021quantum_38}%
  \BibitemOpen
  \bibfield  {author} {\bibinfo {author} {\bibfnamefont {S.}~\bibnamefont {Patranabish}}, \bibinfo {author} {\bibfnamefont {Y.}~\bibnamefont {Wang}}, \bibinfo {author} {\bibfnamefont {A.}~\bibnamefont {Sinha}}, \ and\ \bibinfo {author} {\bibfnamefont {A.}~\bibnamefont {Majumdar}},\ }\href@noop {} {\bibfield  {journal} {\bibinfo  {journal} {Physical Review E}\ }\textbf {\bibinfo {volume} {103}},\ \bibinfo {pages} {052703} (\bibinfo {year} {2021})}\BibitemShut {NoStop}%
\bibitem [{\citenamefont {Chakraborty}\ \emph {et~al.}(2015)\citenamefont {Chakraborty}, \citenamefont {Chakraborty},\ and\ \citenamefont {Das}}]{chakraborty2015effect_39}%
  \BibitemOpen
  \bibfield  {author} {\bibinfo {author} {\bibfnamefont {A.}~\bibnamefont {Chakraborty}}, \bibinfo {author} {\bibfnamefont {S.}~\bibnamefont {Chakraborty}}, \ and\ \bibinfo {author} {\bibfnamefont {M.~K.}\ \bibnamefont {Das}},\ }\href@noop {} {\bibfield  {journal} {\bibinfo  {journal} {Physical Review E}\ }\textbf {\bibinfo {volume} {91}},\ \bibinfo {pages} {032503} (\bibinfo {year} {2015})}\BibitemShut {NoStop}%
\bibitem [{\citenamefont {Mertelj}\ \emph {et~al.}(2012)\citenamefont {Mertelj}, \citenamefont {Cmok}, \citenamefont {{\v{C}}opi{\v{c}}}, \citenamefont {Cook},\ and\ \citenamefont {Evans}}]{mertelj2012critical}%
  \BibitemOpen
  \bibfield  {author} {\bibinfo {author} {\bibfnamefont {A.}~\bibnamefont {Mertelj}}, \bibinfo {author} {\bibfnamefont {L.}~\bibnamefont {Cmok}}, \bibinfo {author} {\bibfnamefont {M.}~\bibnamefont {{\v{C}}opi{\v{c}}}}, \bibinfo {author} {\bibfnamefont {G.}~\bibnamefont {Cook}}, \ and\ \bibinfo {author} {\bibfnamefont {D.~R.}\ \bibnamefont {Evans}},\ }\href@noop {} {\bibfield  {journal} {\bibinfo  {journal} {Physical Review E}\ }\textbf {\bibinfo {volume} {85}},\ \bibinfo {pages} {021705} (\bibinfo {year} {2012})}\BibitemShut {NoStop}%
\bibitem [{\citenamefont {Basu}(2014)}]{basu2014soft_40}%
  \BibitemOpen
  \bibfield  {author} {\bibinfo {author} {\bibfnamefont {R.}~\bibnamefont {Basu}},\ }\href@noop {} {\bibfield  {journal} {\bibinfo  {journal} {Physical Review E}\ }\textbf {\bibinfo {volume} {89}},\ \bibinfo {pages} {022508} (\bibinfo {year} {2014})}\BibitemShut {NoStop}%
\bibitem [{\citenamefont {Hoffman}\ and\ \citenamefont {Davidson}()}]{hoffman2023michel_41}%
  \BibitemOpen
  \bibfield  {author} {\bibinfo {author} {\bibfnamefont {R.}~\bibnamefont {Hoffman}}\ and\ \bibinfo {author} {\bibfnamefont {M.}~\bibnamefont {Davidson}},\ }\href@noop {} {\bibinfo  {journal} {https://www.olympus-lifescience.com/en/microscope resource/primer/techniques/polarized/michel/}\ }\BibitemShut {NoStop}%
\bibitem [{\citenamefont {Garbovskiy}\ and\ \citenamefont {Glushchenko}(2017)}]{garbovskiy2017ferroelectric_42}%
  \BibitemOpen
\bibfield  {journal} {  }\bibfield  {author} {\bibinfo {author} {\bibfnamefont {Y.}~\bibnamefont {Garbovskiy}}\ and\ \bibinfo {author} {\bibfnamefont {A.}~\bibnamefont {Glushchenko}},\ }\href@noop {} {\bibfield  {journal} {\bibinfo  {journal} {Nanomaterials}\ }\textbf {\bibinfo {volume} {7}},\ \bibinfo {pages} {361} (\bibinfo {year} {2017})}\BibitemShut {NoStop}%
\bibitem [{\citenamefont {Haase}\ and\ \citenamefont {Wr{\'o}bel}(2013)}]{haase2013relaxation_43}%
  \BibitemOpen
  \bibfield  {author} {\bibinfo {author} {\bibfnamefont {W.}~\bibnamefont {Haase}}\ and\ \bibinfo {author} {\bibfnamefont {S.}~\bibnamefont {Wr{\'o}bel}},\ }\href@noop {} {\emph {\bibinfo {title} {Relaxation phenomena: liquid crystals, magnetic systems, polymers, high-Tc superconductors, metallic glasses}}}\ (\bibinfo  {publisher} {Springer Science \& Business Media},\ \bibinfo {year} {2013})\BibitemShut {NoStop}%
\bibitem [{\citenamefont {Shanker}\ \emph {et~al.}(2014)\citenamefont {Shanker}, \citenamefont {Prehm}, \citenamefont {Nagaraj}, \citenamefont {Vij}, \citenamefont {Weyland}, \citenamefont {Eremin},\ and\ \citenamefont {Tschierske}}]{shanker2014_44}%
  \BibitemOpen
  \bibfield  {author} {\bibinfo {author} {\bibfnamefont {G.}~\bibnamefont {Shanker}}, \bibinfo {author} {\bibfnamefont {M.}~\bibnamefont {Prehm}}, \bibinfo {author} {\bibfnamefont {M.}~\bibnamefont {Nagaraj}}, \bibinfo {author} {\bibfnamefont {J.~K.}\ \bibnamefont {Vij}}, \bibinfo {author} {\bibfnamefont {M.}~\bibnamefont {Weyland}}, \bibinfo {author} {\bibfnamefont {A.}~\bibnamefont {Eremin}}, \ and\ \bibinfo {author} {\bibfnamefont {C.}~\bibnamefont {Tschierske}},\ }\href@noop {} {\bibfield  {journal} {\bibinfo  {journal} {ChemPhysChem}\ }\textbf {\bibinfo {volume} {15}},\ \bibinfo {pages} {1323} (\bibinfo {year} {2014})}\BibitemShut {NoStop}%
\bibitem [{\citenamefont {Shanker}\ \emph {et~al.}(2012)\citenamefont {Shanker}, \citenamefont {Nagaraj}, \citenamefont {Kocot}, \citenamefont {Vij}, \citenamefont {Prehm},\ and\ \citenamefont {Tschierske}}]{shanker2012nematic_45}%
  \BibitemOpen
  \bibfield  {author} {\bibinfo {author} {\bibfnamefont {G.}~\bibnamefont {Shanker}}, \bibinfo {author} {\bibfnamefont {M.}~\bibnamefont {Nagaraj}}, \bibinfo {author} {\bibfnamefont {A.}~\bibnamefont {Kocot}}, \bibinfo {author} {\bibfnamefont {J.~K.}\ \bibnamefont {Vij}}, \bibinfo {author} {\bibfnamefont {M.}~\bibnamefont {Prehm}}, \ and\ \bibinfo {author} {\bibfnamefont {C.}~\bibnamefont {Tschierske}},\ }\href@noop {} {\bibfield  {journal} {\bibinfo  {journal} {Advanced Functional Materials}\ }\textbf {\bibinfo {volume} {22}},\ \bibinfo {pages} {1671} (\bibinfo {year} {2012})}\BibitemShut {NoStop}%
\bibitem [{\citenamefont {Marino}\ \emph {et~al.}(2012{\natexlab{b}})\citenamefont {Marino}, \citenamefont {Ionescu}, \citenamefont {Marino},\ and\ \citenamefont {Scaramuzza}}]{marino2012dielectric_46}%
  \BibitemOpen
  \bibfield  {author} {\bibinfo {author} {\bibfnamefont {L.}~\bibnamefont {Marino}}, \bibinfo {author} {\bibfnamefont {A.~T.}\ \bibnamefont {Ionescu}}, \bibinfo {author} {\bibfnamefont {S.}~\bibnamefont {Marino}}, \ and\ \bibinfo {author} {\bibfnamefont {N.}~\bibnamefont {Scaramuzza}},\ }\href@noop {} {\bibfield  {journal} {\bibinfo  {journal} {Journal of Applied Physics}\ }\textbf {\bibinfo {volume} {112}},\ \bibinfo {pages} {114113} (\bibinfo {year} {2012}{\natexlab{b}})}\BibitemShut {NoStop}%
\bibitem [{\citenamefont {Guo}\ \emph {et~al.}(2010)\citenamefont {Guo}, \citenamefont {Dhara}, \citenamefont {Sadashiva}, \citenamefont {Radhika}, \citenamefont {Pratibha}, \citenamefont {Shimbo}, \citenamefont {Araoka}, \citenamefont {Ishikawa},\ and\ \citenamefont {Takezoe}}]{guo2010polar_47}%
  \BibitemOpen
  \bibfield  {author} {\bibinfo {author} {\bibfnamefont {L.}~\bibnamefont {Guo}}, \bibinfo {author} {\bibfnamefont {S.}~\bibnamefont {Dhara}}, \bibinfo {author} {\bibfnamefont {B.}~\bibnamefont {Sadashiva}}, \bibinfo {author} {\bibfnamefont {S.}~\bibnamefont {Radhika}}, \bibinfo {author} {\bibfnamefont {R.}~\bibnamefont {Pratibha}}, \bibinfo {author} {\bibfnamefont {Y.}~\bibnamefont {Shimbo}}, \bibinfo {author} {\bibfnamefont {F.}~\bibnamefont {Araoka}}, \bibinfo {author} {\bibfnamefont {K.}~\bibnamefont {Ishikawa}}, \ and\ \bibinfo {author} {\bibfnamefont {H.}~\bibnamefont {Takezoe}},\ }\href@noop {} {\bibfield  {journal} {\bibinfo  {journal} {Physical Review E}\ }\textbf {\bibinfo {volume} {81}},\ \bibinfo {pages} {011703} (\bibinfo {year} {2010})}\BibitemShut {NoStop}%
\bibitem [{\citenamefont {Merkel}\ \emph {et~al.}(2006)\citenamefont {Merkel}, \citenamefont {Kocot}, \citenamefont {Vij}, \citenamefont {Mehl},\ and\ \citenamefont {Meyer}}]{merkel2006orientational_48}%
  \BibitemOpen
  \bibfield  {author} {\bibinfo {author} {\bibfnamefont {K.}~\bibnamefont {Merkel}}, \bibinfo {author} {\bibfnamefont {A.}~\bibnamefont {Kocot}}, \bibinfo {author} {\bibfnamefont {J.}~\bibnamefont {Vij}}, \bibinfo {author} {\bibfnamefont {G.}~\bibnamefont {Mehl}}, \ and\ \bibinfo {author} {\bibfnamefont {T.}~\bibnamefont {Meyer}},\ }\href@noop {} {\bibfield  {journal} {\bibinfo  {journal} {Physical Review E}\ }\textbf {\bibinfo {volume} {73}},\ \bibinfo {pages} {051702} (\bibinfo {year} {2006})}\BibitemShut {NoStop}%
\bibitem [{\citenamefont {Nagaraj}\ \emph {et~al.}(2010)\citenamefont {Nagaraj}, \citenamefont {Panarin}, \citenamefont {Manna}, \citenamefont {Vij}, \citenamefont {Keith},\ and\ \citenamefont {Tschierske}}]{nagaraj2010electric}%
  \BibitemOpen
  \bibfield  {author} {\bibinfo {author} {\bibfnamefont {M.}~\bibnamefont {Nagaraj}}, \bibinfo {author} {\bibfnamefont {Y.~P.}\ \bibnamefont {Panarin}}, \bibinfo {author} {\bibfnamefont {U.}~\bibnamefont {Manna}}, \bibinfo {author} {\bibfnamefont {J.}~\bibnamefont {Vij}}, \bibinfo {author} {\bibfnamefont {C.}~\bibnamefont {Keith}}, \ and\ \bibinfo {author} {\bibfnamefont {C.}~\bibnamefont {Tschierske}},\ }\href@noop {} {\bibfield  {journal} {\bibinfo  {journal} {Applied Physics Letters}\ }\textbf {\bibinfo {volume} {96}} (\bibinfo {year} {2010})}\BibitemShut {NoStop}%
\bibitem [{\citenamefont {Patranabish}\ \emph {et~al.}(2018)\citenamefont {Patranabish}, \citenamefont {Mohiuddin}, \citenamefont {Begum}, \citenamefont {Laskar}, \citenamefont {Pal}, \citenamefont {Rao},\ and\ \citenamefont {Sinha}}]{patranabish2018cybotactic_49}%
  \BibitemOpen
  \bibfield  {author} {\bibinfo {author} {\bibfnamefont {S.}~\bibnamefont {Patranabish}}, \bibinfo {author} {\bibfnamefont {G.}~\bibnamefont {Mohiuddin}}, \bibinfo {author} {\bibfnamefont {N.}~\bibnamefont {Begum}}, \bibinfo {author} {\bibfnamefont {A.~R.}\ \bibnamefont {Laskar}}, \bibinfo {author} {\bibfnamefont {S.~K.}\ \bibnamefont {Pal}}, \bibinfo {author} {\bibfnamefont {N.~V.}\ \bibnamefont {Rao}}, \ and\ \bibinfo {author} {\bibfnamefont {A.}~\bibnamefont {Sinha}},\ }\href@noop {} {\bibfield  {journal} {\bibinfo  {journal} {Journal of Molecular Liquids}\ }\textbf {\bibinfo {volume} {257}},\ \bibinfo {pages} {144} (\bibinfo {year} {2018})}\BibitemShut {NoStop}%
\bibitem [{\citenamefont {Tadapatri}\ \emph {et~al.}(2010)\citenamefont {Tadapatri}, \citenamefont {Hiremath}, \citenamefont {Yelamaggad},\ and\ \citenamefont {Krishnamurthy}}]{tadapatri2010permittivity_50}%
  \BibitemOpen
  \bibfield  {author} {\bibinfo {author} {\bibfnamefont {P.}~\bibnamefont {Tadapatri}}, \bibinfo {author} {\bibfnamefont {U.~S.}\ \bibnamefont {Hiremath}}, \bibinfo {author} {\bibfnamefont {C.}~\bibnamefont {Yelamaggad}}, \ and\ \bibinfo {author} {\bibfnamefont {K.}~\bibnamefont {Krishnamurthy}},\ }\href@noop {} {\bibfield  {journal} {\bibinfo  {journal} {The Journal of Physical Chemistry B}\ }\textbf {\bibinfo {volume} {114}},\ \bibinfo {pages} {1745} (\bibinfo {year} {2010})}\BibitemShut {NoStop}%
\bibitem [{\citenamefont {Ouskova}\ \emph {et~al.}(2003)\citenamefont {Ouskova}, \citenamefont {Buchnev}, \citenamefont {Reshetnyak}, \citenamefont {Reznikov},\ and\ \citenamefont {Kresse}}]{ouskova2003dielectric_51}%
  \BibitemOpen
  \bibfield  {author} {\bibinfo {author} {\bibfnamefont {E.}~\bibnamefont {Ouskova}}, \bibinfo {author} {\bibfnamefont {O.}~\bibnamefont {Buchnev}}, \bibinfo {author} {\bibfnamefont {V.}~\bibnamefont {Reshetnyak}}, \bibinfo {author} {\bibfnamefont {Y.}~\bibnamefont {Reznikov}}, \ and\ \bibinfo {author} {\bibfnamefont {H.}~\bibnamefont {Kresse}},\ }\href@noop {} {\bibfield  {journal} {\bibinfo  {journal} {Liquid Crystals}\ }\textbf {\bibinfo {volume} {30}},\ \bibinfo {pages} {1235} (\bibinfo {year} {2003})}\BibitemShut {NoStop}%
\bibitem [{\citenamefont {Havriliak}\ and\ \citenamefont {Negami}(1966)}]{havriliak1966complex_52}%
  \BibitemOpen
  \bibfield  {author} {\bibinfo {author} {\bibfnamefont {S.}~\bibnamefont {Havriliak}}\ and\ \bibinfo {author} {\bibfnamefont {S.}~\bibnamefont {Negami}},\ }in\ \href@noop {} {\emph {\bibinfo {booktitle} {Journal of Polymer Science Part C: Polymer Symposia}}},\ Vol.~\bibinfo {volume} {14}\ (\bibinfo {organization} {Wiley Online Library},\ \bibinfo {year} {1966})\ pp.\ \bibinfo {pages} {99--117}\BibitemShut {NoStop}%
\bibitem [{\citenamefont {Havriliak}\ and\ \citenamefont {Negami}(1967)}]{havriliak1967complex_53}%
  \BibitemOpen
  \bibfield  {author} {\bibinfo {author} {\bibfnamefont {S.}~\bibnamefont {Havriliak}}\ and\ \bibinfo {author} {\bibfnamefont {S.}~\bibnamefont {Negami}},\ }\href@noop {} {\bibfield  {journal} {\bibinfo  {journal} {Polymer}\ }\textbf {\bibinfo {volume} {8}},\ \bibinfo {pages} {161} (\bibinfo {year} {1967})}\BibitemShut {NoStop}%
\bibitem [{\citenamefont {Chakraborty}\ \emph {et~al.}(2019)\citenamefont {Chakraborty}, \citenamefont {Das}, \citenamefont {Bubnov}, \citenamefont {Weissflog}, \citenamefont {W{\k{e}}g{\l}owska},\ and\ \citenamefont {Dabrowski}}]{chakraborty2019induced_54}%
  \BibitemOpen
  \bibfield  {author} {\bibinfo {author} {\bibfnamefont {S.}~\bibnamefont {Chakraborty}}, \bibinfo {author} {\bibfnamefont {M.~K.}\ \bibnamefont {Das}}, \bibinfo {author} {\bibfnamefont {A.}~\bibnamefont {Bubnov}}, \bibinfo {author} {\bibfnamefont {W.}~\bibnamefont {Weissflog}}, \bibinfo {author} {\bibfnamefont {D.}~\bibnamefont {W{\k{e}}g{\l}owska}}, \ and\ \bibinfo {author} {\bibfnamefont {R.}~\bibnamefont {Dabrowski}},\ }\href@noop {} {\bibfield  {journal} {\bibinfo  {journal} {Journal of Materials Chemistry C}\ }\textbf {\bibinfo {volume} {7}},\ \bibinfo {pages} {10530} (\bibinfo {year} {2019})}\BibitemShut {NoStop}%
\bibitem [{\citenamefont {Miku{\l}ko}\ \emph {et~al.}(2009)\citenamefont {Miku{\l}ko}, \citenamefont {Arora}, \citenamefont {Glushchenko}, \citenamefont {Lapanik},\ and\ \citenamefont {Haase}}]{mikulko2009complementary_55}%
  \BibitemOpen
  \bibfield  {author} {\bibinfo {author} {\bibfnamefont {A.}~\bibnamefont {Miku{\l}ko}}, \bibinfo {author} {\bibfnamefont {P.}~\bibnamefont {Arora}}, \bibinfo {author} {\bibfnamefont {A.}~\bibnamefont {Glushchenko}}, \bibinfo {author} {\bibfnamefont {A.}~\bibnamefont {Lapanik}}, \ and\ \bibinfo {author} {\bibfnamefont {W.}~\bibnamefont {Haase}},\ }\href@noop {} {\bibfield  {journal} {\bibinfo  {journal} {Europhysics Letters}\ }\textbf {\bibinfo {volume} {87}},\ \bibinfo {pages} {27009} (\bibinfo {year} {2009})}\BibitemShut {NoStop}%
\bibitem [{\citenamefont {Dierking}(2003)}]{dierking2003textures_56}%
  \BibitemOpen
  \bibfield  {author} {\bibinfo {author} {\bibfnamefont {I.}~\bibnamefont {Dierking}},\ }\href@noop {} {\emph {\bibinfo {title} {Textures of liquid crystals}}}\ (\bibinfo  {publisher} {John Wiley \& Sons},\ \bibinfo {year} {2003})\BibitemShut {NoStop}%
\bibitem [{\citenamefont {Haller}(1975)}]{haller1975thermodynamic_57}%
  \BibitemOpen
  \bibfield  {author} {\bibinfo {author} {\bibfnamefont {I.}~\bibnamefont {Haller}},\ }\href@noop {} {\bibfield  {journal} {\bibinfo  {journal} {Progress in solid state chemistry}\ }\textbf {\bibinfo {volume} {10}},\ \bibinfo {pages} {103} (\bibinfo {year} {1975})}\BibitemShut {NoStop}%
\bibitem [{\citenamefont {Prasad}\ and\ \citenamefont {Das}(2010)}]{prasad2010refractive_58}%
  \BibitemOpen
  \bibfield  {author} {\bibinfo {author} {\bibfnamefont {A.}~\bibnamefont {Prasad}}\ and\ \bibinfo {author} {\bibfnamefont {M.~K.}\ \bibnamefont {Das}},\ }\href@noop {} {\bibfield  {journal} {\bibinfo  {journal} {Phase Transitions}\ }\textbf {\bibinfo {volume} {83}},\ \bibinfo {pages} {1072} (\bibinfo {year} {2010})}\BibitemShut {NoStop}%
\bibitem [{\citenamefont {Vita}\ \emph {et~al.}(2016)\citenamefont {Vita}, \citenamefont {Hegde}, \citenamefont {Portale}, \citenamefont {Bras}, \citenamefont {Ferrero}, \citenamefont {Samulski}, \citenamefont {Francescangeli},\ and\ \citenamefont {Dingemans}}]{vita2016molecular_59}%
  \BibitemOpen
  \bibfield  {author} {\bibinfo {author} {\bibfnamefont {F.}~\bibnamefont {Vita}}, \bibinfo {author} {\bibfnamefont {M.}~\bibnamefont {Hegde}}, \bibinfo {author} {\bibfnamefont {G.}~\bibnamefont {Portale}}, \bibinfo {author} {\bibfnamefont {W.}~\bibnamefont {Bras}}, \bibinfo {author} {\bibfnamefont {C.}~\bibnamefont {Ferrero}}, \bibinfo {author} {\bibfnamefont {E.~T.}\ \bibnamefont {Samulski}}, \bibinfo {author} {\bibfnamefont {O.}~\bibnamefont {Francescangeli}}, \ and\ \bibinfo {author} {\bibfnamefont {T.}~\bibnamefont {Dingemans}},\ }\href@noop {} {\bibfield  {journal} {\bibinfo  {journal} {Soft Matter}\ }\textbf {\bibinfo {volume} {12}},\ \bibinfo {pages} {2309} (\bibinfo {year} {2016})}\BibitemShut {NoStop}%
\bibitem [{\citenamefont {Ramakrishna}\ \emph {et~al.}(2010)\citenamefont {Ramakrishna}, \citenamefont {Rao}, \citenamefont {Prasad},\ and\ \citenamefont {Pisipati}}]{ramakrishna2010orientational_60}%
  \BibitemOpen
  \bibfield  {author} {\bibinfo {author} {\bibfnamefont {M.}~\bibnamefont {Ramakrishna}}, \bibinfo {author} {\bibfnamefont {N.}~\bibnamefont {Rao}}, \bibinfo {author} {\bibfnamefont {P.~D.}\ \bibnamefont {Prasad}}, \ and\ \bibinfo {author} {\bibfnamefont {V.}~\bibnamefont {Pisipati}},\ }\href@noop {} {\bibfield  {journal} {\bibinfo  {journal} {Molecular Crystals and Liquid Crystals}\ }\textbf {\bibinfo {volume} {528}},\ \bibinfo {pages} {49} (\bibinfo {year} {2010})}\BibitemShut {NoStop}%
\bibitem [{\citenamefont {Francescangeli}\ \emph {et~al.}(2009)\citenamefont {Francescangeli}, \citenamefont {Stanic}, \citenamefont {Torgova}, \citenamefont {Strigazzi}, \citenamefont {Scaramuzza}, \citenamefont {Ferrero}, \citenamefont {Dolbnya}, \citenamefont {Weiss}, \citenamefont {Berardi}, \citenamefont {Muccioli} \emph {et~al.}}]{francescangeli2009ferroelectric_61}%
  \BibitemOpen
  \bibfield  {author} {\bibinfo {author} {\bibfnamefont {O.}~\bibnamefont {Francescangeli}}, \bibinfo {author} {\bibfnamefont {V.}~\bibnamefont {Stanic}}, \bibinfo {author} {\bibfnamefont {S.~I.}\ \bibnamefont {Torgova}}, \bibinfo {author} {\bibfnamefont {A.}~\bibnamefont {Strigazzi}}, \bibinfo {author} {\bibfnamefont {N.}~\bibnamefont {Scaramuzza}}, \bibinfo {author} {\bibfnamefont {C.}~\bibnamefont {Ferrero}}, \bibinfo {author} {\bibfnamefont {I.~P.}\ \bibnamefont {Dolbnya}}, \bibinfo {author} {\bibfnamefont {T.~M.}\ \bibnamefont {Weiss}}, \bibinfo {author} {\bibfnamefont {R.}~\bibnamefont {Berardi}}, \bibinfo {author} {\bibfnamefont {L.}~\bibnamefont {Muccioli}},  \emph {et~al.},\ }\href@noop {} {\bibfield  {journal} {\bibinfo  {journal} {Advanced Functional Materials}\ }\textbf {\bibinfo {volume} {19}},\ \bibinfo {pages} {2592} (\bibinfo {year} {2009})}\BibitemShut {NoStop}%
\bibitem [{\citenamefont {Shen}\ \emph {et~al.}(2011)\citenamefont {Shen}, \citenamefont {Tang},\ and\ \citenamefont {Wang}}]{shen2011spectral_62}%
  \BibitemOpen
  \bibfield  {author} {\bibinfo {author} {\bibfnamefont {J.}~\bibnamefont {Shen}}, \bibinfo {author} {\bibfnamefont {T.}~\bibnamefont {Tang}}, \ and\ \bibinfo {author} {\bibfnamefont {L.-L.}\ \bibnamefont {Wang}},\ }\href@noop {} {\emph {\bibinfo {title} {Spectral methods: algorithms, analysis and applications}}},\ Vol.~\bibinfo {volume} {41}\ (\bibinfo  {publisher} {Springer Science \& Business Media},\ \bibinfo {year} {2011})\BibitemShut {NoStop}%
\bibitem [{\citenamefont {Majumdar}(2010)}]{majumdar2010equilibrium_63}%
  \BibitemOpen
  \bibfield  {author} {\bibinfo {author} {\bibfnamefont {A.}~\bibnamefont {Majumdar}},\ }\href@noop {} {\bibfield  {journal} {\bibinfo  {journal} {European Journal of Applied Mathematics}\ }\textbf {\bibinfo {volume} {21}},\ \bibinfo {pages} {181} (\bibinfo {year} {2010})}\BibitemShut {NoStop}%
\end{thebibliography}%

\end{document}


\maketitle

\section*{Small-angle X-ray scattering (SAXS) data}
For conclusive proof disordered cybotactic cluster, X-ray measurements have now been performed for the pure 8-F-OH and doped 8-F-OH + 0.2 $ wt\%$ BiFeO$_3$ systems, maintaining the experimental condition exactly the same for both the systems, to check the disordered cybotactic clustering in doped systems. The small-angle X-ray scattering (SAXS) patterns show that the intensity of the peak increases upon lowering the temperature in both pure and doped 8-F-OH indicating the presence of ordered clusters in the nematic phase (N$_{cyb}$ phase). However, the intensity of the doped 8-F-OH system is comparatively less than that of the pure compound for the same X-ray exposure time i.e., 15 min and at the same temperature. For instance, at 105 $^{\circ}{\rm C}$, the intensity of the doped system is 14 a.u. and the pure compound is 34 a.u. suggesting that the cybotactic clustering is distorted upon dispersion of BiFeO$_3$ nanoparticles ( Figure S1). 

In both pure and doped 8-F-OH compounds, the layer spacing (d-value) and the correlation length ($\xi$) gradually increases in the N$_{cyb}$ phase region (Figure S2) upon lowering the temperature. It is well documented that the increase in the size of the smectic-type clusters results in relatively long correlation lengths. Interestingly, the correlation length values are comparatively lower for the doped system. For instance, at 135 $^{\circ}{\rm C}$, the $\xi$ value for the doped 8-F-OH ($\xi$ = 61.77 $\AA$) is less than the pure 8-F-OH ($\xi$ = 70.49 $\AA$) plausibly due to the deformation of the cybotactic clusters after the dispersion of BiFeO3 nanoparticles. From this SAXS experiments it is quite evident that the BiFeO$_3$ nanoparticle at higher temperature inevitably destroys the cybotactic clusters via interacting with the bent-core LC systems while upon lowering the temperature some regrowth of clusters was observed

\begin{figure}[!htb]
  \centering
    \includegraphics[width=0.49\textwidth]{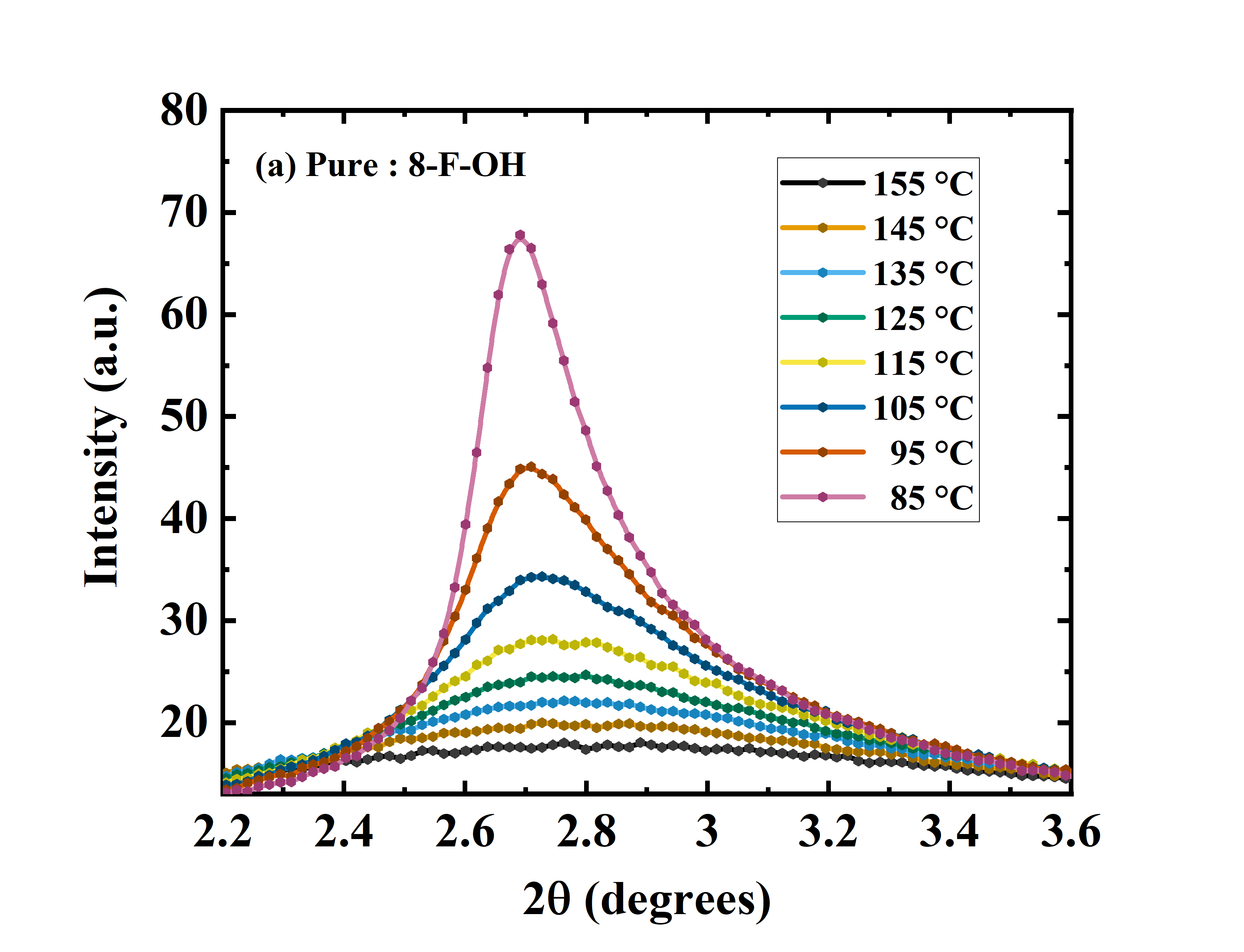}
    \includegraphics[width=0.49\textwidth]{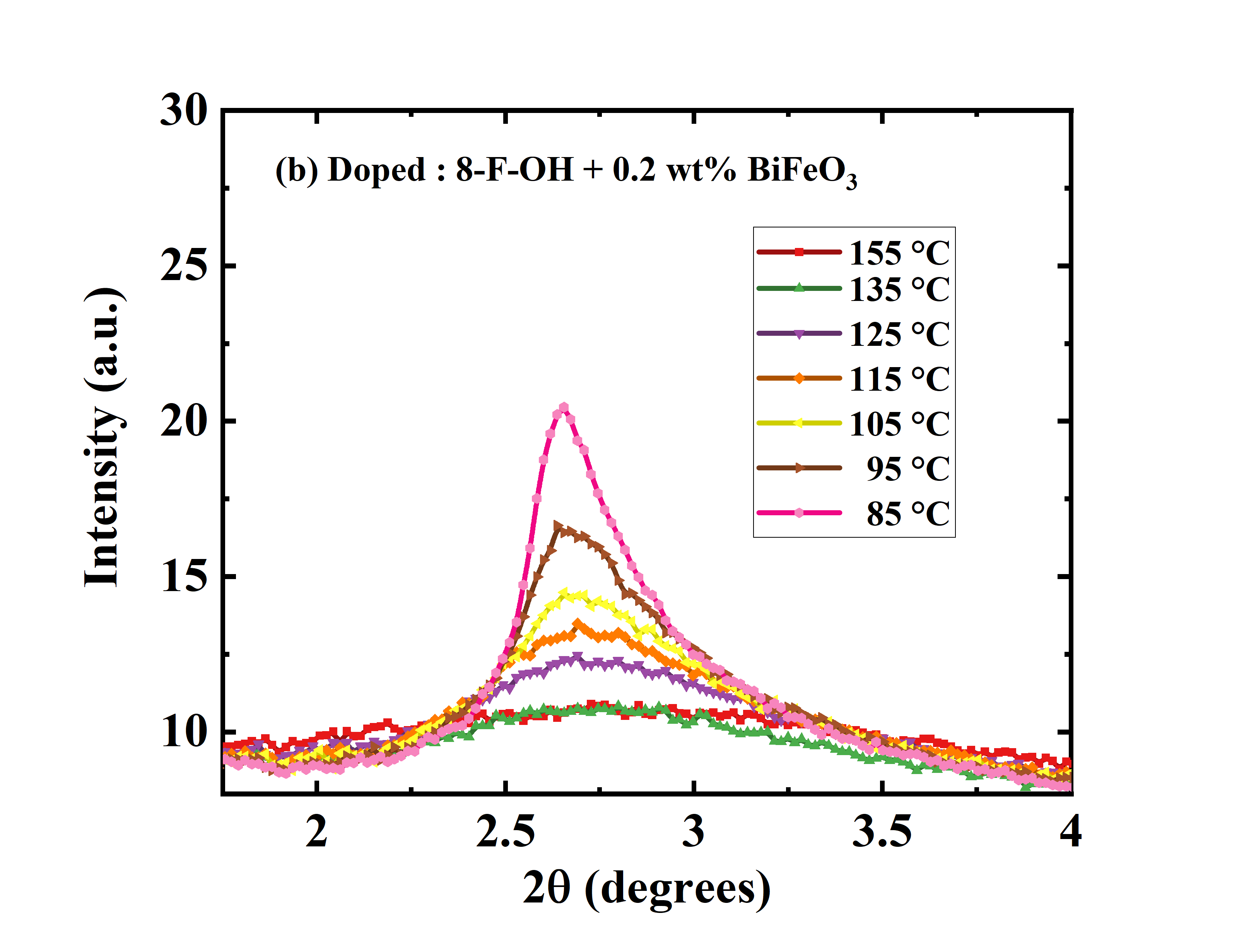} 
  \caption{Temperature-dependent XRD patterns in the small-angle region: Intensity vs. 2$\theta$ plots of the (a) pure compound 8-F-OH and (b) doped 8-F-OH + 0.2 $ wt\%$ BiFeO$_3$ system}
\end{figure}

\begin{figure}[!htb]
  \centering
    \includegraphics[width=0.49\textwidth]{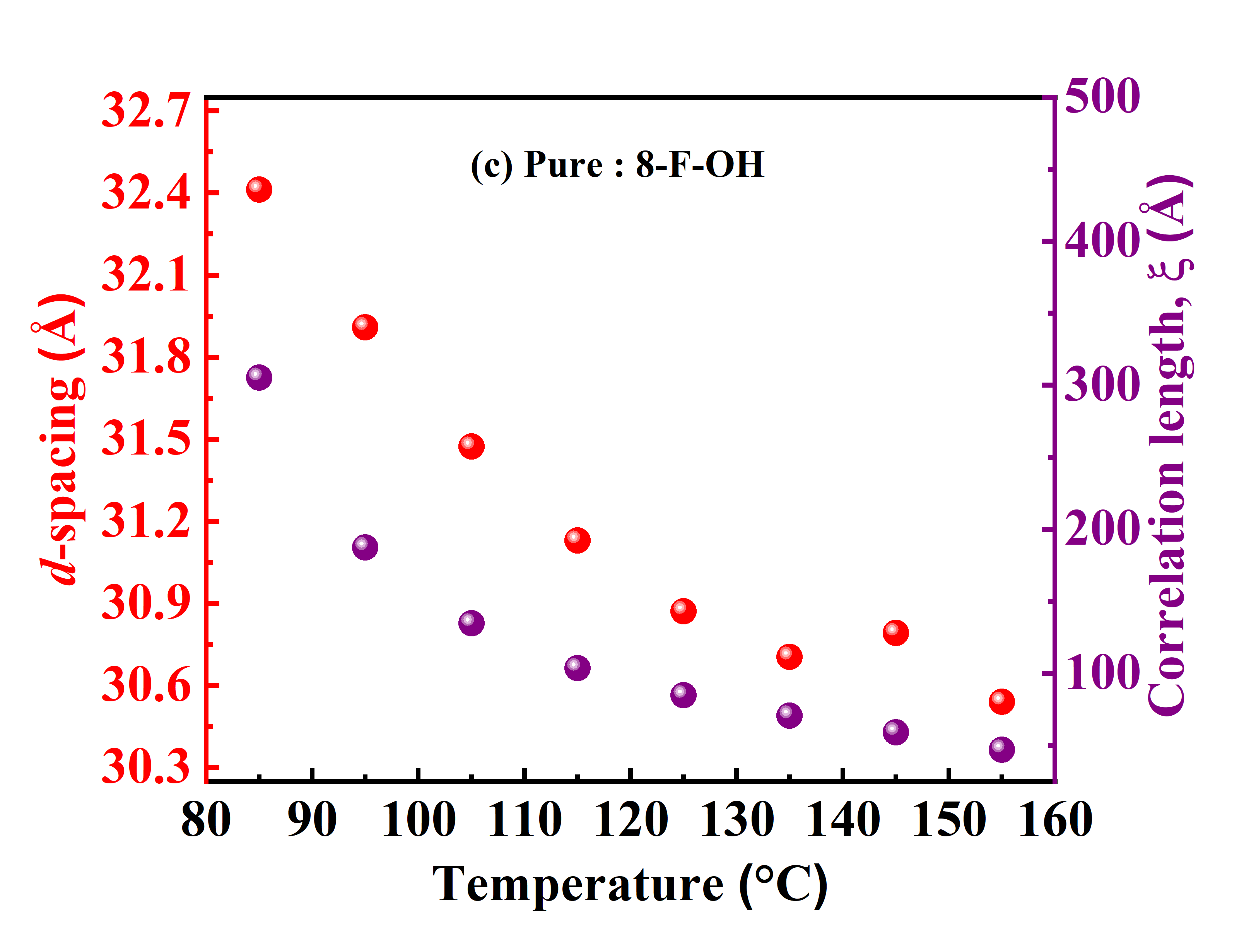}
    \includegraphics[width=0.49\textwidth]{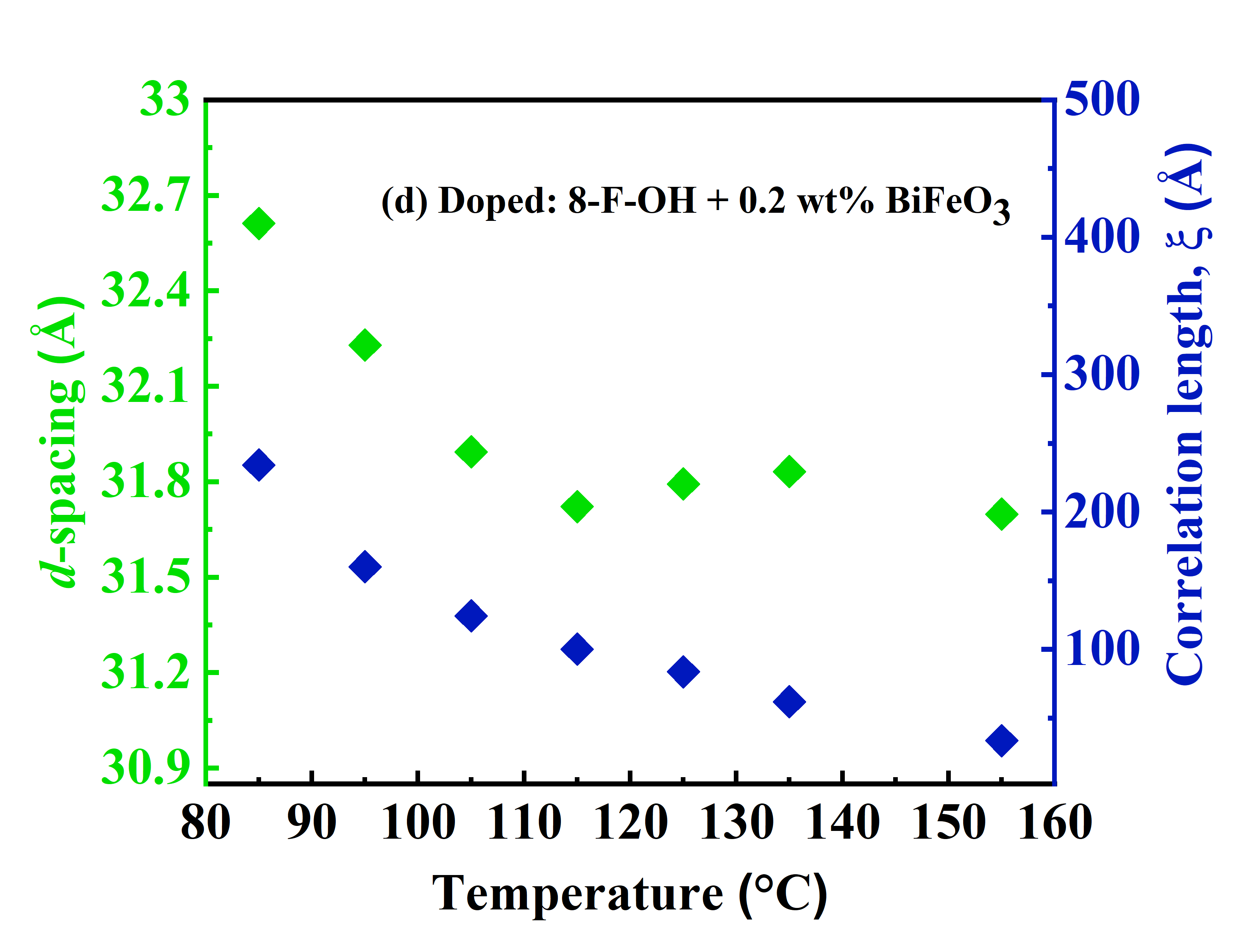} 
  \caption{Variation of correlation length ($\xi$) and d-spacing with temperature of (c) pure 8-F-OH and (d) doped 8-F-OH + 0.2 $ wt\%$ BiFeO$_3$ system.}
\end{figure}